\documentclass[aps,prb,preprint]{revtex4}
\usepackage{amsmath,amssymb,epsfig}
\usepackage{color}

\begin{document}
  

\title{Exact ground states of correlated electrons on
pentagon chains}
\author{Zsolt~Gul\'acsi}
\address{
Department of Theoretical Physics, University of Debrecen, 
H-4010 Debrecen, Hungary }
\date{May 21, 2013}
\begin{abstract}
We construct a class of exact ground states for correlated
electrons on pentagon chains in the high density 
region and discuss their  physical properties. In this 
procedure the Hamiltonian is first cast in a positive 
semidefinite form using composite operators as a linear 
combination of creation operators acting on the sites of 
finite blocks. In the same step, the interaction is
also transformed to obtain terms which require for their 
minimum eigenvalue zero at least one electron on each site.
The transformed Hamiltonian matches the original 
Hamiltonian through a nonlinear system of equations whose 
solutions place the deduced ground states in restricted 
regions of the parameter space. In the second step, 
nonlocal product wave functions in position space are 
constructed. They are proven to be unique ground states 
which describe non-saturated ferromagnetic and correlated 
half metallic states. These solutions emerge when the 
strength of the Hubbard interaction $U_i$ is site dependent 
inside 
the unit cell. In the deduced phases, the interactions 
tune the bare dispersive band structure such to develop an
effective upper flat band. We show that this band 
flattening 
effect emerges for a broader class of chains and is not
restricted to pentagon chains. For the characterization of 
the
deduced solutions, uniqueness proofs, 
exact ground state expectation values 
for long-range hopping amplitudes and correlation functions
are also calculated.  
    
The study of physical reasons which lead to the appearance 
of ferromagnetism has revealed a new mechanism for the 
emergence of an ordered phase, described here in details.
This works as follows: Starting from a completely 
dispersive bare band structure, the interactions quench 
the kinetic energy, hence the ordered phase is obtained
solely by a drastic decrease of the interaction energy. 
Since $U_i$ are site dependent,
this determinative decrease is obtained by a redistribution
of the double occupancy $d_i$ such to attain small $d_i$
where the on-site Coulomb repulsion $U_i$ is high, and 
vice versa. The kinetic energy quench leads to the upper 
effective flat band, whose role is to enhance by its 
degeneracy the switching to the ordered phase dictated and
stabilized by the interactions present. It is shown that 
this phenomenon to occur, a given degree of complexity is
needed for the chain, and the mechanism becomes inactive 
when the $U_i$ interactions are homogeneous, or 
are missing from the ground state wave function.   
\end{abstract}
\pacs{71.10.Fd, 71.27.+a, 71.20.Rv} \maketitle


\section{Introduction}

\subsection{The model systems}

Conducting polymers are an important class of organic 
materials with a wide range of potential applications 
from nanoelectronics \cite{Intr1} to medicine \cite{Intr2}. 
Many of these polymer chains contain five-membered rings 
\cite{Intr1,Intr3,Intr4}. In particular polythiophene 
\cite{Intr6,Intr7,Intr8} was analyzed in the search for 
plastic ferromagnets and for ferromagnetism in systems 
made entirely from nonmagnetic elements. 

The theoretical investigation of ferromagnetism in pentagon
chains was started \cite{Intr9,Intr10} with a focus on 
flat-band ferromagnetism \cite{Intr11}. Suwa et al. 
\cite{Intr12} proposed that the ferromagnetism 
in pentagon chains is related to the hybridization of 
$\sigma$ and $\pi$ bands and hence described these systems by 
a periodic Anderson model (PAM) relying on the previously
established ferromagnetism in the PAM in various conditions
\cite{v1,Intr13}. In the PAM used in 
Ref.11
the electronic interaction acts site selective within the 
unit cell. The model of Suwa et al. \cite{Intr12} was 
therefore an attempt to account for the different atoms 
in the unit cell of the pentagon chains. We will also 
account for this fact by allowing for different on-site 
Coulomb repulsion values on different type of sites inside 
the unit cell. Furthermore, since the pentagon chain 
polymers always have external side groups, the pentagon 
chains we choose to describe also contain such external 
links. For illustration, a schematic view of the 
pentagon chain polyaminotriazole is presented in Fig. 1.

In the early stage of the theoretical work on 
conducting polymers, especially on properties related to
polyacetylene, the electronic correlations were not 
considered important \cite{Intr14}. In recent years it
become however clear that e.g. in pentagon or hexagon 
repeat units in conducting organic materials, the Coulomb 
interaction between the carriers is important; the 
on-site Coulomb repulsion value may even be as large as
$10 eV$ \cite{Intr15}. For n-acenes and n-thiophenes type of
chain structures, Brocks et al. 
emphasized strong correlation effects in acenes and thiophenes
and even conjectured that in the high-density region the
Coulomb interaction should be able to stabilize magnetic 
order \cite{brocks}.  

To account accurately for correlation effects, we will
here use exact methods. Details of the technique are 
presented below.          

\begin{figure}
\centerline{\includegraphics[width=6 cm,height=4 cm]{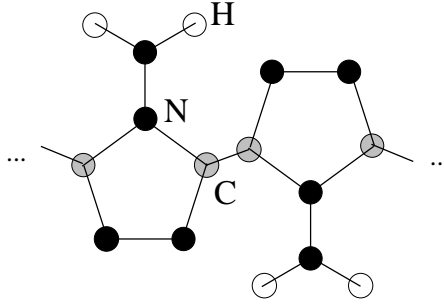}}
\caption{A sequence of the polyaminotriazole chain} 
\label{fig:1}
\end{figure}

\subsection{Technical steps}

\subsubsection{Basic characteristics of the method}

The spectrum of an arbitrary Hamiltonian $\hat H$ 
describing a real physical system is
bounded below by the ground state energy $E_g$. Hence, 
\begin{eqnarray}
\hat H'=\hat H-E_g = \hat O_p.
\label{I1o}
\end{eqnarray}
$\hat O_p$ is a positive semidefinite operator with
only non-negative eigenvalues $o_p \geq 0$, 
$\hat O_p |\phi\rangle = o_p |\phi\rangle$.
Hence, based on (\ref{I1o}), and without any prior 
knowledge about $E_g$, our technique starts by an exact 
rewriting of the starting Hamiltonian in the form
\begin{eqnarray}
\hat H = \sum_i \hat O_{p,i} + C_{g,H} = \hat O_p + C_{g,H},
\label{I1}
\end{eqnarray}
where the c-number $C_{g,H}$ is a function of the
Hamiltonian parameters, and $\hat O_{p,i}$ are positive 
semidefinite operators. The sum $\sum_i \hat O_{p,i}$ is
therfore also a positive semidefinite operator.
In the present case we will use block operators for 
$\hat O_{p,i}$ which are composed of 
linear combinations of  
operators acting on the sites of a finite block, but also 
apply operators which require at least one electron on 
each site of the lattice. Since (\ref{I1o}) always holds, 
the rewriting (\ref{I1}) is in principle always possible,
hence the method used here is independent on the spatial 
dimensions and does not require the integrability of the 
model Hamiltonian. The reason why (\ref{I1}) merits 
special attention is that the deduction of the ground
state $|\Psi_g\rangle$ can be performed by constructing 
the most general wave vector which satisfies the equation 
\begin{eqnarray}
\hat O_p |\Psi_g\rangle =0.
\label{I1oo}
\end{eqnarray}  
For solving (\ref{I1oo}) several techniques have been 
worked out  
\cite{v1,PA1,SC1,Orlik1,gurin,SC2,SC3,g1,m2010,m2007,m2008},
which open a route for deducing exact results for 
non-integrable systems.

Usually the rewriting of the Hamiltonian in positive
semidefinite form (\ref{I1}) can be achieved in several 
different ways depending on the type and the structure of 
the $\hat O_{p,i}$ operators. Each choice provides 
matching equations, which ensure the validity of 
(\ref{I1}), and make explicit how the parameters of $\hat H$
transform into the parameters of the positive semidefinite 
operators and $C_{g,H}$. The solutions of the matching 
equations place each transformation in different domains 
${\cal{D}}$ of the parameter space, where also the 
expression for $C_{g,H}$ follows and where the obtained 
ground state solution $|\Psi_g\rangle$ with the 
ground state energy $E_g=C_{g,H}$ is valid. 

In the process of deducing $|\Psi_g\rangle$, the total
number of particles $N$ is kept fixed. If the ground state 
$|\Psi_g(N)\rangle$ and its energy $E_g(N)$ are obtained
in a particle number dependent fashion, it is also 
possible to obtain information about the low lying 
excitation spectrum, for example via
$\delta \mu =[E_g(N+1)-E_g(N)]-[E_g(N)-E_g(N-1)]$, where
$\delta \mu \ne 0$ ($\delta \mu = 0$) signals the presence 
(absence) of a charge gap.

The construction of the most general ground state means in 
fact to deduce the unique ground state. At this point the 
degeneracy and the uniqueness are different notions. For 
example, the set $|\Psi_g(\gamma)\rangle$ of linearly 
independent wave functions, where $\gamma$ is a possible 
degeneracy index, represents the unique ground state, if 
the linear combination
$|\Psi_g\rangle = \sum_{\gamma} a_{\gamma} |\Psi_g(\gamma)
\rangle$, with arbitrary coefficients $a_{\gamma}$ 
spans the kernel $ker(\hat H')$, where 
$\hat H'=\hat H -C_{g,H}$. The kernel $ker(\hat O)$ of an 
operator $\hat O$, is the Hilbert subspace containing all 
wave vectors $|\Phi\rangle$ with the property 
$\hat O |\Phi\rangle =0$. The set $|\Psi_g(\gamma)\rangle$ 
spans the kernel of $\hat H'$, if two conditions are 
satisfied, namely: i)  $|\Psi_g(\gamma)\rangle
\in ker(\hat H')$ for all $\gamma$, and 
ii) all possible wave functions $|\Psi\rangle$ from 
$ker(\hat H')$ can be expressed in terms of 
$|\Psi_g(\gamma)\rangle$, i.e. 
$|\Psi\rangle = \sum_{\gamma} c_{\gamma} 
|\Psi_g(\gamma)\rangle$ holds with a unique set of 
coefficients $c_{\gamma}$. The proof of the uniqueness 
therfore requires to verify points i) and ii).

The exact ground states $|\Psi_g \rangle$ are deduced
for a fixed Hamiltonian, hence the used
technique is applied to a selected system without 
the need for prior information about $|\Psi_g\rangle$ 
and its physical properties. In these conditions, after 
obtaining $|\Psi_g\rangle$ its physical properties must 
be separately determined. This is done in a final step 
by calculating different expectation values with
the obtained ground state. 

Summarizing the procedure presented above, the applied 
method divides in five consecutive steps: 1) the 
transformation of the Hamiltonian in a positive 
semidefinite form, 2) the solution of the matching 
equations, 3) the construction of the ground states, 
4) the proof of the uniqueness, and 5) the deduction of 
physical properties. 
 
\subsubsection{How the technique evolved}

Although similar methods were applied even earlier \cite{s3},
the deduction of exact ground states of 
non-integrable systems using positive semidefinite 
operators really started after the 
paper of Brandt and Giesekus was published in 1992 
\cite{s8}. But the procedure at that time was 
different: for a decorated lattice described by a Hubbard 
or a periodic Anderson type of model with a $U = \infty$ 
on-site interaction, the transformation of the Hamiltonian 
in positive semidefinite form allowed to deduce an exact 
lower bound $E_{g,l} \leq E_g$ for the ground state energy 
$E_g$. Based on the variational principle 
applied to a properly chosen trial wave function 
$|\phi_v\rangle$, an exact upper bound $E_{g,u} \geq E_g$ 
was obtained in a second step. Pushing these two bonds to
the same value (i.e. $E_{g,l}=E_g=E_{g,u}$) in a 
restricted domain ${\cal{D}}$ of the parameter space led 
the authors to the ground state energy, while the 
uniqueness of the ground state was not proven 
explicitly \cite{s8}. In the same manner the method was 
applied by Tasaki \cite{tas} to deduce a resonating 
valence bond type of ground state at $U=\infty$ in a 
Hubbard model on a decorated lattice, and by Strack 
\cite{s9} to construct ground states
for the periodic Anderson and the extended Emery models 
in $d=1$ and $2$ dimensions on a non-decorated lattice.
The uniqueness of some of these ground states was analyzed  by 
Tasaki \cite{tas1}, too. Other situations with an 
infinitely large on-site interaction were
also studied for the PAM \cite{PA1},
the Falicov-Kimball model \cite{SC1}, and
extended Hubbard models on decorated lattices \cite{UB2}.

Starting in 1993 also results at finite $U$ were obtained
for extended Hubbard models with a paramagnetic insulating
phase, charge density wave \cite{vol1}, or ferromagnetic 
ground states \cite{vol2,vol3}, and also
resonating valence bond type of ground states on 
decorated lattices; for the latter case also the
uniqueness was proven \cite{Tana}.  For the PAM,
the first finite $U$ results were 
published in 2001 for dimensions $d=1$ \cite{Orlik1}
and $d=2$ \cite{gurin}, and later also for $d=3$ 
\cite{v1}, in the latter case also with a uniqueness 
proof. Furthermore, studies of the extended PAM 
\cite{SC2}, superconducting states in 
various circumstances \cite{SC3,MC,Laugh,SC4}, and spin 
systems \cite{s4} were performed, and technical 
advances accomplished \cite{g1}.  

Until say 2000 knowledge was accumulated how to transcribe
a given operator in positive semidefinite form. Despite 
some exceptions \cite{s9,vol1,gurin}, the method 
concentrated on known and fixed wave functions 
$|\Psi_g\rangle$ of superconducting or ferromagnetic 
ground states. This strategy enlarged considerably the 
knowledge of phase diagram regions where these phenomena 
appear (e.g. for ferromagnetism see 
Ref.35), but 
in this framework it was difficult to analyze an a priori 
unknown ground state of a fixed Hamiltonian. This led 
to a misconception about the essence of the procedure 
which followed the sequence of steps \cite{s7}: start 
with a known ground state $|\Psi_N\rangle$ of a Hamiltonian
$\hat H_o$, add operators $\hat P^{\dagger}_i
\hat P_i$ such that $\hat P_i |\Psi_N\rangle=0$, and 
identify $\hat H = \hat H_o +\sum_i \hat P^{\dagger}_i
\hat P_i$ which also has $|\Psi_N\rangle$ as its ground
state. This procedure should be compared with the 
technical steps presented in Sect. I. B.1, which starts 
with the choice of the model Hamiltonian.

Indeed, these developments allow the applicability of the 
technique to a broad spectrum of topics covering 
thermodynamic properties \cite{hon}, topological 
characteristics \cite{top}, or selected Hamiltonians which
describe for example conducting polymers \cite{m2010}.  
The broad possibilities of the method allow also fixed 
system Hamiltonian studies without prior knowledge of the 
ground state, even in disordered case or presence of
textures \cite{disord}.

We use here the technique of positive semidefinite 
operators for a fixed model Hamiltonian, performing the 
characteristic steps as described in subsection I.B.1.
Along this line the method has been successfully applied 
previously to describe the 3D PAM
\cite{v1} and Hubbard chains with 
different geometrical structures \cite{m2007,m2008,m2010}.
The main challenge of this situation is to avoid
arbitrary supplementary contributions in $\hat H$ in order 
to obtain positive semidefinite expressions in the 
Hamiltonian; rather the transformation step is done 
starting from a fixed Hamiltonian $\hat H$.

On the technical level the advantage of a positive 
semidefinite Hamiltonian is utilized in other methods as
well \cite{Laugh,sx5,sy5,s5}. For example the optimal 
ground state method was first
introduced for spin models \cite{s5}, subsequently used
for itinerant systems as well \cite{s7,s6}, and
applied also in various other circumstances
\cite{Alb,Batista}. In this procedure the Hamiltonian $\hat H$ 
is written as a sum over minimal cluster 
contributions taken in the majority of cases over a bond
$\hat H=\sum_{<{\bf i},{\bf j}>}\hat h_{{\bf i},{\bf j}}$, 
where $<{\bf i},{\bf j}>$ are nearest-neighbor lattice 
sites. The operator $\hat h_{{\bf i},{\bf j}}$ is
diagonalized exactly in the Hilbert space of the two-site
cluster. By adding a constant to the Hamiltonian, it is 
achieved that the lowest eigenvalue $\epsilon_0$ of 
$\hat h_{{\bf i},{\bf j}}$ becomes zero ($\epsilon_0=0$), 
hence $\hat h_{{\bf i},{\bf j}}$ is a positive semidefinite
operator, with the ground state 
$|\phi_0({\bf i},{\bf j})\rangle$ satisfying 
$\hat h_{{\bf i},{\bf j}}|\phi_0({\bf i},{\bf j})\rangle=0$
for all bonds. If the global ground state
$|\Phi_0\rangle$ can be constructed from the local ground 
states $|\phi_0({\bf i},{\bf j})\rangle$ (for example
as a product state), one finds the optimal ground state. 
The method is usually applied with some 
{\it a priori} information regarding the ground state 
which might emerge. In principle, the stability conditions 
for the emergence of a given phase can often be improved 
and extended by increasing the minimal cluster size 
\cite{s6}.

In contrast to the optimal ground state method, our 
procedure decomposes the Hamiltonian in several different 
and non-equivalent positive semidefinite block 
contributions. Hence the diagonalization cannot be done at 
the level of one block Hilbert space, but only for the
Hilbert space of the whole system (and in this process 
often operators emerge which cover the whole lattice
\cite{T1}).
This approach allows for a more efficient deduction 
process of unknown ground states without prior
knowledge. The deduction of such type of ground states is
therefore beyond the optimal ground state method.

\subsection{Task and results}

In the present paper we investigate pentagon chain 
polymers (see the example in Fig. 1) where the electrons 
experience local Coulomb interactions on all sites; the 
on-site interactions -- according to the particular 
environment and type of atom -- are permitted to differ on 
individual sites. The hopping parameters and the 
one-particle on-site potentials are prohibited to take 
those special values which lead to flat bands in the bare 
band structure. In these conditions we show rigorously 
that the dispersion of the correlated systems may be 
tuned by the interaction to become flat. Thereby  
ferromagnetic or half metallic states at high electron 
densities emerge. Consequently, Brocks et al.'s 
\cite{brocks} conjecture at the two-particle level, that 
the Coulomb interaction stabilizes magnetic order in acene 
and thiophene is proven at the many-body level in exact 
terms. We demonstrate that the band-flattening effect of 
the interaction can occur in a large class of polymers, 
exceeding considerably the frame of pentagon chains.
The paper not only presents details of the previously
published Letter \cite{m2010}, but also provides further 
information regarding the described procedure for the 
construction of exact ground states. Furthermore, we 
explicitly demonstrate the uniqueness of the solutions 
in different parameter regimes, calculate the long-range 
hopping ground state expectation values, 
present exact results for spin-spin and density-density 
correlation functions, and analyze in details the
physical origin of the deduced ordered phases. 

In different approximations, band-flattening effects given 
by interactions on restricted ${\bf k}$ domains have been 
reported previously in the literature 
\cite{khodel,nozieres}. Band flattening due to
interactions has been observed as well in the case of the 
PAM, where the same $U$ acts on each 
site, but is effective only for one band \cite{gurin,v1}.
Instead, the here reported and exactly described cases
produce the band-flattening effect in polymer structures 
over the whole first Brillouin zone but requires different 
U values on inequivalent lattice sites. This choice accounts  
for the particular environment and type of atom on a paricular
site inside the unit cell.     

\subsection{Physical reasons for the emergence of 
ferromagnetism}

The physical reasons for the emergence of the ordered phase
have been analyzed with full particulars. The results show 
that 
we have found, and in our knowledge, we describe in details 
for the first time a new mechanism leading to the emergence
of ordered phases. The valuable nature of the 
conclusions is guaranted by the unapproximated nature of
the deduced results. 
The mechanism works as follows: starting 
from a completely dispersive bare band structure, the 
interactions quenches the kinetic energy -- roughly at 
the value
present in the non-interacting case -- by producing an 
effective
upper flat band. This has two advantages. First, 
the flat nature
of the effective band, by its huge degeneracy, 
preserves the 
possibility for the system to swich easily in the ordered 
state 
dictated by the interactions present. Second, the kinetic 
energy
quench, allows to reach the ordered phase, exclusively by a 
drastic decrease of the interaction energy. 

Particularly, in the present case when as interactions, 
different on-site Coulomb repulsion values 
$U_{\bf j}$ are present on different type of 
sites ${\bf j}$, the interaction energy
decrease can be performed with extreme 
efficiency (e.g. even 70 \%, see Sect.VIII.C.2). This is
done by a redistribution of the double occupancy 
$d_{\bf j}=\langle \hat n_{{\bf j},\uparrow}\hat n_{{\bf j},
\downarrow} \rangle = \langle \hat d_{\bf j} \rangle$ 
such to obtain high (low) $d_{\bf j}$ 
where $U_{\bf j}$ is low (high).
Such decrease possibility of the interaction energy is not
present when $U_{\bf j}$ 
is homogeneous, and requires as well
besides a given degree of complexity for the chain,
the presence of the interaction in the ground state wave 
function. Since in the present case the determinative 
decrease
of the interaction energy diminishes the average number of
double occupied sites, the ground state becomes 
ferromagnetic at
half (or above half) filled upper effective band. 

The studies made on the lowest unit which
provides the ferromagnetism -- namely a two-cell system with
periodic boundary conditions -- shows that the emergence of
the ferromagnetism as described above, automatically leads 
to a) one electron with fixed spin projection on all sites,
and  b) effective upper flat band at the interpretation 
of the results. We have also shown that given by a), 
when doped, the system becomes a half metal.    

The remaining part of the paper is structured as follows. 
Sect. II presents the studied system, Sect. III describes
the transformation in positive semidefinite form, and
Sect. IV presents the deduced ground states. In Sect. V 
the long-range hopping ground-state expectation values are
calculated. Sect. VI analyzes the correlation functions,
Sect. VII demonstrates the uniqueness of the solutions, and
Sect. VIII analyzes in details the physical reasons of the
emergence of ferromagnetism.
Sect. IX contains the summary and the conclusion.
Mathematical details are summarized in six Appendices A - F.

\begin{figure}
\centerline{
\includegraphics[width=9 cm,height=3 cm]{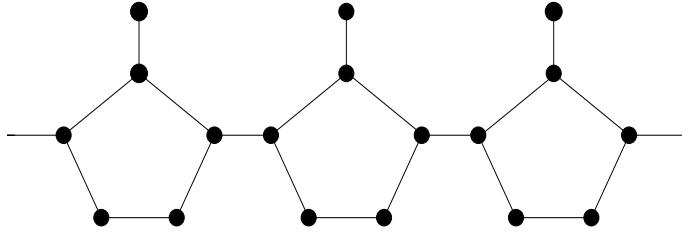}}
\caption{The pentagon chain with external links.} 
\label{fig:2}
\end{figure}

\section{The system studied}

\subsection{The Hamiltonian in ${\bf r}$ space}

We start from a general chain containing $m$ sites
in the unit cell defined at an arbitrary lattice site 
${\bf i}$. These sites are placed at ${\bf i}+{\bf r}_n$, 
$n \leq m$. Hence the Hamiltonian 
$\hat H = \hat H_0 + \hat H_U$ has the form
\begin{eqnarray}
&&\hat H_0= \sum_{{\bf i},\sigma} \sum_{n,n'(n>n')}
(t_{n,n'}
\hat c^{\dagger}_{{\bf i}+{\bf r}_n,\sigma} 
\hat c_{{\bf i}+{\bf r}_{n'},
\sigma} + H.c.) + \sum_{{\bf i},\sigma,n} \epsilon_n 
\hat n_{{\bf i}+
{\bf r}_n, \sigma},
\nonumber\\
&&\hat H_U= \sum_{\bf i} \sum_{n=1}^m U_n \hat n_{
{\bf i}+{\bf r}_n, 
\uparrow} \hat n_{{\bf i}+{\bf r}_n,\downarrow},
\label{EQU1}
\end{eqnarray}
where $t_{n,n'}$ are nearest-neighbor hopping matrix 
elements connecting the sites ${\bf i}+{\bf r}_{n'}$ and 
${\bf i}+{\bf r}_n$. Furthermore $\epsilon_n$ and $U_n>0$ 
are on-site potentials and on-site Coulomb repulsions, 
respectively, defined at the sites ${\bf i}+{\bf r}_n$. 
We analyze chains with $m=m_p+m_e \geq 2$ 
sites in the unit cell with $m_p$ sites  
placed in a closed polygon, and $m_e$ sites in external 
links (i.e. placed in side groups). Neighboring cells 
connect through the single wector
${\bf r}_{m+1}={\bf a}$, where ${\bf a}$ is the 
Bravais vector of the chain, and for simplicity we
choose $|{\bf r}_1|=0$. We note that $t_{n,n'}$ and 
$\epsilon_n$ are arbitrary at this point ($\epsilon_n=0$ 
for all $n$ is also allowed), and are chosen such that
all free-electron bands are dispersive. Furthermore,
$\hat n_{{\bf j},\sigma}=\hat c^{\dagger}_{{\bf j},\sigma}
\hat c_{{\bf j},\sigma}$, where $\hat c^{\dagger}_{{\bf j},
\sigma}$ creates an electron with spin $\sigma$ at the 
site ${\bf j}$.    

In the frame of the Hamiltonian (\ref{EQU1}) we analyze
below pentagon chains as in Fig. 2. The unit cell is 
depicted in detail in Fig. 3. In this specific case
$m_p=5$, $m_e=1$, and $m=6$. Introducing the notations 
$t_c=t_{4,7}, t_h=t_{3,2}, t_f=t_{5,6}, 
t=t_{1,5}=t_{4,5}=t_{2,1}=t_{3,4}$ for the hopping 
amplitudes, the studied Hamiltonian takes the form
\begin{eqnarray}
\hat H_{0} &=& \sum_{\sigma} \sum_{{\bf i}} \: 
\big\{ \: \big[ \: t_f
\hat c^{\dagger}_{{\bf i}+{\bf r}_6,\sigma} 
\hat c_{{\bf i}+{\bf r}_5,\sigma} +
t_c \hat c^{\dagger}_{{\bf i}+{\bf r}_4,\sigma} 
\hat c_{{\bf i}+{\bf a},
\sigma} + t_h \hat c^{\dagger}_{{\bf i}+{\bf r}_2,\sigma} 
\hat c_{{\bf i}+{\bf r}_3,\sigma}+
\nonumber\\
t \big(
\hat c^{\dagger}_{{\bf i}+{\bf r}_5,\sigma} 
\hat c_{{\bf i},\sigma} 
&+&
\hat c^{\dagger}_{{\bf i},\sigma} \hat c_{{\bf i}+
{\bf r}_2,\sigma} +
\hat c^{\dagger}_{{\bf i}+{\bf r}_3,\sigma}\hat c_{
{\bf i}+{\bf r}_4,\sigma} +
\hat c^{\dagger}_{{\bf i}+{\bf r}_4,\sigma}\hat c_{
{\bf i}+{\bf r}_5,\sigma}
\big) + H.c. \big] + \sum_{n=1}^6 \epsilon_n 
\hat n_{{\bf i}+{\bf r}_n,\sigma} \: \big\},
\nonumber\\
\hat H_{U} &=& \sum_{{\bf i}} \sum_{n=1}^6
U_n \hat n_{{\bf i}+{\bf r}_n,\uparrow} \hat n_{
{\bf i}+{\bf r}_n,\downarrow}.
\label{EQU2}
\end{eqnarray}
where $N_c$ represents the number of cells. 
Throughout, $\sum_{\bf i},\prod_{\bf i}$ (or $\sum_{\bf k},
\prod_{\bf k}$ in momentum representation) mean sums and 
products, respectively, over $N_c$ cells.
Periodic boundary conditions are used, the number of 
electrons $N \leq 2 N_{\Lambda}=N_{Max}$ is considered 
fixed, 
where 
$N_{\Lambda}=m N_c$ is the total number of sites. The 
filling is controlled by 
$\rho=N/N_{\Lambda} \leq 2$.

\begin{figure}
\centerline{\includegraphics[width=6 cm,height=6 cm]{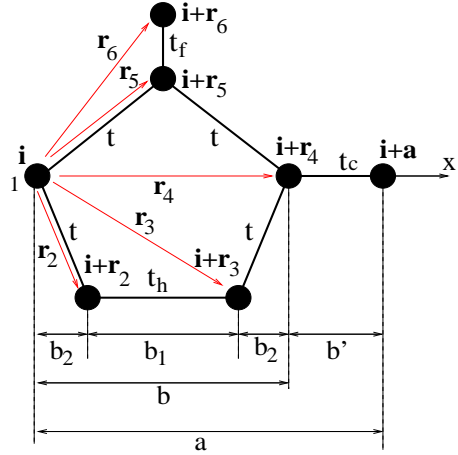}}
\caption{The pentagon unit cell. The site ${\bf i}+{\bf a}$ is
not part of the unit cell containing m=6 sites.} 
\label{fig:3}
\end{figure}

We further note that $m$ provides the number of bands 
as well $N_b=m$, and the maximum number of electrons 
accepted by a given fixed band is 
$N_{1,b}=N_{Max}/N_b=2 N_c$.

\subsection{$\hat {\bf H}_0$ in ${\bf k}$ space}

In order to translate the kinetic energy to ${\bf k}$ 
space, the fermionic operators
$\hat c_{{\bf i}+{\bf r}_n,\sigma}$ are Fourier transformed
via
$\hat c_{{\bf i}+{\bf r}_n,\sigma} = (1/\sqrt{N_c})$ 
$\sum_{{\bf k}} e^{-i{\bf k}\cdot
({\bf i}+{\bf r}_n)} \hat c_{n,{\bf k},\sigma}$,
where ${\bf k}$ is directed along the chain axis
($x$-axis in Fig. 3). Therefore
$|{\bf k}|= k = 2 m \pi/(a N_c)$, $m=0,1,2,...,N_c-1$, and
$|{\bf a}|=a$ is the lattice constant. 
The non-interacting part 
of the Hamiltonian (\ref{EQU2}) then becomes
\begin{eqnarray}
\hat H_0 &=& \sum_{\sigma} \sum_{{\bf k}} 
\: \big\{ \: \big[ \: 
t_f \hat c^{\dagger}_{6,{\bf k},\sigma} \hat c_{
5,{\bf k},\sigma} 
e^{i {\bf k}\cdot({\bf r}_6-{\bf r}_5)} +
t_c \hat c^{\dagger}_{4,{\bf k},\sigma} \hat c_{
1,{\bf k},\sigma} 
e^{i {\bf k}\cdot({\bf r}_4-{\bf a})} +
t_h \hat c^{\dagger}_{2,{\bf k},\sigma} \hat c_{
3,{\bf k},\sigma}
e^{i {\bf k}\cdot({\bf r}_2-{\bf r}_3)} + 
\nonumber\\
&+& t \big( 
\hat c^{\dagger}_{5, {\bf k},\sigma} \hat c_{
1, {\bf k},\sigma} 
e^{i {\bf k}\cdot({\bf r}_5-{\bf r}_1)} +
\hat c^{\dagger}_{1, {\bf k},\sigma} \hat c_{
2, {\bf k},\sigma} 
e^{i {\bf k}\cdot({\bf r}_1-{\bf r}_2)} +
\hat c^{\dagger}_{3, {\bf k},\sigma}\hat c_{
4, {\bf k},\sigma} 
e^{i {\bf k}\cdot({\bf r}_3-{\bf r}_4)} +
\hat c^{\dagger}_{4,{\bf k},\sigma}\hat c_{
5, {\bf k},\sigma}
e^{i {\bf k}\cdot({\bf r}_4-{\bf r}_5)} 
\big) 
\nonumber\\
&+& H.c. \big]
+ \sum_{n=1}^6 \epsilon_n \hat n_{n,{\bf k},\sigma} 
\: \big\} .
\label{EQU3}
\end{eqnarray}
Using the length notations from Fig. 3, the exponents 
in (\ref{EQU3}) are
\begin{eqnarray}
&&{\bf k}\cdot({\bf r}_6-{\bf r}_5) =0, \quad
{\bf k}\cdot({\bf r}_4-{\bf a})= - k b', \quad
{\bf k}\cdot({\bf r}_2-{\bf r}_3)= - k b_1, 
\nonumber\\
&&{\bf k}\cdot({\bf r}_1-{\bf r}_2)=
{\bf k}\cdot({\bf r}_3-{\bf r}_4)= -k b_2, \quad
{\bf k}\cdot({\bf r}_4-{\bf r}_5)= {\bf k}\cdot({\bf r}_5-
{\bf r}_1)=\frac{k b}{2}.
\label{EQU4}
\end{eqnarray}
Hence $\hat H_0$ becomes
\begin{eqnarray}
\hat H_0 &=& \sum_{\sigma} \sum_{\bf k} \: 
\big\{ \: \big[ \: 
t_f \hat c^{\dagger}_{6,{\bf k},\sigma} \hat c_{
5,{\bf k},\sigma} +
t_c \hat c^{\dagger}_{4,{\bf k},\sigma} \hat c_{
1,{\bf k},\sigma} 
e^{- i k b'} +
t_h \hat c^{\dagger}_{2,{\bf k},\sigma} \hat c_{
3,{\bf k},\sigma}
e^{- i k b_1} + 
\nonumber\\
&+& t \big( 
\hat c^{\dagger}_{5, {\bf k},\sigma} \hat c_{
1, {\bf k},\sigma} 
e^{+ i \frac{k b}{2}} +
\hat c^{\dagger}_{1, {\bf k},\sigma} \hat c_{
2, {\bf k},\sigma} 
e^{- i k b_2 } +
\hat c^{\dagger}_{3, {\bf k},\sigma}\hat c_{
4, {\bf k},\sigma} 
e^{- i k b_2 } +
\hat c^{\dagger}_{4,{\bf k},\sigma}\hat c_{
5, {\bf k},\sigma}
e^{+ i \frac{ k b}{2}} 
\big) 
\nonumber\\
&+& H.c. \big] + \sum_{n=1}^6 \epsilon_n \hat n_{
n,{\bf k},\sigma} \big\}. 
\label{EQU5}
\end{eqnarray}

\subsection{ The bare band structure}

The bare band structure is obtained by diagonalizing 
$\hat H_0$
from (\ref{EQU5}), 
\begin{eqnarray}
\hat H_0= \sum_{\sigma} \sum_{\bf k}
(\hat c^{\dagger}_{1,{\bf k},\sigma},
\hat c^{\dagger}_{2,{\bf k},\sigma}, ..., 
\hat c^{\dagger}_{6,{\bf k},\sigma} ) \tilde M
\left( \begin{array}{c}
\hat c_{1,{\bf k},\sigma} \\
\hat c_{2,{\bf k},\sigma} \\
.....               \\
\hat c_{6,{\bf k},\sigma} \\
\end{array} \right) ,
\label{EQU6}
\end{eqnarray}
where the $6\times 6$ matrix $\tilde M$ has the form
\begin{eqnarray}
\tilde M =
\left( \begin{array}{cccccc}
\epsilon_1 & te^{-ikb_2} & 0 & t_c e^{+ikb'} & 
t e^{-ik\frac{b}{2}} 
& 0 \\
t e^{+ikb_2} & \epsilon_2 & t_h e^{-ikb_1} & 0 & 0 & 0 \\
0 & t_h e^{+ikb_1 } & \epsilon_3 & t e^{-i k b_2} & 0 & 0 \\
t_c e^{-ikb'} & 0 & t e^{+ i k b_2} & \epsilon_4 & 
t e^{+ i k 
\frac{b}{2}} & 0 \\
t e^{+ik\frac{b}{2}} & 0 & 0 & t e^{- i k \frac{b}{2}} & 
\epsilon_5 & t_f
\\
0 & 0 & 0 & 0 & t_f & \epsilon_6
\end{array} 
\right) .
\label{EQU7}
\end{eqnarray}
The diagonalized energies (i.e. the bare band structure)
follows from the secular equation of $\tilde M$.
The orthonormalized eigenvectors of  $\tilde M$ determine
the diagonalized canonical Fermi operators.

Taking into account $b=b_1+2b_2, a=b+b'$ (see Fig. 3), and 
adopting the symmetry  $\epsilon_1=\epsilon_4$ and 
$\epsilon_2=\epsilon_3$, from (\ref{EQU7}) one obtains the  
bare band structure $\epsilon=E_{\nu}(k)$, 
$\nu \leq m=6$ from the solutions of
\begin{eqnarray}
&& 2 t_c t^2 \Big\{ (\epsilon_6-\epsilon )
[(\epsilon_2-\epsilon)^2-t_h^2]
- t_h [(\epsilon_6-\epsilon )(\epsilon_5-\epsilon ) 
- t_f^2] \Big\} \cos a k 
\nonumber\\
&&+[(\epsilon_6-\epsilon)(\epsilon_5-\epsilon)- t_f^2]
\Big\{[(\epsilon_2-\epsilon)
(\epsilon_1-\epsilon) - t^2]^2 - (\epsilon_1-\epsilon)^2 
t_h^2
-(\epsilon_2-\epsilon)^2 t_c^2 + t_h^2 t_c^2 \Big\}
\nonumber\\
&&+ 2 (\epsilon_6-\epsilon) t^2 \Big\{ 
(\epsilon_1-\epsilon) 
[t_h^2-
(\epsilon_2-\epsilon)^2] + t^2 (\epsilon_2-\epsilon + 
t_h) \Big\} = 0.
\label{EQU8}
\end{eqnarray}

\subsection{Conditions for flat bands}

Equation (\ref{EQU8}) is of the form $A(\epsilon) \cos ak 
+ B(\epsilon) =0$, where $A(\epsilon)$ and 
$B(\epsilon)$ do not depend on $k$. Hence, in order to
obtain a $k$-independent solution for $\epsilon$, 
simultaneously $A(\epsilon)=B(\epsilon)=0$ is required. 
Flat bands therefore emerge in the bare band structure when 
the following equations hold:
\begin{eqnarray}
[(\epsilon_6-\epsilon)(\epsilon_5-\epsilon)-t^2_f] 
&=& \frac{(\epsilon_6-
\epsilon)}{t_h}[(\epsilon_2-\epsilon)^2-t_h^2] 
\nonumber\\
&=& \frac{2 (\epsilon-
\epsilon_6) t^2 \{ (\epsilon_1-\epsilon) [t_h^2-
(\epsilon_2-\epsilon)^2] + t^2 (\epsilon_2-\epsilon + t_h) 
\} }{
\{[(\epsilon_2-\epsilon)
(\epsilon_1-\epsilon) - t^2]^2 - (\epsilon_1-\epsilon)^2 
t_h^2
-(\epsilon_2-\epsilon)^2 t_c^2 + t_h^2 t_c^2 \} }.
\label{EQU9}
\end{eqnarray}

The two equations (\ref{EQU9}) provide solutions of the 
form $\epsilon=C$, where $C=0$ can be chosen as the 
origin for the energy. In this particular case, 
(\ref{EQU9}) leads to the two necessary flat-band 
conditions
\begin{eqnarray}
&&\frac{\epsilon_6 \epsilon_5-t_f^2}{\epsilon_6}=\frac{
\epsilon_2^2-
t_h^2}{t_h},
\nonumber\\
&&\frac{\epsilon_2^2-t_h^2}{t_h}=- 2t^2\frac{\epsilon_1(
t_h^2-
\epsilon_2^2)+t^2(\epsilon_2+t_h)}{(\epsilon_2 \epsilon_1
-t^2)^2-
\epsilon_1^2t_h^2-\epsilon_2^2t_c^2+t_c^2t_h^2}.
\label{EQU10}
\end{eqnarray}
In the remainder of this paper we exclude those sets of 
$\hat H_0$ parameters which provide solutions for 
(\ref{EQU9}) or (\ref{EQU10}), hence in the analyzed cases 
flat bands are absent in the bare band structure.

\section{Transformation of $\hat {\bf H}$ in positive 
semidefinite form}

\subsection{The transformation for the pentagon chain}

\subsubsection{The construction of the transformed Hamiltonian}

In order to transform the Hamiltonian (\ref{EQU2})
in positive semidefinite form we introduce five block 
operators $\hat G^{\dagger}_{\alpha,{\bf i},
\sigma}=\sum_{\ell \in B_{\alpha}} a_{\alpha,\ell} 
\hat c^{\dagger}_{{\bf i}+{\bf r}_{\ell},\sigma},$ 
for each unit cell at the lattice site ${\bf i}$, which act
on the blocks $B_{{\bf i},\alpha}$ for $\alpha=1,...,5$
(see Fig. 4) as follows

\begin{figure}
\centerline{\includegraphics
[width=6 cm,height=6 cm]{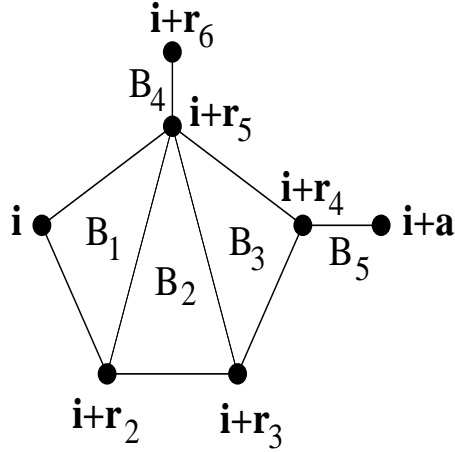}}
\caption{The blocks $B_{{\bf i},\alpha}$, $\alpha=1,...,5$ 
defined inside the unit cell at the lattice site 
${\bf i}$.} 
\label{fig:4}
\end{figure}

\begin{eqnarray}
&&\hat G_{1,{\bf i},\sigma}= a_{1,1} \hat c_{{\bf i}
+{\bf r}_1,
\sigma} + a_{1,2} \hat c_{{\bf i}+{\bf r}_2,\sigma} +
a_{1,5} \hat c_{{\bf i}+{\bf r}_5,\sigma} ,
\nonumber\\
&&\hat G_{2,{\bf i},\sigma}= a_{2,2} \hat c_{{\bf i}+
{\bf r}_2,
\sigma} + a_{2,3} \hat c_{{\bf i}+{\bf r}_3,\sigma} +
a_{2,5} \hat c_{{\bf i}+{\bf r}_5,\sigma} ,
\nonumber\\  
&&\hat G_{3,{\bf i},\sigma}= a_{3,3} \hat c_{{\bf i}+
{\bf r}_3,
\sigma} + a_{3,4} \hat c_{{\bf i}+{\bf r}_4,\sigma} +
a_{3,5} \hat c_{{\bf i}+{\bf r}_5,\sigma} ,
\nonumber\\ 
&&\hat G_{4,{\bf i},\sigma}= a_{4,5} \hat c_{{\bf i}+
{\bf r}_5,
\sigma} + a_{4,6} \hat c_{{\bf i}+{\bf r}_6,\sigma} ,
\nonumber\\
&&\hat G_{5,{\bf i},\sigma}= a_{5,4} \hat c_{{\bf i}+
{\bf r}_4,
\sigma} + a_{5,7} \hat c_{{\bf i}+{\bf a},\sigma} ,
\label{EQU11}
\end{eqnarray}
where $a_{n,n'}$ are numerical coefficients, and  
$|{\bf r}_1|=0$, ${\bf r}_7={\bf a}$. 
As seen from (\ref{EQU11}), 
$B_{{\bf i},1}, B_{{\bf i},2}, B_{{\bf i},3}$ are 
blocks defined on triangles, while $B_{{\bf i},4}$ and
$B_{{\bf i},5}$ are defined on individual bonds (see 
also Fig. 4). Based on (\ref{EQU11}), we define the 
operator
\begin{eqnarray}
\hat H_G=\sum_{\sigma}\sum_{{\bf i}}
\sum_{\alpha=1}^5
\hat G_{\alpha,{\bf i},\sigma} \hat G^{\dagger}_{
\alpha,{\bf i},\sigma}.
\label{EQU12}
\end{eqnarray} 
We now consider the positive semidefinite operator
\begin{eqnarray}
\hat P_{{\bf i},n}=\hat n_{{\bf i}+{\bf r}_n,\uparrow} 
\hat n_{{\bf i}+{\bf r}_n,\downarrow}- (\hat n_{{\bf i}+
{\bf r}_n,\uparrow}+\hat n_{{\bf i}+{\bf r}_n,\downarrow})
+1, \quad n=1,...,6,
\label{EQU12a}
\end{eqnarray}
which requires at least one electron on the site ${\bf i}+
{\bf r}_n$ for its minimum eigenvalue zero, and define
the operator
$\hat P_n = \sum_{{\bf i}}\hat P_{{\bf i},n}$. 
Using these notations, the Hamiltonian (\ref{EQU2}) 
transforms exactly to
\begin{eqnarray}
\hat H = \hat H_G + \hat H_P +C_g,
\label{EQU13}
\end{eqnarray}
where $C_g =q_U N- N_c (2K + \sum_{n=1}^6U_n)$ is a scalar, 
$\hat H_P=\sum_{n=1}^6 U_n \hat P_{n}$, 
$\hat H'=\hat H-C_g=\hat H_G + \hat H_P$ are 
positive semidefinite operators, and $q_U$ is obtained from
the solution of (\ref{EQU14}).

Equality (\ref{EQU13}) provides relations between the 
parameters of $\hat H$, and the numerical coefficients 
$a_{\alpha,\ell}$ of the $\hat G_{\alpha,{\bf i},\sigma}$
operators. These relations are called 
matching equations, which in the present case
have the form
\begin{eqnarray}
&&-t_f=a^*_{4,6} a_{4,5}, \quad -t_h=a^*_{2,2} a_{2,3}, 
\quad
-t_c=a^*_{5,4} a_{5,7},
\nonumber\\
&&-t = a^*_{1,5} a_{1,1} = a^*_{1,1} a_{1,2} =  a^*_{3,3} 
a_{3,4} =  
a^*_{3,4} a_{3,5},
\nonumber\\
&&0 = a^*_{2,5} a_{2,3} + a^*_{3,5} a_{3,3} 
= a^*_{2,5} a_{2,2} + a^*_{1,5} a_{1,2} ,
\nonumber\\
&&q_U-(U_6+\epsilon_6) = |a_{4,6}|^2,
\nonumber\\
&&q_U-(U_5+\epsilon_5) = |a_{4,5}|^2 + 
|a_{1,5}|^2 + |a_{3,5}|^2 + |a_{2,5}|^2,
\nonumber\\
&&q_U-(U_2+\epsilon_2)= |a_{1,2}|^2 + |a_{2,2}|^2 =  
|a_{2,3}|^2 + |a_{3,3}|^2,
\nonumber\\
&&q_U-(U_1+\epsilon_1)= |a_{1,1}|^2 + |a_{5,7}|^2 =  
|a_{3,4}|^2 + |a_{5,4}|^2 ,
\label{EQU14}
\end{eqnarray}
while for the constant $K$, obtained from 
$K = \sum_{\alpha=1}^5 z_{\alpha}$, 
$z_{\alpha}=\{\hat G_{\alpha,{\bf i},\sigma},
\hat G^{\dagger}_{\alpha,{\bf i},\sigma} \} \ne 0$ one has
\begin{eqnarray}
K &=& (|a_{1,5}|^2+|a_{1,1}|^2 +|a_{1,2}|^2)+(|a_{2,5}|^2
+|a_{2,2}|^2 + |a_{2,3}|^2)+(|a_{3,5}|^2+|a_{3,3}|^2 
+|a_{3,4}|^2)
\nonumber\\
&+&(|a_{4,6}|^2+|a_{4,5}|^2) + (|a_{5,4}|^2+|a_{5,7}|^2 ).
\label{EQU15}
\end{eqnarray}
Because of the symmetrical placement in the unit cell of sites
$1,4$ and $2,3$ respectively, see Fig. 3, 
the equalities 
$U_1=U_4$, $U_2=U_3$, and $\epsilon_1=\epsilon_4$, 
$\epsilon_2=\epsilon_3$, have been considered in
(\ref{EQU14}). This choice is used also in the rest of the
paper.

In the derivation of the matching equations 
(\ref{EQU14},\ref{EQU15}), one first calculates
$\sum_{\alpha=1}^5
\hat G_{\alpha,{\bf i},\sigma} \hat G^{\dagger}_{
\alpha,{\bf i},\sigma}$ from (\ref{EQU12}), and deduces the
expression for $\hat H_G$. Then, using (\ref{EQU12a}) and
$\hat H_P$, the transformed Hamiltonian is explicitly
obtained from (\ref{EQU13}). After this step  
the obtained result is compared with the starting Hamiltonian 
(\ref{EQU2}) and the coefficients of the identical 
operator contributions in (\ref{EQU13}) and (\ref{EQU2})
are identified. 
This comparison provides the matching equations 
(\ref{EQU14},\ref{EQU15}).

Following this strategy one realizes that the
$\hat G_{\alpha,{\bf i},\sigma} \hat G^{\dagger}_{
\alpha,{\bf i},\sigma}$ terms essentially produce the 
kinetic energy terms of (\ref{EQU2}), while the 
$\hat P_{{\bf i},n}$ contributions provide the interaction 
terms of the starting Hamiltonian. But the $\hat H_G$ and
$\hat H_P$ terms contribute together in the matching equations 
[see the 4th - 7th rows of (\ref{EQU14})], leading to a 
positive semidefinite decomposition in (\ref{EQU13}) which 
mixes the interaction and the kinetic terms of the 
starting Hamiltonian. The matching equations 
(\ref{EQU14},\ref{EQU15}) for the unknown variables 
$a_{\alpha,\ell}$, are nonlinear and coupled 
complex algebraic equations.

\subsubsection{Solutions of the matching conditions}

Since the matching equations (\ref{EQU14}) demand 
$t_h=-|a_{1,1}|^2|a_{2,2}|^2/|a_{3,4}|^2$, solutions 
exist only for $t_h < 0$. Furthermore, introducing the 
notations
\begin{eqnarray}
\hspace*{-1cm}
Q_1=q_U-U_2-\epsilon_2-|t_h|, \quad
Q_2=q_U-U_1-\epsilon_1-|t_c|, \quad Q_3=|a_{4,6}|=
\sqrt{q_U-U_6-\epsilon_6},
\label{EQU16}
\end{eqnarray}
(\ref{EQU14},\ref{EQU16}) require $Q_1,Q_2,Q_3 > 0$. 
With these 
observations, the following solutions for (\ref{EQU14}) 
are obtained 
\begin{eqnarray}
&&a_{1,2}=a_{1,5}=\sqrt{Q_1}e^{i\phi_1}, \quad \hspace*{0.4cm}
a_{1,1}= -\frac{t}{\sqrt{Q_1}}e^{i\phi_1},
\nonumber\\
&&a_{2,2}=a_{2,3}=- \sqrt{ |t_h|}e^{i\phi_2}, \quad
a_{2,5}=\frac{Q_1}{\sqrt{|t_h|}}e^{i\phi_2},
\nonumber\\
&&a_{3,3}=a_{3,5}=\sqrt{Q_1}e^{i\phi_3}, \quad \hspace*{0.4cm}
a_{3,4}=-\frac{t}{\sqrt{Q_1}}e^{i\phi_3},
\nonumber\\
&&a_{4,6}=Q_3e^{i\phi_4}, \quad \hspace*{2cm}
a_{4,5}=-\frac{t_f}{Q_3},
\nonumber\\
&&a_{5,7}=\sqrt{|t_c|}e^{i\phi_5}, \quad \hspace*{1.55cm}
a_{5,4}=-sign(t_c) \sqrt{|t_c|} e^{i\phi_5},
\label{EQU17}
\end{eqnarray}
where $\phi_{\alpha}$, $\alpha=1,...,5$ are arbitrary 
phases, which therefore can be set to zero.
Besides, two supplementary conditions are found 
from the 5th and 7th line of (\ref{EQU14}), namely
\begin{eqnarray}
&&(Q_2+|t_c|)=\frac{t^2}{Q_1}+|t_c|,
\nonumber\\
&&\frac{(Q_1+|t_h|)^2-t_h^2}{|t_h|}= (q_U-U_5-
\epsilon_5)
-\frac{t_f^2}{Q_3^2}.
\label{EQU18}
\end{eqnarray}
Taking into account $Q_1,Q_2 > 0$, from the first 
(second) line of (\ref{EQU18}) one finds the parameters
$q_U$ ($Q_3$) in the form
\begin{eqnarray}
&&q_U=\frac{1}{2}\{ (U_1+\epsilon_1+|t_c|)+(U_2+\epsilon_2
+|t_h|)+
\sqrt{[(U_2+\epsilon_2+|t_h|)-(U_1+\epsilon_1+|t_c|)]^2+
4t^2} \},
\nonumber\\
&&Q_3=\frac{|t_f|\sqrt{|t_h|}}{\sqrt{[ |t_h|(q_U-U_5-
\epsilon_5)-(q_U-U_2-\epsilon_2)^2+t_h^2]}}.
\label{EQU19}
\end{eqnarray}
Thereby the constant $K$ is determined as
\begin{eqnarray}
K=2(|t_h|+|t_c|)+4Q_1+Q_3^2 +\frac{Q_1^2}{|t_h|}+
\frac{2t^2}{Q_1}+
\frac{t_f^2}{Q_3^2}.
\label{EQU20}
\end{eqnarray}

\subsubsection{The parameter-space region where the
solution holds}

The solution (\ref{EQU17},\ref{EQU19},\ref{EQU20})
is valid, if the expression under the square root
in $Q_3$ is positive. Introducing 
$W=q_U-[(q_U-U_2-\epsilon_2)^2-t_h^2]/|t_h|$, 
one has $Q_3=|t_f|[W-\epsilon_5-U_5]^{-1/2}$, hence
$W-\epsilon_5 > U_5$ must hold. $U_5 > 0$, therefore implies 
$W > \epsilon_5$. Furthermore,
introducing $Z=U_6+\epsilon_6$, from the 4th line of
(\ref{EQU17}), and the 4th line of (\ref{EQU14}) one also 
has $Z=q_U-Q_3^2$. Since $U_6 > 0$, it follows that 
$Z-\epsilon_6 > 0$, consequently also 
$Z=q_U-Q_3^2 > \epsilon_6$ holds. 
One further observes that all $U_n$ values appear in 
the expression for $U_6=Z-\epsilon_6$. Consequently, a 
restriction regarding the value of $U_6$ emerges from this 
equality.
The inequalities $q_U-(U_n+\epsilon_n) > 0$ required by 
(\ref{EQU14}) are automatically satisfied for all 
$n=1,...,6$. 

Summarizing the necessary conditions for the solubility of
the matching equations, the parameter-space region where
solutions exist is defined by
\begin{eqnarray}
&&t_h < 0, \quad Z=q_U-Q_3^2 > \epsilon_6, \quad W=q_U - 
\frac{(q_U-U_2
-\epsilon_2)^2-t_h^2}{|t_h|} > \epsilon_5,
\nonumber\\
&&W-\epsilon_5 > U_5 > 0, \quad U_6 =Z-\epsilon_6.
\label{EQU21}
\end{eqnarray}  

\subsection{The transformation for a general chain}

The presented transformation
can be performed for more general chain structures as well.
In the general case 
[see the Hamiltonian (\ref{EQU1})], a
chain structure is characterized by a cell build up from a
closed polygon which may be connected also to side groups.
The unit cell contains $m=m_p+m_e \geq 2$ sites
placed at ${\bf i}+{\bf r}_n$, $n=1,...,m$, where 
$m_p$ denotes the number of sites in the closed 
polygon, and $m_e$ represents the number of sites in the 
side groups. Furthermore, the unit cells must be connected
through a single site. In this respect Sect. III. A. 
represents in fact an example for the transformation of
$\hat H$ in the particular case $m=6, m_p=5, m_e=1$.

In defining the $\hat G^{\dagger}_{\alpha,{\bf i},\sigma}$ 
operators for the general case, one takes into account 
that the closed polygon in the unit cell is constructed 
from $m_p$ sites. Inside the polygon, $m_p-2$ triangle 
domains $B_{{\bf i},\alpha}$ can be chosen such that
they share one common site  (see in Fig. 4 the domains 
$B_1,B_2,B_3$ for the case $m_p=5$). On these 
$m_p-2$ triangle domains $m_p-2$ block operators 
$\hat G^{\dagger}_{\alpha,{\bf i},\sigma}$ are constructed.
Because of the presence of the side groups, $m_e+1$ bond 
domains must be considered in addition (see in Fig. 4 the 
domains $B_4,B_5$ for the case $m_e=1$). From these, $m_e$ 
bonds contain one side-group site (the bond domain $B_4$ 
in Fig. 4), and the remaining bond contains the 
cell-interconnection point ${\bf r}_{m+1}$ (the bond 
domain $B_5$ in Fig. 4). On these $m_e+1$ bonds 
further $m_e+1$ block operators 
$\hat G^{\dagger}_{\alpha,{\bf i},\sigma}$ are constructed. 
Hence, one has in total
$(m_p-2)+(m_e+1)=(m_p+m_e)-1=m-1$ operators 
$\hat G^{\dagger}_{\alpha,{\bf i},\sigma}$ 
for the unit cell constructed at the 
lattice site ${\bf i}$, in the general case.
With these block operators one obtains for the
expression of the transformed Hamiltonian the relation
(\ref{EQU13}), which in the general case has the form
\begin{eqnarray}
&&\hat H_G=\sum_{\sigma}\sum_{{\bf i}} 
\sum_{\alpha=1}^{m-1}
\hat G_{\alpha,{\bf i},\sigma} \hat G^{\dagger}_{
\alpha,{\bf i},
\sigma}, \quad
\hat H_P=\sum_{n=1}^m U_n \hat P^{(n)},
\nonumber\\
&&C_g =q_U N- N_c (2\sum_{\alpha=1}^{m-1}z_{\alpha} + 
\sum_{n=1}^{m}U_n), 
\label{EQU30}
\end{eqnarray} 
where $z_{\alpha}$ has been defined below (\ref{EQU14}). 
With these notations the expression for the transformed 
Hamiltonian (\ref{EQU13}) remains unchanged, and provides
similar results as for the pentagon chain. For this reason
we focus below on the pentagon chain as a general example.

\section{The deduced ground states}

\subsection{Upper band at half-filling}

\subsubsection{The structure of the ground state}

Since $2N_{\Lambda}$ represents the maximum electron number
we now define $N=2N_{\Lambda}-N_c=11N_c=N^{*}$. This $N^{*}$
value corresponds to the half-filled upper band case. The 
ground-state wave function $|\Psi_g(N^*)\rangle$ for this 
situation is given by
\begin{eqnarray}
|\Psi_g(N^*)\rangle = [\prod_{\sigma} \hat G^{\dagger}_{
\sigma}]
\hat F^{\dagger}|0\rangle,
\label{EQU31}
\end{eqnarray}
where $\hat G^{\dagger}_{\sigma}=\prod_{{\bf i}} 
\prod_{\alpha=1}^5 \hat G^{\dagger}_{\alpha,{\bf i},
\sigma}$, and the operator $\hat F^{\dagger}=\prod_{
{\bf i}} \hat c^{\dagger}_{{\bf i}+{\bf r}_{
n_{\bf i}},\sigma}$ introduces one electron with fixed spin 
$\sigma$ in each unit cell; $|0\rangle$ represents the 
bare vacuum. The ground state energy is $E_g=C_g$
from (\ref{EQU13}).

\subsubsection{The deduction procedure for the ground state}

The reason why (\ref{EQU31}) represents the ground state
is as follows. In (\ref{EQU13}) the system Hamiltonian has
been transformed to the form presented in (\ref{I1}) where 
\begin{eqnarray}
\hat O_{p,1}=\hat H_G, \quad \hat O_{p,2}=\hat H_P, \quad
\hat O_p=\hat O_{p,1}+\hat O_{p,2}=\hat H_G + \hat H_P.
\label{EQU31a}
\end{eqnarray}

As explained in (\ref{I1oo}), the ground state is deduced 
via $\hat O_p |\Psi_g\rangle =0$. In the present
case this equation implies
\begin{eqnarray}
&&\hat O_{p,1}|\Psi_g\rangle = \hat H_G |\Psi_g\rangle =0,
\nonumber\\
&&\hat O_{p,2}|\Psi_g\rangle = \hat H_P |\Psi_g\rangle =0.
\label{EQU31b}
\end{eqnarray}

In order to satisfy the first equation in (\ref{EQU31b}) 
we note that due to the structure of $\hat H_G$ in 
(\ref{EQU12}) and $\hat G^{\dagger}_{\alpha,{\bf i},
\sigma} \hat G^{\dagger}_{\alpha,{\bf i},\sigma} =0$,
a Hilbert space vector
of the form $|\Psi_g\rangle = [\prod_{\sigma} \hat G^{
\dagger}_{\sigma}]\hat F^{\dagger}|0\rangle$ satisfies
$\hat H_G |\Psi_g\rangle =0$, where as shown below
(\ref{EQU31}), $\hat G^{\dagger}_{\sigma}=
\prod_{{\bf i}} \prod_{\alpha=1}^5 
\hat G^{\dagger}_{\alpha,{\bf i},\sigma}$. 
At this stage $\hat F^{\dagger}$  is an arbitrary operator 
which preserves a nonzero value for the norm.
$\hat F^{\dagger}$ is determined such that the second
equation in (\ref{EQU31b}) is satisfied as well.

In deducing $\hat F^{\dagger}$ we recall that 
the positive semidefinite operator $\hat H_P$ defined in 
(\ref{EQU13}) requires for its minimum eigenvalue zero at 
least one electron on each site. There are six sites in 
each unit cell (see Fig. 3), and 5 electrons with fixed 
spin are introduced in average in each cell
by $\hat G^{\dagger}_{\sigma}$. Consequently, since 
$\prod_{\alpha=1}^5 \hat G^{\dagger}_{\alpha,{\bf i},
\sigma}$ can introduce one electron also in the
$(i+1)$-th cell at the contact site between the cells,
maximally two sites can remain empty in a given cell. 
This situation is overcome by choosing
\begin{eqnarray}
\hat F^{\dagger}=\prod_{{\bf i}} 
\hat c^{\dagger}_{{\bf i}+{\bf r}_{n_{\bf i}},\sigma}
\label{EQU31c}
\end{eqnarray}
which adds one electron with fixed spin $\sigma$ in each 
cell at an arbitrary site. As a consequence, 
$\hat G^{\dagger}_{\sigma}\hat F^{\dagger}$ introduces 
$6N_c$ electrons in the system containing $N_{\Lambda}
=6N_c$ sites. So each site contains one spin $\sigma$ 
electron, hence $\hat H_P |\Psi_g\rangle=0$ holds.
Therefore also the second equation in (\ref{EQU31b}) is 
satisfied, and (\ref{I1oo}) is fulfilled. Finally we note
that $\prod_{\sigma}\hat G^{\dagger}_{\sigma}$ introduces 
$2*(5N_c)=10N_c$ electrons into the system ($5N_c$ electrons
for each spin projection), while $\hat F^{\dagger}$ 
introduces $N_c$ electrons with spin
$\sigma$, hence the number of electrons is $N=11N_c$. This
number of electrons for $N_{\Lambda}=6N_c$ sites (6 bands)
represents indeed the half-filled upper band situation. 

It is important to note, that if even one spin is reversed 
in $\hat F^{\dagger}$ from (\ref{EQU31c}), the emergence 
of empty sites in $|\Psi_g\rangle$ cannot be excluded. This
is shown by an elementary counting of the possible states.
Indeed, for one spin flip in $\hat F^{\dagger}$ from
(\ref{EQU31c}), $\prod_{\sigma}\hat G^{\dagger}_{\sigma}$ 
introduces $5N_c$ spin $\sigma$ and $5N_c$ spin $-\sigma$ 
electrons, while $\hat F^{\dagger}$ provides $N_c-1$ spin 
$\sigma$ and  one spin $-\sigma$ electron. Hence 
$N_{\sigma}=6N_c-1$ ($N_{\sigma}$ being the total number of
electrons with spin $\sigma$),
and $N_{-\sigma}=5N_c+1$ is the number
of electrons with fixed spin indices. From these electrons, 
$5N_c+1$ sites with double occupancy can be created 
from $N_{\Lambda}=6N_c$ total number of sites.
Consequently, one remains with 
$6N_c-(5N_c+1)=N_c-1$ empty sites, on 
which the remaining $(6N_c-1)-(5N_c+1)=N_c-2$ electrons with
spin $\sigma$ must be placed. This necessarily leaves one 
site empty.

\subsubsection{Transcription of the ground state}
 
Since there is one spin $\sigma$ electron on each site in 
the ground state (\ref{EQU31}), all spin $\sigma$ 
electrons are localized, and (\ref{EQU31}), as an 
unnormalized wave function, can be identically rewritten as
\begin{eqnarray}
|\Psi_g(N^*)\rangle = [\prod_{{\bf i}} 
[(\prod_{n=1}^6
\hat c^{\dagger}_{{\bf i}+{\bf r}_n,\sigma})
(\prod_{\alpha=1}^5
\hat G^{\dagger}_{\alpha,{\bf i},-\sigma})]|0\rangle.
\label{EQU32}
\end{eqnarray}
The number of mobile $-\sigma$ electrons is
$5N_c$, and the ground state represents a state 
with total spin $S=S_z^{Max}=N_c/2$. Apart from the
trivial $(2S+1)$ degeneracy related to the orientation 
of the total spin, the ground state is non-degenerate 
and unique (see Sect. VII. A). The ground state 
expectation value of the long-range hopping terms in the
$N_c \to \infty$ limit yields an exponential decay 
(see Sect. V. D.2), hence the system is a nonsaturated 
ferromagnet which is localized in the thermodynamic limit.  

\subsection{Upper band above half filling}

\subsubsection{The structure of the ground state}

In the present situation one has $N > N^*=11N_c$, 
$N=N^*+\bar N$, 
which means that $\bar N$ $-\sigma$ spin electrons 
have been added to the system. The ground state wave
function in the $S_z=S_z^{Max}$ spin sector becomes
\begin{eqnarray}
|\Psi_g(N^*+\bar N)\rangle=\hat Q^{\dagger}_{\bar N} 
|\Psi_g(N^*)\rangle,
\label{EQU33}
\end{eqnarray} 
where $\hat Q^{\dagger}_{\bar N}=\prod_{\gamma=1}^{\bar N}
\hat c^{\dagger}_{n_{\gamma},{\bf k}_{\gamma},-\sigma}$ is 
a product of $\bar N$ arbitrary, but different 
$\hat c^{\dagger}_{n,{\bf k},-\sigma}$ operators. 

\subsubsection{The deduction of the ground state}

In deducing the ground state at $N>N^*$, we again follow 
the steps to satisfy (\ref{EQU31b}). The first equation in 
(\ref{EQU31b}) is indeed fulfilled because
$\hat G^{\dagger}_{\alpha,{\bf i},\sigma}$ 
anticommutes with the fermionic creation operators 
$\hat c^{\dagger}_{n,{\bf k},-\sigma}$, hence 
\begin{eqnarray}
\hat G^{\dagger}_{
\alpha,{\bf i},\sigma}\hat Q^{\dagger}_{\bar N}= 
(-1)^{\bar N}
\hat Q^{\dagger}_{\bar N} \hat G^{\dagger}_{\alpha,{\bf i},
\sigma},
\label{eq32}
\end{eqnarray}
and consequently 
\begin{eqnarray}
\hat H_G |\Psi_g(N^*+\bar N)\rangle &=&
\sum_{\sigma}\sum_{{\bf i}} \sum_{\alpha=1}^5
\hat G_{\alpha,{\bf i},\sigma} 
\hat G^{\dagger}_{\alpha,{\bf i},\sigma}
\hat Q^{\dagger}_{\bar N}
|\Psi_g(N^*)\rangle 
\nonumber\\
&=& \sum_{\sigma}\sum_{{\bf i}}
\sum_{\alpha=1}^5 \hat G_{\alpha,{\bf i},\sigma} 
(-1)^{\bar N}
\hat Q^{\dagger}_{\bar N} \hat G^{\dagger}_{\alpha,{\bf i},
\sigma}|\Psi_g(N^*)\rangle =0
\label{eq33}
\end{eqnarray}
because $\hat G^{\dagger}_{\alpha,{\bf i},\sigma}
|\Psi_g(N^*)\rangle =0$ holds. Therefore the first line in
equation (\ref{EQU31b}) is satisfied.

In order to prove the second equation in (\ref{EQU31b}), 
it is sufficient to note that the
$\hat Q^{\dagger}_{\bar N}$ operator 
does not alter the presence of at least one electron per 
site, hence $\hat H_P |\Psi_g(N^*+\bar N)\rangle=0$ also 
holds. Therefore $(\hat H_G + \hat H_P)|\Psi_g(N^*+\bar N)
\rangle=0$ and (\ref{EQU33}) indeed provides the ground 
state for $N=N^*+ \bar N$ electrons.

\subsubsection{The nature of the ground state}

Because of the itinerant electrons introduced by 
$\hat Q^{\dagger}_{\bar N}$, (\ref{EQU33}) possesses 
extended electronic states. $E_g=C_g$ depends linearly
on $N$, consequently $\delta \mu=E_g(N+1)-2E_g +E_g(N-1)=0$
for $\bar N > 1$, hence the charge excitations are gapless. 
The ground state is a non-saturated ferromagnet with
localized spin $\sigma$ electrons and mobile
$-\sigma$ spin electrons (see Sect. V. D.1), and therefore
describes a correlated half metallic state.

\subsection{The flat nature of the effective band}

Before proceeding further, we emphasize the physical conditions
for which the deduced ground states in the previous two 
subsections are obtained. Since the ground states 
$|\Psi_g\rangle$ presented in (\ref{EQU31},\ref{EQU33}) 
satisfy $\hat H_P |\Psi_g\rangle=0$, the transformed 
Hamiltonian (\ref{EQU13}) acting on the ground state 
becomes $\hat H_G+C_g$, $\hat H_G=\hat H_{kin}+K_G$ with 
the kinetic energy operator 
$\hat H_{kin}=-\sum_{\sigma} \sum_{{\bf i}}
\sum_{\alpha=1}^5 \hat G^{\dagger}_{\alpha,{\bf i},\sigma}
\hat G_{\alpha,{\bf i},\sigma}$ and $K_G=2N_c 
\sum_{\alpha=1}^{5}z_{\alpha}$. Hence, the Hamiltonian 
acting on the ground state is in fact of the form 
$\hat H_{kin}+C$, with the constant $C=C_g+K_G$. 
Furthermore, the kinetic part $\hat H_{kin}$ is quadratic 
in the original fermionic operators $\hat c_{{\bf i}+
{\bf r}_n,\sigma}$, has the same hopping matrix elements 
as $\hat H$ in (\ref{EQU2}), but the on-site energies are 
renormalized to
\begin{eqnarray}
\epsilon^R_n=\epsilon_n+U_n-q_U,
\label{EQU22}
\end{eqnarray}
where $q_U$, being non-linear in $U_n$, is given in 
(\ref{EQU19}). We demonstrate 
below, that the $U_n$ dependent $\hat H_{kin}$ possesses
an interaction induced upper flat band. This is done by 
showing that the flat band emergence conditions 
(\ref{EQU10}) are satisfied for (\ref{EQU22}), when the 
required conditions for the existence of the presented 
solutions (see Sect. III. A.2) hold. 

Indeed, with $t_h=-|t_h|$ and $Q_3^2=q_U-U_6-\epsilon_6$
one finds from the second line of (\ref{EQU18})
\begin{eqnarray}
\frac{(q_U-U_5-\epsilon_5)(q_U-U_6-\epsilon_6)-t^2_f}{
q_U-U_6-\epsilon_6}= \frac{(q_U-U_2-\epsilon_2)^2-t_h^2}{
-t_h}. 
\label{EQU23}
\end{eqnarray}
Using (\ref{EQU22}), the equation (\ref{EQU23}) is 
equivalent to 
\begin{eqnarray}
\frac{\epsilon^R_5 \epsilon^R_6-t_f^2}{\epsilon^R_6}=
\frac{{\epsilon^R_2}^2-t_h^2}{t_h}.
\label{EQU24}
\end{eqnarray}
For the renormalized energies $\epsilon_n^R$, 
Eq. (\ref{EQU24}) is identical to the first equation of 
(\ref{EQU10}). To prove the flatband nature, we need to
verify that also the second equation of (\ref{EQU10})
is fulfilled with the renormalized on-site energies
$\epsilon_n^R$.

In order to show this, we employ the first relation from 
(\ref{EQU18}), which becomes $Q_1Q_2=t^2$, and 
consequently, can be rewritten  with (\ref{EQU22}) as
\begin{eqnarray}
(\epsilon_1^R+|t_c|)(\epsilon_2^R+|t_h|)=t^2.
\label{EQU25}
\end{eqnarray}
The second equation of (\ref{EQU10}) with $t_h=-|t_h|$ 
and the renormalized energies (\ref{EQU22}) takes the form
\begin{eqnarray}
\frac{{\epsilon_2^R}^2-t_h^2}{|t_h|}=2t^2
\frac{\epsilon_1^R(t_h^2-
{\epsilon_2^R}^2)+t^2(\epsilon_2^R-|t_h|)}{(\epsilon^R_2 
\epsilon^R_1-t^2)^2-{\epsilon_1^R}^2t_h^2
-{\epsilon_2^R}^2t_c^2+t_c^2t_h^2}.
\label{EQU26}
\end{eqnarray}
Equation (\ref{EQU26}) is satisfied by (\ref{EQU25}),
because using (\ref{EQU25}) in the numerator of 
(\ref{EQU26}) on the right side one finds
$\epsilon_1^R(t_h^2-{\epsilon_2^R}^2)+t^2(\epsilon_2^R
-|t_h|)=|t_c|({\epsilon_2^R}^2-t_h^2)$. Hence (\ref{EQU26})
takes the form
\begin{eqnarray}
(\epsilon^R_2 
\epsilon^R_1-t^2)^2-{\epsilon_1^R}^2t_h^2
+t_c^2(t_h^2-{\epsilon_2^R}^2)=2t^2|t_c||t_h|,
\label{EQU27}
\end{eqnarray}
which is identically rewritten as
\begin{eqnarray}
t^4-2t^2(\epsilon_1^R \epsilon_2^R+|t_c||t_h|)
+({\epsilon_2^R}^2-
|t_h|^2)({\epsilon_1^R}^2-|t_c|^2)=0.
\label{EQU28}
\end{eqnarray}
Because of (\ref{EQU25}), the last term in (\ref{EQU28}) 
is equal to
$t^2({\epsilon_2^R}-|t_h|)({\epsilon_1^R}-|t_c|)$, and
therefore (\ref{EQU28}) turns into
\begin{eqnarray}
t^2 - 2(\epsilon_1^R \epsilon_2^R+|t_c||t_h|)+
({\epsilon_2^R}-|t_h|)({\epsilon_1^R}-|t_c|)=0.
\label{EQU29}
\end{eqnarray}
The non-trivial solution of (\ref{EQU29}) is indeed 
(\ref{EQU25}). Consequently, the renormalized on-site 
energies (\ref{EQU22}) lead to a flat band in the 
effective band structure. From the structure of 
the secular equation (\ref{EQU8}) it follows that 
the flat band is the upper band, and since the renormalized
energies (\ref{EQU22}) contain the $U_n$ values, the 
flattening effect originates from the interactions. 
This conclusion remains valid also if $\epsilon_n=0$ for 
all $n=1,...,6$.

Importantly, the described flattening effect emerges only
for non-uniform $U_n$ values. Indeed, when
$U_n=U$ for all $n=1,...,6$, the $q_U$ parameter from
(\ref{EQU19}) becomes $q_U= U + \bar C$, where $\bar C$ is 
a $U$ independent constant. Consequently, $q_U-U$ in the 
renormalized energies (\ref{EQU22}) cancels the entire $U$ 
dependence, and the interaction has no influence on the 
effective band structure. We note that the non-uniform $U_n$
values account for the particular environment and type of 
atom on a particular site inside the unit cell. Furthermore,
even one U value different from others is sufficient to 
maintain the effect.

\section{The long-range hopping ground state 
expectation value}

\subsection{Description of the problem}

In order to collect more information about the 
ground state (\ref{EQU33}), we consider in this section 
for simplicity $\bar N=1$, and calculate the long-range 
hopping ground state expectation value
\begin {eqnarray}
\Gamma_{\bf i}({\bf r})=\langle (\hat c^{\dagger}_{{\bf i}
+{\bf r}_{
n_{\gamma}},-\sigma} \hat c_{{\bf i}+{\bf r}_{n_{\gamma}}
+{\bf r},
-\sigma}+ H.c.)\rangle,
\label{EQU34}
\end{eqnarray}
where $\langle...\rangle=\langle \Psi_g|...|\Psi_g\rangle 
/ \langle \Psi_g|\Psi_g\rangle$, and 
\begin{eqnarray}
|\Psi_g\rangle= \hat c^{\dagger}_{n_{\gamma},
{\bf k}_{\gamma},-\sigma} |\Psi_g(N^*)\rangle. 
\label{EQU35}
\end{eqnarray}
In (\ref{EQU34}) the site ${\bf i}$ is a fixed, but 
arbitrary site, and ${\bf r}=N_a {\bf a}$, where 
$N_a$ is an arbitrary integer number.
Furthermore $n_{\gamma}$ is fixed but 
arbitrary too, and for example we consider below
$n_{\gamma}=5$. The momentum ${\bf k}_{\gamma}$ in 
(\ref{EQU35}) is an arbitrary momentum of the electron 
injected into the system with $N=N^*$ electrons, (i.e. the
half filled upper band case). 
 
\subsection{The ground state in ${\bf k}$ space}

The calculations are performed in ${\bf k}$ space. 
With the Fourier transformation for 
$\hat c_{{\bf i}+{\bf r}_n,\sigma}$ used in (\ref{EQU3}) in
Sect. II. B., the Fourier transformed $\hat G_{\alpha,{
\bf i},\sigma}$ operators, based on 
(\ref{EQU11}, \ref{EQU17}), become of the form
\begin{eqnarray}
&&\hat G_{1,{\bf k},\sigma}= \sqrt{Q_1} (\hat c_{2,{\bf k},
\sigma}
e^{-i{\bf k}\cdot{\bf r}_2} + \hat c_{5,{\bf k},\sigma} 
e^{-i{\bf k}\cdot{\bf r}_5})- \frac{t}{\sqrt{Q_1}}
\hat c_{1,{\bf k},\sigma} e^{-i{\bf k}\cdot{\bf r}_1},
\nonumber\\
&&\hat G_{2,{\bf k},\sigma}= -\sqrt{|t_h|}(\hat c_{2,{
\bf k},\sigma}
e^{-i{\bf k}\cdot{\bf r}_2}+ \hat c_{3,{\bf k},\sigma}
e^{-i{\bf k}\cdot{\bf r}_3}) + \frac{Q_1}{\sqrt{|t_h|}}
\hat c_{5,{\bf k},\sigma} e^{-i{\bf k}\cdot{\bf r}_5},
\nonumber\\  
&&\hat G_{3,{\bf k},\sigma}= \sqrt{Q_1}
(\hat c_{3,{\bf k},\sigma}
e^{-i{\bf k}\cdot{\bf r}_3}+ \hat c_{5,{\bf k},\sigma}
e^{-i{\bf k}\cdot{\bf r}_5}) -\frac{t}{\sqrt{Q_1}} 
\hat c_{4,{\bf k},\sigma}e^{-i{\bf k}\cdot{\bf r}_4},
\nonumber\\ 
&&\hat G_{4,{\bf k},\sigma}= Q_3 \hat c_{6,{\bf k},\sigma} 
e^{-i{\bf k}\cdot{\bf r}_6} - \frac{t_f}{Q_3}
\hat c_{5,{\bf k},\sigma}e^{-i{\bf k}\cdot{\bf r}_5},
\nonumber\\
&&\hat G_{5,{\bf k},\sigma}= \sqrt{|t_c|} 
\hat c_{4,{\bf k},\sigma}
e^{-i{\bf k}\cdot{\bf r}_4} - sign(t_c)\sqrt{|t_c|}
\hat c_{1,{\bf k},\sigma} e^{-i{\bf k}\cdot
({\bf r}_1+{\bf a})},
\label{EQU37}
\end{eqnarray}
where the arbitrary $\phi_{\alpha}$ phases in 
(\ref{EQU17}) have been neglected because they have no
influence on the result.

The fermionic nature of the operators ensures
$\hat c_{n,{\bf k},\sigma}\hat c_{n,{\bf k},\sigma}=0$, 
$\hat G_{\alpha,{\bf k},\sigma}\hat G_{\alpha,{\bf k},
\sigma}=0$, for all $n,\alpha$. Therefore
\begin{eqnarray}
\prod_{{\bf i}} \hat c_{{\bf i}+{\bf r}_n,\sigma}
= Z_n
\prod_{{\bf k}} \hat c_{n,{\bf k},\sigma}, \quad
\prod_{{\bf i}} \hat G_{\alpha,{\bf i},\sigma}= Z
\prod_{{\bf k}} \hat G_{\alpha,{\bf k},\sigma},
\label{EQU38}
\end{eqnarray}
where $Z_n=\sum_P(-1)^{p_P}
\exp[i({\bf k}_1\cdot{\bf j}^{(n)}_1 +
{\bf k}_2\cdot{\bf j}^{(n)}_2 +...+ 
{\bf k}_{N_c}\cdot{\bf j}^{(n)}_{N_c})]$,
${\bf j}_{\beta}^{(n)}={\bf i}_{\beta}+{\bf r}_n$, 
$\beta=1,...,N_c$ covers all sites of the Bravais lattice 
(in ${\bf r}$ space for ${\bf i}_{\beta}$, and ${\bf k}$ 
space for ${\bf k}_{\beta}$). The sum $\sum_P$ extends
over all permutations $P$ of
$(1,2,3,...,N_c)$ to $(\beta_1,\beta_2,...,\beta_{N_c})$, 
and $p_P$ is the parity of P. Specifically 
$Z=Z_n(|{\bf r}_n|=0)$.

With the Fourier transformed operators the unnormalized 
$|\Psi_g(N^*)\rangle$ takes the form
\begin{eqnarray}
|\Psi_g(N^*)\rangle=\prod_{{\bf k}} \prod_{n=1}^6 
\hat c^{\dagger}_{n,{\bf k},\sigma} \prod_{\alpha=1}^5 
\hat G^{\dagger}_{\alpha,{\bf k},-\sigma} |0\rangle.
\label{EQU39}
\end{eqnarray}

\subsection{Calculation of $\Gamma_{\bf i}({\bf r})$ }

\subsubsection{The starting relation}

Starting from (\ref{EQU34}, \ref{EQU35}, \ref{EQU39}) and 
taking into account that $|\Psi_g(N^*)\rangle$ contains all 
$\hat c^{\dagger}_{n,{\bf k},\sigma}$ operators, 
one obtains for $\Gamma_{\bf i}({\bf r})= \bar \Gamma_{
\bf i}({\bf r}) + c.c.$
\begin{eqnarray}
&&\bar \Gamma_{\bf i}({\bf r})= \frac{1}{N_c}
\sum_{{\bf k}_1} \sum_{{\bf k}_2} 
\bar \Gamma(
{\bf k}_1,{\bf k}_2)e^{i{\bf k}_1\cdot({\bf i}+{\bf r}_5)}
e^{-i{\bf k}_2\cdot({\bf i}+{\bf r}_5+{\bf r})},
\label{EQU40}\\
&&\bar \Gamma({\bf k}_1,{\bf k}_2) =
\frac{ \langle 0 |
[\prod_{{\bf k}} \prod_{\alpha=1}^5 \hat G_{\alpha,
{\bf k},
-\sigma}] \hat c_{5,{\bf k}_{\gamma},-\sigma} 
[\hat c^{\dagger}_{5,{\bf k}_1,-\sigma} \hat c_{5,{\bf k}_2,
-\sigma}
] \hat c^{\dagger}_{5,{\bf k}_{\gamma},-\sigma}
[\prod_{{\bf k}} \prod_{\alpha=1}^5 
\hat G^{\dagger}_{
\alpha,{\bf k},-\sigma}]|0\rangle }{ 
\langle 0 |
[\prod_{{\bf k}} \prod_{\alpha=1}^5 \hat G_{\alpha,
{\bf k},
-\sigma}]
\hat c_{5,{\bf k}_{\gamma},-\sigma}
\hat c^{\dagger}_{5,{\bf k}_{\gamma},-\sigma}
[\prod_{{\bf k}} \prod_{\alpha=1}^5 
\hat G^{\dagger}_{\alpha,{\bf k},-\sigma}]|0 \rangle},
\nonumber
\end{eqnarray}
where ${\bf k}_1,{\bf k}_2$ are introduced by the Fourier 
transforms of the $\hat c^{\dagger}_{{\bf i}+
{\bf r}_{n_{\gamma}},-\sigma}$ and $\hat c_{{\bf i}
+{\bf r}_{n_{\gamma}}+{\bf r},-\sigma}$ operators
in (\ref{EQU34}). However, the analysis of (\ref{EQU40}) shows 
that
\begin{eqnarray}
\bar \Gamma({\bf k}_1,{\bf k}_2)= \bar \Gamma({\bf k}_1) 
\delta_{{\bf k}_1,{\bf k}_2}.
\label{EQU41}
\end{eqnarray}
Therefore (\ref{EQU40}) is ${\bf i}$ independent, and via 
$\Gamma_{\bf i}({\bf r})=(1/N_c)\sum_{{\bf i}}
\Gamma_{\bf i}({\bf r}) \equiv \Gamma({\bf r})$ one finds
\begin{eqnarray}
\Gamma({\bf r})&=& \frac{1}{N_c}\sum_{
{\bf k}}[e^{-i{\bf k}\cdot{\bf r}} 
\bar \Gamma({\bf k}) + c.c.],
\quad \bar \Gamma({\bf k})=\frac{\langle \Psi_{-\sigma}| 
\hat n_{5,
{\bf k},-\sigma}|\Psi_{-\sigma}\rangle}{\langle 
\Psi_{-\sigma}|\Psi_{-\sigma}\rangle},
\nonumber\\
|\Psi_{-\sigma}\rangle
&=&\hat c^{\dagger}_{5,{\bf k}_{\gamma},-\sigma}
|\Psi_G\rangle, \quad \hspace*{2.2cm} |\Psi_G\rangle =
[\prod_{{\bf k}'}\prod_{\alpha=1}^5 
\hat G^{\dagger}_{\alpha,{\bf k}',-\sigma}]|0\rangle.
\label{EQU42}
\end{eqnarray}

\subsubsection{Calculation of $\bar \Gamma({\bf k})$ }

In evaluating $\bar \Gamma({\bf k})$ defined in 
(\ref{EQU42}),
we exploit that ${\bf k}_{\gamma}$ in $|\Psi_{-\sigma}
\rangle$ 
is fixed. Furthermore, if ${\bf k}={\bf k}_{\gamma}$ holds, 
$\bar \Gamma({\bf k})=1$ naturally follows from 
(\ref{EQU42}). Hence
\begin{eqnarray}
\bar \Gamma({\bf k})=\delta_{{\bf k},{\bf k}_{\gamma}} + 
(1- \delta_{{\bf k},{\bf k}_{\gamma}}) \frac{\langle 
\Psi_G|(1 -
\hat n_{5,{\bf k}_{\gamma},-\sigma}) \hat n_{5,{\bf k},-\sigma}
|\Psi_G \rangle}{\langle \Psi_G|(1-\hat n_{5,{\bf k}_{\gamma},
-\sigma})|\Psi_G\rangle}.
\label{EQU43}
\end{eqnarray}
In the last term of (\ref{EQU43}), all momentum contributions
appear in factorized form in $|\Psi_G\rangle$. Consequently, 
the contribution of the average of $(1 -\hat n_{5,{\bf k}_{
\gamma},-\sigma})$ cancels out, and one is left with 
\begin{eqnarray}
&&\frac{\langle \Psi_G|(1-\hat n_{5,{\bf k}_{\gamma},-\sigma}) 
\hat n_{5,{\bf k},-\sigma}|\Psi_G \rangle}{\langle \Psi_G|(1-
\hat n_{5,{\bf k}_{\gamma},-\sigma})|\Psi_G\rangle} =
\frac{A({\bf k})}{B({\bf k})},
\nonumber\\
&&A({\bf k})= \langle 0 | \hat G_{5,{\bf k},-\sigma} ... 
\hat G_{1,{\bf k},-\sigma} \hat n_{5,{\bf k},
-\sigma} \hat G^{\dagger}_{1,{\bf k},-\sigma} ...
\hat G^{\dagger}_{5,{\bf k},-\sigma}|0\rangle,
\nonumber\\
&&B({\bf k})= \langle 0 | \hat G_{5,{\bf k},-\sigma} ... 
\hat G_{1,{\bf k},-\sigma} \hat G^{\dagger}_{
1,{\bf k},-\sigma} ...
\hat G^{\dagger}_{5,{\bf k},-\sigma}|0\rangle.
\label{EQU44}
\end{eqnarray}
Equations (\ref{EQU43}) and (\ref{EQU44}) are therefore
summarized as
\begin{eqnarray}
\bar \Gamma({\bf k})=\delta_{{\bf k},{\bf k}_{\gamma}} + 
(1- \delta_{{\bf k},{\bf k}_{\gamma}})\frac{A({\bf k})}{
B({\bf k})}.
\label{EQU45}
\end{eqnarray}

Due to its definition in (\ref{EQU44})
$B({\bf k})$ is of the form
\begin{eqnarray}
B({\bf k})=Det(d_{\alpha,\alpha'}), \quad d_{\alpha,\alpha'}=
\{\hat G_{\alpha,{\bf k},-\sigma}, \hat G^{\dagger}_{\alpha',
{\bf k},-\sigma} \}.
\label{EQU46}
\end{eqnarray}
For the calculation of $A({\bf k})$ one uses 
$\hat n_{5,{\bf k},-\sigma} = 1 - \hat c_{5,{\bf k},-\sigma} 
\hat c^{\dagger}_{5,{\bf k},-\sigma}$, and the second
equation in (\ref{EQU44}) therefore implies
$A({\bf k})=B({\bf k})-F({\bf k})$, where
\begin{eqnarray}
F({\bf k}) =\langle 0 | \hat G_{5,{\bf k},-\sigma}... 
\hat G_{1,{\bf k},-\sigma} [ \hat c_{5,{\bf k},-\sigma} 
\hat c^{\dagger}_{5,{\bf k},-\sigma} ] \hat G^{\dagger}_{
1,{\bf k},-\sigma} ...
\hat G^{\dagger}_{5,{\bf k},-\sigma}|0\rangle.
\label{EQU47}
\end{eqnarray}
Based on this definition, $F({\bf k})$ is calculated as 
$F({\bf k})= Det(f_{\beta,\beta'})$ where the 
$6\times 6$ matrix $(f_{\beta,\beta'})$ is obtained from the 
$5\times 5$ matrix $(d_{\alpha,\alpha'})$ by extending it with 
the matrix elements $f_{1,1}=1, f_{1,1+\alpha}=\{\hat c_{5,
{\bf k},-\sigma},\hat G^{\dagger}_{\alpha,{\bf k},-\sigma} \}=
f^{*}_{1+\alpha,1}$ for $\alpha=1,...,5$, while 
$f_{\beta,\beta'}=d_{\beta-1,\beta'-1}$ for $\beta,\beta' 
\geq 2$ [see also (\ref{A6})]. The result is
\begin{eqnarray}
A({\bf k})=A_1+A_2\cos ({\bf a}\cdot{\bf k}), \quad
B({\bf k})=B_1+B_2\cos ({\bf a}\cdot{\bf k}) > 0, 
\label{EQU48}
\end{eqnarray}
where $A^*({\bf k})=A({\bf k})$, $B^*({\bf k})=B({\bf k})$, and
\begin{eqnarray}
A_1&=&2|t_c| \: \{ \: Q_3^2 \: [ \: 2Q_1^2(2|t_h|+Q_1)
+2t^2(Q_1+|t_h|) +
Q_1(t^2+2Q_1^2) + \frac{Q_1^4+Q_1^2t^2}{|t_h|} \: ] 
\nonumber\\
&+&
\frac{t_f^2}{Q_3^2} \: [ t^2(|t_h|+Q_1) + Q_1^3 - 
\frac{Q_1^4+Q_1^2t^2}{|t_h|}\: ] \: \},
\nonumber\\
A_2&=&2t_c t^2 \: [ \: \frac{|t_h| t_f^2}{Q_3^2} - 
2 Q_3^2(Q_1+|t_h|) \: ],
\nonumber\\
B_1&=&2|t_c| \: \{ \: Q_3^2 \: [ \: (2|t_h|+\frac{Q_1^2}{
|t_h|})(Q_1^2+t^2) +
(t^2+2Q_1^2)(|t_h|+2Q_1) \: ] + \frac{|t_h|t^2t_f^2}{
Q_3^2} \: \},
\nonumber\\
B_2&=&2t^2t_c \: [ \: \frac{|t_h|t_f^2}{Q_3^2} -
Q_3^2(|t_h|+2Q_1) \: ] .
\label{EQU49}
\end{eqnarray}
The last two equations in (\ref{EQU49}) ensure $B_1 > B_2$
and therefore $B({\bf k}) > 0$.

From (\ref{EQU42},\ref{EQU45}) one finally obtains
\begin{eqnarray}
\Gamma({\bf r})= \frac{2}{N_c} \cos ( {\bf k}_{\gamma}\cdot
{\bf r} )  + \frac{2}{N_c} 
\sum_{{\bf k}\ne {\bf k}_{\gamma}} 
\frac{A({\bf k})}{B({\bf k})}\cos({\bf k}\cdot{\bf r}),
\label{EQU50}
\end{eqnarray}
or alternatively
\begin{eqnarray}
\Gamma({\bf r})=\frac{2}{N_c} (1- \frac{A({\bf k}_{\gamma})}{
B({\bf k}_{\gamma})})
\cos ({\bf k}_{\gamma}\cdot{\bf r}) + I({\bf r}), \quad
I({\bf r})= \frac{2}{N_c}\sum_{{\bf k}}
\frac{A({\bf k})}{B({\bf k})} \cos({\bf k}\cdot{\bf r}). 
\label{EQU51}
\end{eqnarray}
 
\subsubsection{Plane wave contribution to $\Gamma({\bf r})$}

In order to analyze the result for $\Gamma({\bf r})$ we define
the planewave state with wave vector ${\bf k}_{\gamma}$
\begin{eqnarray}
|\phi\rangle=\hat c^{\dagger}_{5,{\bf k}_{\gamma},-\sigma}
|0\rangle,
\label{EQU51a}
\end{eqnarray}
and calculate its long-range hopping expectation
value denoted by $\Gamma^{0}_{\bf i}({\bf r})$ 
\begin{eqnarray}
\Gamma^0_{\bf i}({\bf r}) = \frac{\langle \phi| [ 
\hat c^{\dagger}_{{\bf i}+{\bf r}_5,-\sigma} \hat c_{{\bf i}+
{\bf r}_5+{\bf r},-\sigma} + H. c. ]|\phi \rangle}{\langle
\phi |\phi \rangle}.
\label{EQU52}
\end{eqnarray}
One obtains the standard planewave result
\begin{eqnarray}
\Gamma^0_{\bf i}({\bf r}) &=& \Gamma^0({\bf r})=
\frac{2}{N_c} \sum_{{\bf k}_1}
\langle \phi | \hat n_{2,{\bf k}_1,-\sigma} | \phi \rangle 
\cos ({\bf k}_1 \cdot {\bf r}) = \frac{2}{N_c} 
\sum_{{\bf k}_1} 
\delta_{{\bf k}_1,{\bf k}_{\gamma}} 
\cos ({\bf k}_1\cdot{\bf r})
\nonumber\\ 
&=& \frac{2}{N_c} \cos ({\bf k}_{\gamma}\cdot{\bf r}).
\label{EQU53}
\end{eqnarray}
Incorporating this result in (\ref{EQU51}) leads to
\begin{eqnarray}
\Gamma({\bf r})=\Gamma^0({\bf r}) \Big( 1- 
\frac{A({\bf k}_{\gamma})}{B({\bf k}_{\gamma})} \Big) 
+ I({\bf r}).
\label{EQU54}
\end{eqnarray}

The function $I({\bf r})$ defined in (\ref{EQU51}) can be 
explicitly calculated in the thermodynamic limit replacing the
discrete sum $(1/N_c)\sum_{{\bf k}}$ by the integral
$(1/2\pi)\int_{-\pi}^{+\pi}dk$ with $k={\bf k}\cdot{\bf a}$.
In this limit we obtain $I({\bf r})=0$ for $B_2=0$, while  
$B_2 \ne 0$ leads to the relation
\begin{eqnarray}
I({\bf r}=N_a{\bf a})
=\frac{2}{\pi B_2} \frac{(A_1B_2-B_1A_2)}{B_1} \int_0^{\pi} 
\frac{\cos (N_ak) \: dk}{1+ q \cos k},
\label{EQU55}
\end{eqnarray}
where $0 < q=B_2/B_1 <1$, and $N_a$ is an arbitrary and 
nonzero integer. The integral can be explicitly calculated 
\cite{GR1}: 
\begin{eqnarray}
\int_0^{\pi} \frac{\cos (N_a x)}{1+q \cos x} dx 
= \frac{\pi}{\sqrt{1-q^2}}
\: \Big( \frac{\sqrt{1-q^2} -1}{q} \Big)^m , \quad q^2 < 1 .
\label{EQU56}
\end{eqnarray}
Introducing the notation $p=|\sqrt{1-q^2}-1|/|q| < 1$, one 
therefore obtains
\begin{eqnarray}
I({\bf r}=N_a {\bf a}) \sim p^{N_a} = e^{-{N_a} {\bar K}}, 
\quad \bar K = ln \frac{1}{p} > 0.
\label{EQU57}
\end{eqnarray}
Hence, $I({\bf r})$ decreases exponentially in the 
thermodynamic limit.

\subsection{ Conclusions for $\Gamma({\bf r})$ }

For fillings above a half filled upper band, e.g. 
for $N=N^*+1$,
Eqs. (\ref{EQU54},\ref{EQU57}) tell in the thermodynamic limit 
that for long distances
\begin{eqnarray}
\Gamma({\bf r})=\Gamma^0({\bf r}) \Big( 1- \frac{A({\bf k}_{
\gamma})}{B({\bf k}_{\gamma})} \Big).
\label{EQU58}
\end{eqnarray}
Hence the behavior of the system is governed by a free plain 
wave contribution which is renormalized by a ${\bf k}_{\gamma}$
dependent factor. Consequently, the system is itinerant, the 
ground state has extended nature, and since there is no
charge excitation gap, we conclude that the system is
conducting. This conclusion holds also for other choices of
$n_{\gamma}$ and also increasing $\bar N$.
For $N=N^*$ the expression for $\Gamma({\bf r})$ does not 
contain the ${\bf k}_{\gamma}$ contribution, hence it is given 
only by the $I({\bf r})$ term. Consequently, the ground state 
(\ref{EQU31}) is localized in the thermodynamic limit.

\section{Correlation functions}

\subsection{Introduction}

In order to further characterize the deduced ground states, we
calculate in this section the spin-spin 
($\Gamma^{SS}_{{\bf i},{\bf j}}$)
and the density-density ($\Gamma^{NN}_{{\bf i},{\bf j}}$) 
correlation functions at $N=N^*$.
The ground state is $|\Psi_g(N^*)\rangle$ presented
in (\ref{EQU31},\ref{EQU32},\ref{EQU39}), and one has
\begin{eqnarray}
&&\Gamma^{SS}_{{\bf i},{\bf j}}({\bf r}={\bf r}_{\bf i} - 
{\bf r}_{\bf j})= \langle
\hat {\bf S}_{{\bf i}} \cdot \hat {\bf S}_{{\bf j}} \rangle 
- \langle \hat {\bf S}_{{\bf i}}
\rangle \cdot \langle \hat {\bf S}_{{\bf j}} \rangle,
\nonumber\\
&&\Gamma^{NN}_{{\bf i},{\bf j}}({\bf r}={\bf r}_{\bf i} - 
{\bf r}_{\bf j})= \langle
\hat n_{\bf i}  \hat n_{\bf j} \rangle 
- \langle \hat n_{\bf i}
\rangle \langle \hat n_{\bf j} \rangle,
\label{EQU59}
\end{eqnarray}
where $\langle ... \rangle = \langle \Psi({N^*})|...|
\Psi({N^*})\rangle/\langle \Psi({N^*})|\Psi({N^*}) \rangle$, 
and ${\bf i},{\bf j}$ are arbitrary. The notations used in
(\ref{EQU59}) are standard, 
$\hat n_{\bf i}=\sum_{\sigma} \hat n_{{\bf i},\sigma}$, 
$\hat n_{{\bf i},\sigma}= \hat c^{\dagger}_{{\bf i},\sigma}
\hat c_{{\bf i},\sigma}$, and 
\begin{eqnarray}
&&\hat S_{{\bf i},x}=\frac{1}{2}(\hat S_{{\bf i},+} 
+ \hat S_{{\bf i},-}), \quad
\hat S_{{\bf i},y}=\frac{1}{2i}(\hat S_{{\bf i},+} 
- \hat S_{{\bf i},-}), \quad
\hat S_{{\bf i},z}=\frac{1}{2}(\hat n_{{\bf i},\uparrow} 
- \hat n_{{\bf i},
\downarrow}),
\nonumber\\
&&\hat S_{{\bf i},+}=\hat c^{\dagger}_{{\bf i},\uparrow} 
\hat c_{{\bf i},
\downarrow}, \quad \hat S_{{\bf i},-}=\hat c^{\dagger}_{
{\bf i},\downarrow} 
\hat c_{{\bf i},\uparrow}.
\label{EQU60}
\end{eqnarray} 
In the subsequent calculations $\sigma=\uparrow$ is used in
in the expression of the ground state (\ref{EQU32}).

\subsection{Starting relations regarding the expectation values}

The expectation values in (\ref{EQU59}) are essentially 
determined by the properties of $|\Psi({N^*}) \rangle$,
which provide at an arbitrary site ${\bf j}$
\begin{eqnarray}
&&\hat n_{{\bf j},\uparrow} |\Psi({N^*})\rangle = 
|\Psi({N^*})\rangle, \quad
\hat N_{\downarrow}|\Psi({N^*})\rangle = 5N_c, 
\quad \hat S_{{\bf j},+}|\Psi({N^*})\rangle = 0, 
\nonumber\\
&&\langle \Psi({N^*}) |\hat S_{{\bf j},+}|\Psi({N^*})\rangle =
(\langle \Psi({N^*}) |\hat S_{{\bf j},-}|\Psi({N^*})\rangle)^* 
= 0,
\label{EQU61}
\end{eqnarray}
where the operator of the total
number of down spin electrons $\hat N_{\downarrow}$ 
is a constant of motion. As mentioned in connection with 
(\ref{EQU40}), all $\hat c_{n,{\bf k},\sigma=\uparrow}$
operators are present in the ground state, and
in the $-\sigma=\downarrow$ part all ${\bf k}$ values are
multiplicatively separated; the norm is therefore provided by
$\langle \Psi_g(N^*) | \Psi_g(N^*) \rangle = 
\langle \Psi_G | \Psi_G \rangle = \prod_{{\bf k}=1}^{
N_c} B({\bf k})$, where
\begin{eqnarray}
B({\bf k}) = \langle 0|\prod_{\alpha=1}^5 
\hat G_{\alpha,{\bf k},
\downarrow}|\prod_{\alpha=1}^5 \hat G^{\dagger}_{\alpha,
{\bf k},\downarrow}|0\rangle.
\label{EQU62}
\end{eqnarray}
The explicit expression of $B({\bf k})$ is
presented in the last line of (\ref{EQU44}). 

Based on 
(\ref{EQU62}), for an arbitrary operator 
$\hat A_{{\bf k},\downarrow}$ one obtains
\begin{eqnarray}
\langle \hat A_{{\bf k},\downarrow}\rangle = 
\frac{1}{B({\bf k})}
\langle 0|\prod_{\alpha=1}^5 \hat G_{\alpha,{\bf k},
\downarrow}| \hat A_{{\bf k},\downarrow}|\prod_{\alpha=1}^5
\hat G^{\dagger}_{\alpha,{\bf k},\downarrow}|0\rangle.
\label{EQU63}
\end{eqnarray}
One further notes that for an arbitrary index $n$
one has
\begin{eqnarray}
\langle \hat c_{n,{\bf k}_1,\downarrow} \hat c^{\dagger}_{
n,{\bf k}_2,\downarrow}\rangle = \delta_{{\bf k}_1,
{\bf k}_2} \langle \hat c_{n,{\bf k}_1,\downarrow} 
\hat c^{\dagger}_{n,{\bf k}_1,\downarrow}\rangle.
\label{EQU64}
\end{eqnarray}
For the following calculational steps we
introduce the notations
\begin{eqnarray}
&&Y_n=\frac{1}{N_c}\sum_{{\bf k}} Y_{n,{\bf k}}, 
\quad Y_{n,{\bf k}}= \langle \hat c_{n,{\bf k},\downarrow}
\hat c^{\dagger}_{n,{\bf k},\downarrow} \rangle,
\nonumber\\
&&Z_n({\bf r})=\frac{1}{N_c}\sum_{{\bf k}}
e^{-i {\bf k}\cdot {\bf r}} Y_{n,{\bf k}},
\label{EQU65}
\end{eqnarray}
where $0 \leq Y_{n,{\bf k}} \leq 1$ and
$0 \leq Y_{n} \leq 1$ holds.

\subsection{On-site expectation values}

Using (\ref{EQU61},\ref{EQU65}), for arbitrary $n$ and
${\bf i}$, in the spin case one finds
\begin{eqnarray}
&&\langle \hat S_{{\bf i}+{\bf r}_n,z} \rangle =
\frac{Y_n}{2},
\nonumber\\
&&\langle {\hat {\bf S}_{{\bf i}+{\bf r}_n}}^2 \rangle =
3 \langle {\hat S_{{\bf i}+{\bf r}_n,z}}^2 \rangle =
\frac{3}{4} \frac{1}{N_c} \sum_{{\bf k}} 
\langle \hat c_{n,{\bf k},\downarrow} \hat c^{\dagger}_{
n,{\bf k},\downarrow} \rangle = \frac{3}{4} Y_n,
\label{EQU66}
\end{eqnarray}
while in the density case one has
\begin{eqnarray}
\langle \hat n_{{\bf i}+{\bf r}_n} \rangle = 2 - Y_n.
\label{EQU67}
\end{eqnarray}

Using the presented results, for example
the average spin per unit cell can be calculated. 
For the unit cell at site
${\bf i}$, using the first line of (\ref{EQU66}) one finds
\begin{eqnarray}
\sum_{n=1}^6 \langle \hat S_{{\bf i}+{\bf r}_n,z}
\rangle = \frac{1}{2} \sum_{n=1}^6 Y_n.
\label{EQU68}
\end{eqnarray}
As seen, the average spin per cell is ${\bf i}$ independent,
and as shown below (\ref{EQU65}) the $Y_n$ terms are 
positive. Consequently, the ferromagnetic character of the 
analyzed ground state is
underlined by the constant value of (\ref{EQU68}) along the
chain.

\subsection{The two-site contributions}

\subsubsection{The spin-spin correlation case}

For the spin case at ${\bf j}_1 \ne {\bf j}_2$ and
${\bf j}_1={\bf i}_1
+{\bf r}_n$, ${\bf j}_2={\bf i}_2+{\bf r}_n$
one finds
\begin{eqnarray}
\langle \hat {\bf S}_{{\bf j}_1} \cdot\hat {\bf S}_{{\bf j}_2}
\rangle &=& \frac{1}{4N_c^2} \sum_{{\bf k}_1,{\bf k}_2} 
\sum_{{\bf k}_3,{\bf k}_4}
e^{-i({\bf k}_1 \cdot {\bf j}_2 + {\bf k}_2\cdot {\bf j}_1 -
{\bf k}_3 \cdot {\bf j}_1 -{\bf k}_4 \cdot
{\bf j}_2)} \langle \hat c_{n,{\bf k}_1,\downarrow} 
\hat c_{n,{\bf k}_2,\downarrow}
\hat c^{\dagger}_{n,{\bf k}_3,\downarrow} 
\hat c^{\dagger}_{n,{\bf k}_4,\downarrow})
\rangle
\nonumber\\
&=&\frac{1}{4N_c^2} \sum_{{\bf k}_1,{\bf k}_2;{\bf k}_1 
\ne {\bf k}_2} 
(1-e^{-i({\bf j}_2-{\bf j}_1)\cdot ({\bf k}_1-{\bf k}_2)})
\langle \hat c_{n,{\bf k}_1,\downarrow}
\hat c^{\dagger}_{n,{\bf k}_1,\downarrow} 
\hat c_{n,{\bf k}_2,\downarrow} 
\hat c^{\dagger}_{n,{\bf k}_2,\downarrow} \rangle,
\label{EQU69}
\end{eqnarray} 
where one notes that only  the ${\bf k}_2={\bf k}_3 \ne 
{\bf k}_1={\bf k}_4$ and ${\bf k}_1={\bf k}_3 \ne 
{\bf k}_2={\bf k}_4$ terms from the first line 
of (\ref{EQU69}) provide nonzero contributions. 
Introducing the notation ${\bf r}=
{\bf i}_2-{\bf i}_1$, one finds for these two cases
\begin{eqnarray}
&&\frac{1}{N_c^2}\sum_{{\bf k}_1,{\bf k}_2;{\bf k}_1 
\ne {\bf k}_2} 
\langle \hat c_{n,{\bf k}_1,\downarrow}
\hat c^{\dagger}_{n,{\bf k}_1,\downarrow} 
\hat c_{n,{\bf k}_2,\downarrow} 
\hat c^{\dagger}_{n,{\bf k}_2,\downarrow} \rangle =
 Y_n^2  - \frac{1}{N_c^2}
\sum_{{\bf k}_1} Y^2_{n,{\bf k}_1},
\nonumber\\
&&\frac{1}{N_c^2}\sum_{{\bf k}_1,{\bf k}_2;{\bf k}_1 \ne 
{\bf k}_2}
e^{-i{\bf r}\cdot({\bf k}_1-{\bf k}_2)} 
\langle \hat c_{n,{\bf k}_1,\downarrow}
\hat c^{\dagger}_{n,{\bf k}_1,\downarrow} 
\hat c_{n,{\bf k}_2,\downarrow} 
\hat c^{\dagger}_{n,{\bf k}_2,\downarrow} \rangle =
|Z_n({\bf r})|^2 - \frac{1}{N_c^2}
\sum_{{\bf k}_1} Y^2_{n,{\bf k}_1}.
\label{EQU70}
\end{eqnarray} 
Hence,
using (\ref{EQU70}) together with the first line of
(\ref{EQU66}) one obtains
\begin{eqnarray}
\langle {\hat {\bf S}}_{{\bf i}_1+{\bf r}_n} \cdot
{\hat {\bf S}}_{{\bf i}_2+{\bf r}_n} \rangle = 
\langle {\hat {\bf S}}_{{\bf i}_1 +{\bf r}_n} \rangle \cdot
\langle {\hat {\bf S}}_{{\bf i}_2 + {\bf r}_n} 
\rangle - \frac{1}{4}
|Z_n({\bf r})|^2, \quad 
\langle {\hat {\bf S}}_{{\bf i}_1 +{\bf r}_n} \rangle \cdot
\langle {\hat {\bf S}}_{{\bf i}_2 + {\bf r}_n} 
\rangle = \frac{1}{4}Y_n^2.
\label{EQU71}
\end{eqnarray}

\subsubsection{The density-density correlation case}

Given the specific expression of the ground state
$|\Psi_g(N^*)\rangle$ it follows that
\begin{eqnarray}
\langle \hat n_{{\bf j}_1} \hat n_{{\bf j}_2} \rangle =
4 \langle \hat {\bf S}_{{\bf j}_1} \cdot \hat {\bf S}_{
{\bf j}_2} \rangle - 8 (\langle {
\hat S_{{\bf j}_1,z}}^2 \rangle + \langle {\hat S_{
{\bf j}_2,z}}^2 \rangle)  + 4.
\label{EQU73}
\end{eqnarray}
Using the second line of (\ref{EQU66}), and  
(\ref{EQU67},\ref{EQU71}) one finds
\begin{eqnarray}
\langle \hat n_{{\bf i}_1+{\bf r}_n} \hat n_{{\bf i}_2+
{\bf r}_n} \rangle = \langle \hat n_{{\bf i}_1+{\bf r}_n} 
\rangle \langle\hat n_{{\bf i}_2+{\bf r}_n} \rangle 
- |Z_n({\bf r})|^2, \quad
\langle \hat n_{{\bf i}_1+{\bf r}_n} \rangle 
\langle\hat n_{{\bf i}_2+{\bf r}_n} \rangle = (Y_n-2)^2.
\label{EQU74}
\end{eqnarray}

\subsection{The deduced expression of the correlation functions}

Based on (\ref{EQU71},\ref{EQU74}) the correlation functions
from (\ref{EQU59}) become
\begin{eqnarray}
\Gamma^{SS}_{{\bf i}_1+{\bf r}_n,{\bf i}_2+{\bf r}_n}(
{\bf r})=\frac{1}{4} \Gamma^{NN}_{{\bf i}_1+{\bf r}_n,
{\bf i}_2+{\bf r}_n}({\bf r})= - \frac{1}{4}|Z_n({\bf r})
|^2.
\label{EQU75}
\end{eqnarray}
These expressions are valid for ${\bf i}_1 \ne
{\bf i}_2$, i.e. $|{\bf r}| \ne 0$. The case $|{\bf r}|=0$ must
be separately deduced and based on (\ref{EQU66},\ref{EQU67})
it leads to
\begin{eqnarray}
\Gamma^{SS}_{{\bf i}_1+{\bf r}_n,{\bf i}_1+{\bf r}_n}(0)
=\frac{Y_n}{4}(3-Y_n), \quad
\Gamma^{NN}_{{\bf i}_1+{\bf r}_n,
{\bf i}_1+{\bf r}_n}({\bf r})= Y_n(1-Y_n).
\label{EQU76}
\end{eqnarray}

\subsection{Characteristics}

\subsubsection{The $Y_n$ expression}

The $Y_n$ term is calculated based on the first line of
(\ref{EQU65}). Starting from $Y_{n,{\bf k}}$ 
in (\ref{EQU65}),
one finds $Y_{n,{\bf k}}=F_n({\bf k})/B({\bf k})$, where in
analogy to (\ref{EQU47})
\begin{eqnarray}
F_n({\bf k}) =\langle 0 | \hat G_{5,{\bf k},\downarrow} 
\hat G_{4,{\bf k},\downarrow}... 
\hat G_{1,{\bf k},\downarrow} 
[ \hat c_{n,{\bf k},\downarrow} \hat c^{\dagger}_{n,
{\bf k},\downarrow} ] \hat G^{\dagger}_{1,{\bf k},\downarrow} 
\hat G^{\dagger}_{2,{\bf k},\downarrow} ...
\hat G^{\dagger}_{5,{\bf k},\downarrow}|0\rangle.
\label{EQU77}
\end{eqnarray}
With the same strategy as described for (\ref{EQU47}) one 
finds 
\begin{eqnarray}
F_n({\bf k})=Y_{1,n}+Y_{2,n}\cos ({\bf k}\cdot{\bf a}),
\label{EQU78}
\end{eqnarray}
where, as an example for the case $n=2$, the scalars
$Y_{1,n},Y_{2,n}$ are presented in the Appendix A. Finally
one obtains
\begin{eqnarray}
Y_{n,{\bf k}}=\frac{Y_{1,n}+Y_{2,n} \cos({\bf k}\cdot{\bf a})}{
B_1+B_2 \cos({\bf k}\cdot{\bf a})},
\label{EQU79}
\end{eqnarray}
where $B_1,B_2$ have been calculated in (\ref{EQU49}). 
Inserting (\ref{EQU79}) in (\ref{EQU65}), the integration
in the thermodynamic limit proceeds as described for 
(\ref{EQU55}), and
taking into account $B_1> |B_2|$, yields
\begin{eqnarray}
Y_n= \frac{Y_{2,n}}{B_2} + \frac{Y_{1,n}B_2-B_1Y_{2,n}}{B_2
\sqrt{B_1^2-B_2^2}}.
\label{EQU80}
\end{eqnarray}

\subsubsection{The $Z_n({\bf r})$ expression}

For the calculation of $Z_n({\bf r})$, (\ref{EQU79}) is 
inserted in the second line of (\ref{EQU65}). In the
thermodynamic limit the calculational steps described in
(\ref{EQU55}-\ref{EQU57}) are applied which yields 
\begin{eqnarray}
Z_n(r=N_a a)= \frac{(Y_{1,n}B_2-B_1Y_{2,n})}{B_2} \: 
\frac{[-sign(B_2)]^{N_a}}{\sqrt{B_1^2-B_2^2}}
\: e^{-N_a \bar K },
\label{EQU81}
\end{eqnarray}
where $\bar K$ has been defined in (\ref{EQU57}), and $N_a$
is an arbitrary integer.

\subsection{Physical considerations}

First we recall that the system contains
$N_{\Lambda}=6N_c$ sites, hence the maximum
total number of electrons is $N_{Max}=12N_c$. In 
the analyzed case one has $N=11N_c$ electrons, $6N_c$
with spin $\uparrow$, and $5N_c$ with spin $\downarrow$.
Since $N_{\uparrow}=N_{\Lambda}$, the $\sigma=\uparrow$
electrons are localized and frozen. Only the $\sigma=
\downarrow$ electrons are able to move on $N_c$ empty 
positions (not empty sites). 
$N_{\uparrow}-N_{\downarrow}=N_c$ 
creates a nonsaturated ferromagnetic state, whose total 
spin per cell (with 6 sites) is 
relatively small ($1/2$ in $\hbar$ units); the total spin
per site is therefore $1/12$ in the same units. Besides, 
there is 
only one spin-down electron per cell, and all these 
circumstances provide a specific correlation behavior.
The long-range ferromagnetic order
$\lim_{{\bf r} \to \infty} \langle \hat {\bf S}_{{\bf i}+
{\bf r}_n} \cdot \hat {\bf S}_{{\bf i}+{\bf r}+{\bf r}_n} 
\rangle = Y_n^2/4 \ne 0$ is not altered at long distances 
(where, see (\ref{EQU81}), $Z_n(r) \sim \exp[-(r/a)\bar K] 
\to 0$). But, on short
distances covering several cells [note that $\bar K$ in 
(\ref{EQU57}), because of the logarithm, is usually a small
number], the correlations are diminished relative to their
long range value, i.e $\langle \hat {\bf S}_{{\bf i}+
{\bf r}_n} \cdot \hat {\bf S}_{{\bf i}+{\bf r}+{\bf r}_n} 
\rangle < Y_n^2/4$. 

Concerning the density correlations, at long distances (
$|{\bf r}|>>1$) one has
$\langle \hat n_{{\bf i}+{\bf r}_n} \hat n_{{\bf i}+{\bf r}_n
+{\bf r}} \rangle \to (Y_n-2)^2$ which represents usually a 
nonzero, ${\bf i}$ and ${\bf r}$ independent value. This value
is diminished for small $|{\bf r}|$, which shows that short
range fluctuations are present into the system.

\section{Uniqueness Proof}

\subsection{Uniqueness at ${\bf N}={\bf N}^*$}

\subsubsection{Introduction}

We prove below the
uniqueness of the solution (\ref{EQU31},\ref{EQU32}) which 
represents the ground state $|\Psi_g(N^*)\rangle$
at $N=N^*$. This means that $|\Psi_g(N^*)\rangle$
spans the $S=S_z=S_z^{Max}=N_c/2$ sector of $ker(\hat H')$, 
and apart from the trivial $(2S+1)$fold degeneracy related 
to the orientation of the total spin, the ground state is 
non-degenerate.

To prove the uniqueness, we follow the strategy already 
described in Sect. I.B.1. Based on (\ref{EQU13}), one
starts from the positive semidefinite Hamiltonian 
$\hat H'=\hat H-C_g=\hat H_G + \hat H_P$. By 
introducing $\hat H_G(\alpha,{\bf i},\sigma)=\hat G_{\alpha,
{\bf i},\sigma} \hat G^{\dagger}_{\alpha,{\bf i},\sigma}$,
the Hamiltonian $\hat H'$
becomes a sum of positive semidefinite terms of the form
\begin{eqnarray}
\hat H' = \sum_{\sigma} \sum_{\alpha=1}^5 \sum_{
{\bf i}} \hat H_G(\alpha,{\bf i},\sigma) + \hat H_P,
\quad \hat H_G=\sum_{\sigma} \sum_{\alpha=1}^5 \sum_{
{\bf i}} \hat H_G(\alpha,{\bf i},\sigma). 
\label{EQU82}
\end{eqnarray} 
Consequently
\begin{eqnarray}
&&ker(\hat H')=ker(\hat H_G) \bigcap ker(\hat H_P),
\nonumber\\
&&ker(\hat H_G)=\prod_{{\bf i}}\prod_{\sigma} [
ker(\hat H_G(1,{\bf i},\sigma))\bigcap \hat H_G(2,{\bf i},
\sigma) \bigcap ....\bigcap ker(\hat H_G(5,{\bf i},
\sigma))].
\label{EQU83}
\end{eqnarray}

The proof of the uniqueness of a ground
state $|\Psi_g\rangle$  proceeds in two steps, namely first
(step I.) one demonstrates that $|\Psi_g\rangle$ is contained 
in $ker(\hat H')$, and second (step II.) one shows that all 
elements of $ker(\hat H')$ can be expressed in terms of
$|\Psi_g\rangle$.

For $N=N^*$ and $|\Psi_g(N^*)\rangle$ presented
in (\ref{EQU31},\ref{EQU32}), it was already shown [see
below (\ref{EQU32})] that $\hat H_G |\Psi_g(N^*)\rangle=
\hat H_P |\Psi_g(N^*)\rangle=0$, hence 
$|\Psi_g(N^*)\rangle \in ker(\hat H')$. Consequently, for
the proof of the uniqueness of $|\Psi_g(N^*)\rangle$ we
focus on step II, but also take into account the 
consequences arising from step I. 
The demonstration follows a sequence of seven steps
(for details see Appendix B) as follows:

First (Lemma 1) we show that in the presence of 
N particles, all vectors
from $ker[\hat H_G(\alpha,{\bf i},\sigma)]$ can be written
in the form $|V_{G,\alpha,{\bf i},\sigma}\rangle = 
\hat G^{\dagger}_{\alpha,{\bf i},\sigma} \hat W^{\dagger}_{
G,\alpha}|0\rangle$, where $\hat W^{\dagger}_{G,\alpha}$ is
an arbitrary operator which introduces $(N-1)$ electrons 
into the system and preserves a nonzero norm for
$|V_{G,\alpha,{\bf i},\sigma}\rangle$. 

The second step (Lemma 2) verifies that
in the presence of $N \geq 10N_c$ particles, all vectors 
from $ker(\hat H_G)$ can be written in the form 
$|V_G\rangle = [ \prod_{\sigma} \prod_{\alpha=1}^5 
\prod_{{\bf i}} \hat G^{\dagger}_{\alpha,{\bf i},
\sigma} ] \hat W^{\dagger}_G |0\rangle$,
where $\hat W^{\dagger}_G$ is an arbitrary operator which 
introduces $(N-10N_c)$ particles in the system, and 
preserves a nonzero norm for $|V_G\rangle$.

In the following three steps, first (Lemma 3) we show that
the choice 
$\hat W^{\dagger}_G(N=11N_c=N^*)=
\hat F^{\dagger}=\prod_{{\bf i}} 
\hat c^{\dagger}_{{\bf i}+{\bf r}_{n_{\bf i}},\sigma}$
i.e. a product over $N_c$ creation operators with the same 
fixed spin index with one creation operator taken from 
each cell, leads to the state (\ref{EQU31},\ref{EQU32})
in the form $|V_{G,P}\rangle = [ \prod_{\sigma} \prod_{
\alpha=1}^5 \prod_{{\bf i}} \hat G^{\dagger}_{
\alpha,{\bf i},\sigma} ] \hat F^{\dagger} |0\rangle$ 
inside $ker(\hat H')$. 

In the fourth step
(Lemma 4) we prove that the kernel $ker(\hat H')$ at fixed 
$S=S_z^{Max}=N_c/2$ and $N=N^*$ contains
only the unique vector $|V_{G,P}\rangle$; in the subsequent 
step (Lemma 5) we underline that states with
total spin $S < N_c/2$ are not contained in $ker(\hat H')$.

For the last two steps it remains to show (Lemma 6)
that at total spin
$S=S_z^{Max}=N_c/2$, all $S_z\in [-S,+S]$, components of 
$|V_{G,P}\rangle =|\Psi_g(N^*)\rangle$ are contained in 
$ker(\hat H')$, and (Lemma 7) that $ker(\hat H')$ at 
$N=N^{*}$ is $(N_c+1)$dimensional. Hence it is concluded
that $|\Psi_g(N^*)\rangle$, apart from 
the trivial $(2S+1)$fold degeneracy related to the 
orientation of the total spin, is the unique ground state.

\subsection{Uniqueness at ${\bf N}>{\bf N}^*$}

The proof of the uniqueness at $N=N^*+{\bar N}$ proceeds in a
similar way as in the previous subsection for the
case $N=N^*$. Below we present the proof for
${\bar N}=1$. We do not see reasons why the presented 
procedure to not work for arbitrary $1 < {\bar N} < N_c$.

Using the ground state from (\ref{EQU33})
\begin{eqnarray}
|\Psi_g(N^*+1)\rangle = \hat c^{\dagger}_{n_{\gamma},
{\bf k}_{\gamma},\downarrow}|\Psi_g(N^*)\rangle,
\label{EQU102}
\end{eqnarray}
where $n_{\gamma}$ and ${\bf k}_{\gamma}$ are fixed, but 
arbitrary. Lemma 1 and Lemma 2 work analogously, 
$ker(\hat H_G)$ remains as shown in (\ref{EQU86}), taking into 
account that the particle number has been increased by one. 
For Lemma 3,  $|V_{G,P}\rangle$ in (\ref{EQU89}) is replaced by
\begin{eqnarray}
|V_{G,P}(N^*+1)\rangle = \hat c^{\dagger}_{n_{\gamma},
{\bf k}_{\gamma},\downarrow}[ \prod_{\sigma} 
\prod_{\alpha=1}^5 
\prod_{{\bf i}}\hat G^{\dagger}_{\alpha,{\bf i},\sigma} ] 
\hat F^{\dagger} |0\rangle.
\label{EQU103}
\end{eqnarray}
which via $\hat F^{\dagger}$ from (\ref{EQU90}) is contained
in $ker(\hat H')$. 

Differences in demonstration appear at the 
level of Lemma 4. This is because the state (\ref{EQU103})  
corresponding to $S=S_z^{Max}=(N_c-1)/2$, acquires
a supplementary $N_c$fold degeneracy due to the ${\bf k}$
value of the additional $\bar N=1$ electron with
reversed spin. This $N_c$fold degeneracy in (\ref{EQU103}) is
provided by $(n_{\gamma},{\bf k}_{\gamma})$. But
for a fixed spin projection we start with $6N_c$ canonical 
Fermi operators, from which $5N_c$ linearly independent 
$\hat G^{\dagger}_{\alpha,{\bf i},\downarrow}$ operators 
are created, so $N_c$ linearly independent 
contributions remain for the $(n_{\gamma},
{\bf k}_{\gamma})$ degeneracy.

Differences are encountered also at the level of Lemma 5. 
Since by flipping one spin in 
$\hat F^{\dagger}$ relative to (\ref{EQU90}) in an arbitrary
unit cell at ${\bf i}_1$,
at the level of the state $|\chi_{\sigma}\rangle$
in (\ref{EQU93}), a doubly occupied site and 
an empty site appear in this cell. In order 
to place the ground state inside $ker(\hat H_P)$, the 
,,in principle'' possibility to fill up the empty site by the 
additional $\bar N=1$ electron appears. This, if 
possible, can be done with arbitrary spin, hence destroys
the structure of (\ref{EQU103}) and therefore also 
(\ref{EQU102}). But even if the doubly
occupied site remains fixed at
the place where the spin flip in $\hat F^{\dagger}$ has been
introduced, the position of the empty site 
in different terms of $|\chi_{\sigma}\rangle$ is different. 
Consequently, a multiplicative 
$\hat c^{\dagger}_{{\bf i}_1+{\bf r}_{n_1,e},\sigma_1}$ 
operator introduced in $|V'_{G,P}\rangle$ below 
(\ref{EQU93}) of the form
\begin{eqnarray}
|V'_{G,P}(N^*+1)\rangle = \hat c^{\dagger}_{{\bf i}_1+
{\bf r}_{n_1,e},\sigma_1}
[ \prod_{{\bf i}} 
\prod_{\alpha=1}^5 \hat G^{\dagger}_{\alpha,{\bf i},
-\sigma} ] |\chi_{\sigma}\rangle 
\label{EQU104}
\end{eqnarray}
does not place this vector in $ker(\hat H_P)$. The unique 
correct form is given by (\ref{EQU102}). This remains true
even if one increases the number of spin flips in 
$\hat F^{\dagger}$ relative to (\ref{EQU90}). 

The unique possible total spin for $ker(\hat H_G)
\bigcap ker(\hat H_P)$ is therefore $S=(N_c-1)/2$ and states
in $ker(\hat H')$ have now always this $S$ value. The fact 
that all spin sectors corresponding to $S=(N_c-1)/2$, 
$S_z < S_z^{Max}=S$ are also contained in $ker(\hat H')$ is 
demonstrated identically as shown in Lemma 6 at $N=N^{*}$. 
With the changes mentioned above, Lemma 7 also remains valid. 
In conclusion, since (\ref{EQU102}) belongs to
the spin sector $S=S_z^{Max}=(N_c-1)/2$, it is unique
apart from the trivial $(2S+1)=N_c$ degeneracy related to the 
orientation of the total spin, and apart from the $N_c$-fold 
degeneracy provided by the additional 
electron with reversed spin. Hence the total degeneracy,
and the dimension of $ker(\hat H')$ is $(2S+1)N_c=N_c^2$.

\section{The physical origin of ferromagnetism}

\subsection{Starting characteristics}

In this section we analyze the physical reason
of the emergence of ferromagnetism in the described 
pentagon chain case with half filled upper band. 

Because of the Bravais periodicity,
$N_c$ can be decreased up to $N_c=2$ without to destroy the 
observed ferromagnetic phase. Indeed,
at half filled upper band, using the notations introduced
below (\ref{EQU2}), the number of electrons present in the 
system is 
\begin{eqnarray}
N=(N_b-1)N_{1b}+
N_{1b}/2=2N_c(N_b-1)+N_c=2N_cN_b-N_c=2N_{\Lambda}-N_c.
\label{eqN1}
\end{eqnarray}
If one has $N_c=1$, one obtains $N_{\Lambda}=N_b$ and 
$N=2N_b-1$. Consequently, only one single occupied site is
present in the system, and the rest of $N_b-1$ sites are 
double occupied, hence inert. The remaining one electron 
has an arbitrary spin orientation, i.e. at $N_c=1$ one 
cannot discuss about ferromagnetism in this case.

In conclusion, the $N_c=2$ two cells ring is the smallest
system providing the observed ferromagnetism. Hence, in 
order to understand its emergence reasons, below one 
concentrates on the two cells ring case. 

\subsection{Two cell rings}

\subsubsection{The electron configuration leading to 
ferromagnetism}

At $N_c=2$, according to (\ref{eqN1}), the number of 
electrons in the system is $N=2(N_{\Lambda}-1)$. This means
that $C_1:$ 1 empty site and the rest of $N_{\Lambda}-1$ 
sites doubly occupied holding $N=2(N_{\Lambda}-1)$ 
electrons, or $C_2:$ 2 singly occupied sites
and the rest of $N_{\Lambda}-2=2(N_b-1)$ sites doubly 
occupied holding $4(N_b-1)$ electrons, are the only 
possible electron configurations that can appear into the 
system. On its turn, the configuration $C_2$ can be of
$C_{2;\sigma,-\sigma}$, or $C_{2,\sigma,\sigma}$ type, 
depending on the opposite [i.e. $(\sigma,-\sigma)$], or 
parallel [i.e. $(\sigma,\sigma)$] orientation of the spin of
the two electrons present on the two single occupied sites.

We choose to describe the ferromagnetism in the total spin
$S_z=Max(S_z)=1$ spin sector (note that now, this is also 
the unique $S_z \ne 0$ spin sector). The unique electron 
configuration leading to $S_z=1$ is $C_{2,\sigma,\sigma}$, 
hence this is the unique possible electron configuration 
which emerges in our case in the system when ferromagnetism
appears.   

\subsubsection{The presence of one electron with fixed spin
$\sigma$ on all sites in the ferromagnetic case}

As presented at the subsection VIII.B.2 above, the 
ferromagnetism is 
obtained by the unique configuration $C_{2,\sigma,\sigma}$.
In this configuration, if we take $\sigma=\uparrow$,
the $N_{\Lambda}-2$ double occupied sites provide
$N_{\Lambda}-2$ electrons with spin $\uparrow$, 
furthermore the 2 supplementary electrons on the single 
occupied sites have also $\uparrow$ spin orientation. 
Consequently the number of  spin 
$\uparrow$ electrons is equal to the number of sites in
the system ($N_{\Lambda}$), hence on each site of the 
system one has one electron with spin $\uparrow$, i.e. 
$\langle \hat n_{{\bf i}+{\bf r}_n,\uparrow}\rangle =1$ 
(and also $\hat n_{{\bf i}+{\bf r}_n,\uparrow}|\Psi_g\rangle
=|\Psi_g\rangle$, where $|\Psi_g\rangle$ is the interacting
ground state) for all lattice sites ${\bf i}$ and all 
in-cell site indices $n \leq m= N_b$. 

\subsubsection{The emergence of the upper effective  flat band in 
the ferromagnetic case}

If we interpret the obtained results in a one-particle 
picture (with effective bands), the half filled upper band 
and the above deduced relation $\langle \hat n_{
{\bf i}+{\bf r}_n,\uparrow}\rangle =1$ automatically lead
to an upper flat band. Indeed, taking into account that for
$N_c=2$ one has $N_{1b}=4$, in this case $N_b-1$ bands are
completely filled [providing $N_{1b}(N_b-1)=4(N_b-1)$ 
electrons with zero total spin], and the upper band 
is half filled [providing
$N_{1b}/2=2$ electrons]. As seen from (\ref{eqN1}), since
the total number of electrons is $N=4(N_b-1)+2$, above the
completely filled $N_b-1$ bands, only 2 electrons remain 
available for the upper band.

Taking now into account
that in one fixed band, for a fixed spin projection, 
$N_{1b}/2=2$ places are present, in order to have 
ferromagnetism (i.e. $\langle \hat n_{{\bf i}+{\bf r}_n,
\uparrow}\rangle =1$), our upper half filled band must 
contain the remaining 2 electrons with the same spin 
projection, providing the state with minimum energy. 
This is 
possible to be fulfilled only if the upper band is flat. 

\subsubsection{The average number of double occupied sites in
r-space}

If in an eigenstate of the system $|\phi\rangle$, the 
electron configuration $C_{2,\sigma,-\sigma}$ occurs, since
the action of $\hat H$ on $C_{2,\sigma,-\sigma}$ leads to
the configuration $C_1$ and vice versa, it results that
$|\phi\rangle$, satisfying $\hat H |\phi\rangle=E |\phi\rangle$, (excepting special cases with rare accidental 
cancellations)
will automatically contain also the electron configuration
$C_1$. This in not true for the configuration
$C_{2,\sigma,\sigma}$. Consequently, in ${\bf r}$-space, 
the average number of double occupied sites decreases when 
the configuration  
$C_{2,\sigma,\sigma}$ is present, i.e. see Sect. VIII.B.1, 
the system becomes ferromagnetic.

\subsection{The importance of different U values on different
type of sites}

We demonstrate below that the obtained ferromagnetic state
emerges via the minimization of the interacting energy, and
in this process, the different U values on different type of
sites plays an important role. 
We analyze below $N_c=2$ case with periodic boundary
conditions.

\begin{figure} [h]                                        
\centerline{\includegraphics[width=12 cm,height=6 cm]{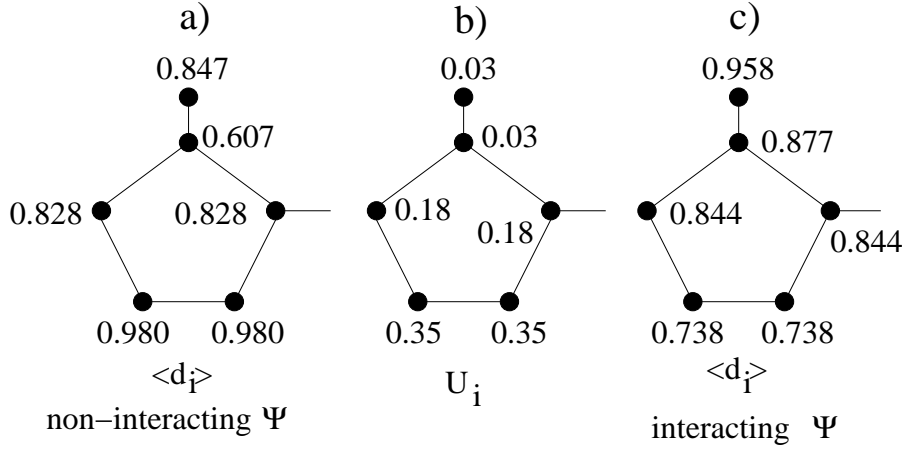}} 
\caption{Used site dependent on-site Coulomb repulsion 
values (Fig.5.b) together with the site dependent double 
occupancies in the non-interacting case (Fig.5.a) and site
dependent double occupancies (Fig.5.c) provided
by the exact interacting ground state in (\ref{EQU32}). 
The $\hat H_0$ parameters are those used in Fig.3 
of Ref. [\cite{m2010}]. With notations used in (5), and
expressed in $t=1$ units, one has
$t_c=0.5, t_h=-1.1, t_f=1.2, \epsilon_1=\epsilon_4=-2.5,
\epsilon_2=\epsilon_3=-2.0, \epsilon_5=\epsilon_6=-2.1,
U_1=U_4=0.18, U_2=U_3=0.35, U_5=U_6=0.03$.}     
\end{figure} 

\subsubsection{The case of the pentagon chain with external 
links}

First we analyze the two cells system holding the 
Hamiltonian parameters present in Fig.3 of Ref. 
[\cite{m2010}]. 
The below presented Fig.5 presents one (arbitrary) cell of 
this case.

Fig.5.a. shows the double occupancy 
$d_{\bf j}= \langle \hat d_{\bf j}\rangle =\langle 
\hat n_{{\bf j},
\uparrow} \hat n_{{\bf j},\downarrow}\rangle$ calculated 
in the
non-interacting ground state ($U_n=0$) case. As can be 
observed, on different type of sites, the double occupancy 
is
different. Hence, even the non-interacting state offers the
possibility to introduce different $U_n$ values on different
type of sites since this possibility gives a guarancy for
the better minimization of the interaction energy when the
interaction term is turned on.

Fig.5.b. presents the $U_n$ values on different type of 
sites inside the unit cell used for the parameter set in 
Fig.3 of Ref. [\cite{m2010}].
These being introduced, the exact
interacting ferromagnetic ground state is obtained at half
filled upper band ($N_c=2, N_b=6, N_{\Lambda}=12, 
N=11N_c=22$), which produces the double occupancies 
presented in Fig.5.c.

Compairing Fig.5.a and Fig.5.c one observes that 
given by the presence of different $U_n$ values on different
type of sites, the system has the possibility to better
minimize its interaction energy. In order to do that, the
non-interacting $d_{\bf j}$ values are reorganized 
such to obtain in the interacting case the
maximum (minimum) double occupancy where $U_{\bf j}$ has its
minimum (maximum) value. A such type of restructuration of
the double occupancy cannot be done at homogeneous 
$U_{\bf j}=U$ for all ${\bf j}$, because in this case the 
change of the double
occupancy at different type of sites has no effect on the
interaction energy. Indeed, in this case 
at $U_{\bf j}=U$, since 
$\hat n_{{\bf j},\uparrow}|\Psi_g\rangle = |\Psi_g\rangle$ 
hence $\langle \hat n_{{\bf j},\uparrow}\rangle =1$ holds 
in the ground state, one obtains
\begin{eqnarray}
E_{int}=\sum_{\bf j} U_{\bf j}\langle \hat n_{{\bf j},
\uparrow} \hat n_{{\bf j},\downarrow}\rangle =U  
\sum_{\bf j} \langle \hat n_{{\bf j},\downarrow}\rangle
=10 U.
\label{eqN2}
\end{eqnarray}
This value is a constant, because in the present case 
$N_{\uparrow}=N_{\Lambda}=12$, and $N_{\downarrow}=
\sum_{\bf j} \langle \hat n_{{\bf j},\downarrow}\rangle=
N-N_{\uparrow}=10=$ constant.

\subsubsection{The decrease in the interaction energy drives the
transition}

In order to obtain an image about different energy 
variations at the transition to the ferromagnetic state 
and see which term drives the transition, we calculated
in the presence of the interaction:  
a) the exact 
kinetic ($E_{kin}$), interaction ($E_{int}$), and total
($E_g=E_{kin}+E_{int}$) energy provided by the exact ground
state in the interacting case, and b) the kinetic 
($E_{0,kin}$), interaction ($E_{0,int}$), and total 
($E_{0,g}=E_{0,kin}+E_{0,int}$) energy deduced from the
non-interacting ground state used as a trial wave function
(at $U_n \ne 0$). 
After this step we deduce the energy variations
$\Delta E_{\mu}=E_{\mu}-E_{0,\mu}$, where $\mu=$kin, int, g,
and express $\delta E_{\mu}=\Delta E_{\mu}/E_{0,\mu}$ in 
percents (the mathematical details of the calculation technique are presented in the Appendix D).

In the case of the pentagon chain described in 
Fig.5, one finds (in $t=1$ units) $E_{0,kin}=-52.597,
E_{0,int}=2.055, E_{0,g}=-50.541$, while for the exact
ground state $E_{kin}=-51.223, E_{int}=0.664, E_g=-50.558$.
Consequently, when the ferromagnetic phase emerges, the
interaction energy decreases $69.5 \%$, the kinetic energy 
increases $2.6\%$, and as a consequence the total energy
decreases $0.03\%$. As can be seen, clearly, the transition
is driven by the strong decrease of the interaction energy. 
In the same time, the kinetic energy is practically 
quenched in
the close vicinity of the kinetic energy $E_{0,kin}$ 
present 
before the interactions were turned on (note that 
$E_{0,kin}$ remains the same at $U_n=0$ for all $n$, 
see Appendix D).

\begin{figure} [h]                                        
\centerline{\includegraphics[width=7 cm,height=5 cm]{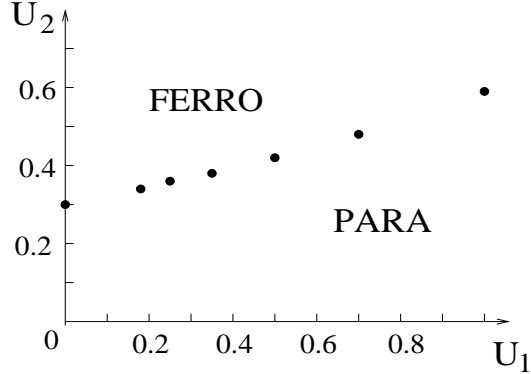}} 
\caption{The ferromagnetic and paramagnetic regions in the
phase diagram for the system analyzed in Fig.5, 
taking into account variable $U_n$. One notes that $U_2$ is
on the bottom site of the pentagon while $U_1$ is at the 
site ${\bf i}$ on the line of the chain (see Fig.3). 
Excepting $U_n$,
the other $\hat H$ parameters have been maintained
unchanged.}   
\end{figure} 

\subsubsection{The $U_n$ dependence of  $|\Psi_g\rangle$ }

We must underline here a main characteristics of the exact
interacting ground state. In order to produce a 
redistribution of the double occupancy such to introduce
high (small) $d_{\bf j}$ on ${\bf j}$
sites where $U_{\bf j}$ is small (high) and to obtain in 
this manner the extremely important strong decrease of the
interaction energy, the interacting ground state 
$|\Psi_g\rangle$ must depend on the $U_n$ interaction 
strength values (otherwise, $|\Psi_g\rangle$ not ``knows''
where $U_n$ is high and where is small). Without this 
information, the mentioned redistribution of the double 
occupancy cannot be done.

\subsubsection{Lower bound for the interaction at the
emergence of ferromagnetism}

Another characteristics of the emerging ferromagnetism is
the presence of the lower bound for the interaction when 
the ferromagnetism appears (see exemplification in Fig.6).
The motivations for this behavior are the inequalities 
(\ref{EQU21}) which are conditioning the solution of the 
matching equations (\ref{EQU14},\ref{EQU15}).

We note that given by the complex structure of the
inequalities (\ref{EQU21}), the behavior seen in
Fig.6 differs from that obtained for example during the 
particle-hole transformed Mielke-Tasaki behavior to the 
upper band for the simple triangular case 
[see (\ref{eqN7})], where the ferromagnetic 
solution, at fixed $\hat H_0$ parameters, and a coordinate
system as used in Fig.6, exists only on a line.

\subsubsection{The case of the pentagon chain without external 
links and antennas}

In order to see in what extent the obtained results for the
pentagon chain analyzed in Fig.5. are influenced by external
bonds, antennas, and $\epsilon_n$ on-site one-particle
potentials, we describe below the $N_c=2$ case of a pentagon
chain without such characteristics
(see mathematical details in Appendix E). 
The system and the used notations are presented in
Fig.7. We note that the results obtained at half filled
upper band are similar to those obtained for the pentagon
chain with external links described in Fig.5. In the present
case $N_b=4$, $N_{\Lambda}=8$ and $N=14$ corresponds to the
half filled upper band.

\begin{figure} [h]                                        
\centerline{\includegraphics[width=6 cm,height=3 cm]{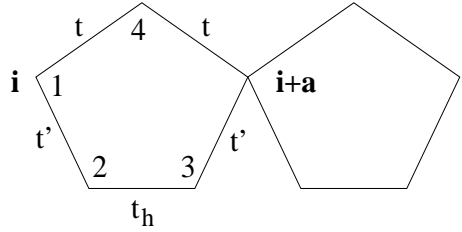}} 
\caption{The pentagon chain without external links and
antennas. The numbers represent the in-cell notation of 
sites $n$. The hopping matrix elements connected to 
different bonds are also specified.}   
\end{figure} 

First we present the results for the Hamiltonian parameter
values
\begin{eqnarray}
&&t=1.0, \: t'=1.0, \: t_h=-1.1, U_2=U_3=0.5, \: U_1=0.18, 
\: U_4=0.035,
\nonumber\\
&&\epsilon_1=\epsilon_2=\epsilon_3=\epsilon_4=0.
\label{eqN3}
\end{eqnarray}

\begin{figure} [h]                                        
\centerline{\includegraphics[width=12 cm,height=6 cm]{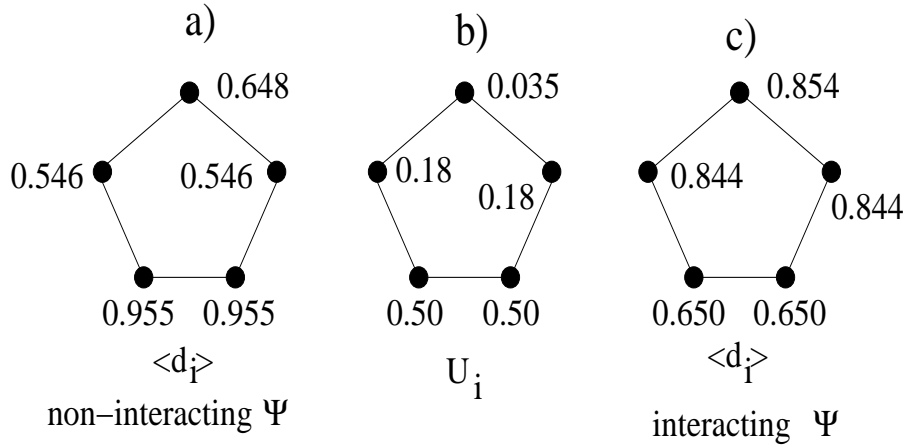}} 
\caption{For the pentagon chain in Fig.7 at $N_c=2$,  
parameters in (\ref{eqN3}), and half filled upper band: 
the on-site Coulomb repulsion 
values (Fig.8.b) together with the site dependent double 
occupancies in the non-interacting ground state (Fig.8.a), 
and the site dependent double occupancies provided
by the exact interacting ground state (Fig.8.c). }     
\end{figure} 

Fig.8 shows in the presence of the $U_n$ Hubbard 
repulsions from (\ref{eqN3}) reproduced in Fig.8.b, 
the double occupancies calculated for this
case for the non-interacting (Fig.8.a) and interacting 
(Fig.8.c) ground states. As seen, the same strong 
redistribution of the double occupancies appears in the
interacting case as observed in Fig.5. for the chain with
external links, antennas and $\epsilon_n$ values.

We calculated also for this case presented in Fig.8
the different characteristic energy values (the technique is
presented in details in Appendix D) obtaining 
(in $t=1$ units) for the non-interacting trial wave function
$E_{0,kin}=-4.628, E_{0,int}=2.154, E_{0,g}=-2.474$, 
while for the exact ground state 
$E_{kin}=-4.179, E_{int}=1.665, E_g=-2.513$.
Consequently, when the ferromagnetic phase emerges, the
interaction energy decreases $22.7 \%$, the kinetic energy 
increases $9.7\%$, and as a consequence the total energy
decreases $1.5\%$. As seen, similarly to the results in
Sect. VIII.C.2, the transition
is driven by the strong decrease of the interaction energy. 

So even if the ratio $|\delta E_{int}|/|\delta E_{kin}|$ 
can be increased by the presence of external links, 
antennas, and $\epsilon_n$ values, the main driving force of
the transition to the ferromagnetic phase remains the strong
decrease observed in $\delta E_{int}$.

We note that similar results for other parameter values
have been also found.
In order to exclude the possibility of the influence of
accidental special parameter values in the results, we 
analyzed instead of the parameter set in (\ref{eqN3}),
also the set of input Hamiltonian parameters
\begin{eqnarray}
&&t=1.0, \: t'=2.5, \: t_h=-0.5, U_2=U_3=0.52, \: U_1=0.1, 
\: U_4=0.045,
\nonumber\\
&&\epsilon_1=\epsilon_2=\epsilon_3=\epsilon_4=0.
\label{eqN4}
\end{eqnarray}
The obtained results are as follows:
For the non-interacting trial wave function
$E_{0,kin}=-7.765, E_{0,int}=2.004, E_{0,g}=-5.760$, 
while for the exact ground state 
$E_{kin}=-7.610, E_{int}=1.731, E_g=-5.878$.
In this case the interaction energy decreases 13.65 \%, the
kinetic energy increases 1.9 \%, these leading to a total
energy decrease of 2 \%. As can be seen, the behavior is
similar to that observed above.

In order to test the importance of different $U_n$ values,
we also analyzed a case when the $U_n$ parameters are 
almost uniformized
\begin{eqnarray}
&&t=1.0, \: t'=2.5, \: t_h=-0.5, U_2=U_3=5.0, \: U_1=4.5, 
\: U_4=4.57,
\nonumber\\
&&\epsilon_1=\epsilon_2=\epsilon_3=\epsilon_4=0.
\label{eqN5}
\end{eqnarray}
The results deduced for this case are as follows:
For the non-interacting trial wave function
$E_{0,kin}=-7.765, E_{0,int}=29.569, E_{0,g}=21.804$, 
while for the exact ground state 
$E_{kin}=-7.584, E_{int}=28.584, E_g=21.00$.
In this case the interaction energy decreases 3.34 \%, the
kinetic energy increases 2.3 \%, these leading to a total
energy decrease of 3.6 \%.

It can be observed that when the $U_n$ values are almost the
same, as explained in (\ref{eqN2}), the redistribution of
the double occupancy in the interacting ground state is 
without substantial effect in decreasing the interaction
energy. As a consequence $|\delta E_{int}|$ and 
$|\delta E_{kin}|$ a) decrease in value and become small, 
and b) besides, one obtains $|\delta E_{int}|\sim 
|\delta E_{kin}|$. As can be seen, the behavior at the
transition to the ferromagnetic state changes completely.

We further note that when for all $n$, $U_n=U$ uniformized
on-site Coulomb repulsions are present, the renormalized 
on-site energies $\epsilon_{n,R}$ become $U_n$ independent,
and the $U_n$ values disappear completely from the 
interacting ground state wave function $|\Psi_g\rangle$
[see (\ref{EQU11},\ref{EQU16}-\ref{EQU20},\ref{EQU22}) for the 
pentagon case with external links,
and (E15-E24) for the pentagon case without external links].
As a consequence, in the interacting case, the 
redistribution of the site dependent double occupancies 
[such to have small (high) $d_{\bf j}$ where $U_{\bf j}$
is high (small)]
effectively cannot be done, hence the strong decrease in
the interaction energy cannot be achieved.  

\subsubsection{Further observations}

By studying the uniformized on-site Coulomb repulsion case,
as a main observation we note that, the presence of the 
$U_n$ values in the interacting ground state 
$|\Psi_g\rangle$, see also Sect. VIII.C.3, is a major 
requirement in the emergence of the described ferromagnetic
state. In the described pentagon cases (with or without
links, antennas, or $\epsilon_n$ value), this property 
disappears at $U_n=U$ for all $n$, i.e. homogeneous 
interaction. 

We further underline that the presence of the $U_n$ terms 
in $|\Psi_g\rangle$ is guaranteed by the non-linear 
renormalization factor $p$ in the expression of the 
renormalized one-particle on-site energies $\epsilon_{n,R}$.
Indeed in this case $U_n-p$ present in $\epsilon_{n,R}$
remains $U_n$ dependent [see (\ref{EQU22}),(E24)]. 
Since $\epsilon_{n,R}$ enters in
$|\Psi_g\rangle$, the exact interacting ground state remains
interaction dependent for the described chains.

\subsection{The example of the triangle chain}

\subsubsection{Introduction}

We show below, that the presence of $U_n$ in the interacting
ground state wave function $|\Psi_g\rangle$ for 
non-homogeneous $U_n$ interactions requires a given degree
of complexity for the structure of the chain, which is 
present in the pentagon chain cases
(with or without external links, antennas and $\epsilon_n$
values). But if we simplify the chain structure, for example
taking a simple triangular chain as shown in Fig.9, such 
type of property is no more present even if $U_n$ is
different at different type of sites.

\begin{figure} [h]                                        
\centerline{\includegraphics[width=8 cm,height=3 cm]{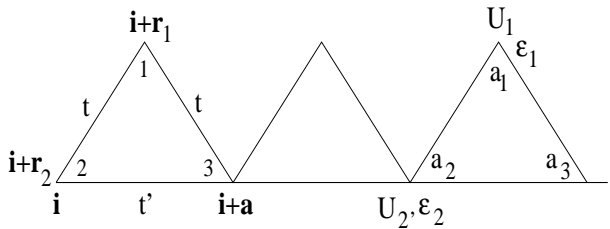}} 
\caption{Triangle chain without links and antennas. The 
first cell at ${\bf i}$ presents the in-cell notation of
sites, the in-cell two sites ${\bf r}_1, {\bf r}_2$, and
hopping matrix elements. The last cell presents the block
operator koefficients, the $U_n$ terms, and on-site one-particle
potentials $\epsilon_n$.}     
\end{figure} 

The mathematical description of the chain together with the
deduction of its exact ground state at half filled upper
band in similar conditions as has been done for the pentagon
chain case, is presented in Appendix F. From (F17,F18) is
seen that at first view the ground state, at least in its
mathematical form, resambles to the ground states obtained
in the pentagon chain cases. But a study of operators
entering into the ground state wave function [see (F15,F16)]
shows that $|\Psi_g\rangle$ not depends on $U_n$, hence the
simple triangular case, even if it is ferromagnetic, will
semnificatively differ from the ground state deduced in the
pentagon case.

\subsubsection{The reason of the differences in the simple
triangular case}

For the simple triangular case, the condition for the flat 
band emergence [see the second line of (F9)] reads
\begin{eqnarray}
\frac{t^2}{t'}+(\epsilon_2-\epsilon_1)-2t'=0.
\label{eqN6}
\end{eqnarray}
At $t'>0$ and $(\epsilon_2-\epsilon_1) < 2t'$ this 
condition provides a lowest flat band, which if half filled,
give rise to a Mielke-Tasaki type of flat band 
ferromagnetism. This last, possesses a ground state wave
function which is interaction independent since provides 
the same behavior at arbitrary (but non-zero) on-site
and homogeneous repulsive interaction.

A particle-hole transformation for this chain maps the
behavior from the lowest band to the same physical behavior
on the upper band, but the description uses holes instead of
particles. The same physical behavior means that on the 
half filled upper band the emerging ferromagnetism is of
Mielke-Tasaki type, hence possesses an interaction 
independent ground state wave function.

The particle-hole transformation means in this case
(see pg. 4057 of Ref.[\cite{KP}]) $t \to -t$, $t' \to -t'$,
$\epsilon_n \to -\epsilon_n -U_n$, which applied to
(\ref{eqN6}) gives
\begin{eqnarray}
U_2=U_1+\epsilon_1-\epsilon_2 + \frac{t^2}{-t'}+ 2t'=0.
\label{eqN7}
\end{eqnarray}
For the $t'=-|t'|$ requirement that one has for the upper
half filled ground state wave function deduced for the
triangular case with our method, (\ref{eqN7}) represents
exactly our condition (F19) for the existence of our 
solution. Furthermore, if (\ref{eqN7}) is satisfied at
$t'<0$, we indeed find un upper flat band in the system.

Consequently, with our procedure, in the simple
triangular case, we deduce a ferromagnetic solution at
half filled upper band, which is similar to the 
particle-hole transformed Mielke-Tasaki solution to the
half filled upper band, hence the ground state must be 
interaction independent. 

\subsubsection {The mathematical reasons for differences}

At mathematical level, the reason of the $U_n$ independence
of the ground state is related to the linear expression of 
the renormalization parameter $p$ [see (F14)] in terms of 
the $U_n$ interactions. Because of this reason, the $U_n-p$ 
differences present in renrmalized on-site one-particle
potentials $\epsilon_{n,R}$, [see (F20)], become $U_n$
independent, hence the interaction not influences nor the
renormalized band structure, nor the resulting ground state
wave function. 

\subsubsection{The degree of complexity}

Ultimately, at the level of the mathematical description,
and related to our technique, 
this result is connected to the simple structure of the 
matching conditions obtained in the triangular case
[see (F12)]. In other words, the simple triangular chain 
case not reaches that degree of complexity, which is able to
place the $U_n$ terms everywhere, so olso inside the ground
state wave function.

\subsubsection{Energy variations at the emergence of the
ferromagnetic state}

Using the technique described in Sect. VIII.C.2 
(see Appendix D), we calculated
also here at $N_c=2$ the
energy variations at the emergence of the 
ferromagnetic state. The results are as follows:

Using the parameter set
\begin{eqnarray}
t=1.0, \: t'=-1.5, \: \epsilon_1=\epsilon_2=0, \: U_1=2.4,
\: U_2=0.066, 
\label{eqN8}
\end{eqnarray} 
the following results are obtained:
\begin{eqnarray}
&&E_{0,kin}=-6.0, E_{0,int}=4.833, E_{0,g}=-1.167, 
\nonumber\\ 
&&E_{kin}=-5.151, E_{int}=3.951, E_g=-1.20.
\label{eqN9}
\end{eqnarray}
Consequently, the interaction energy decreases 18.2 \%, the
kinetic energy increases 14.14 \%, all these leading to a 
total energy decrease of 2.82 \%.

In order to not be influenced by accidental results, we have
tested also other parameter set by increasing the $U_n$
values:
\begin{eqnarray}
t=1.0, \: t'=-1.5, \: \epsilon_1=\epsilon_2=0, \: U_1=5.0,
\: U_2=2.666. 
\label{eqN10}
\end{eqnarray}
This situation leads to the following results
\begin{eqnarray}
&&E_{0,kin}=-6.0, E_{0,int}=11.333, E_{0,g}=5.333, 
\nonumber\\ 
&&E_{kin}=-5.151, E_{int}=9.151, E_g=4.00.
\label{eqN11}
\end{eqnarray}
In this case the interaction energy decreases 19.2 \%, the
kinetic energy increases 14.14 \%, all these leading to a 
total energy decrease of 24.9 \%. Note that $U_n$ even if 
not enters in the interacting ground state, as the
coupling constant influences the interaction energy, and 
also is present in the ground state energy. 

As can be seen, the determinative interaction energy 
decrease much higher in absolute value than the kinetic 
energy increase is no more present here. The redistribution
of double occupancy such to increase $d_{\bf j}$ where 
$U_{\bf j}$ is small and vice versa, is not
effective in this case since the ground state wave 
function not contains the $U_n$ values. Consequently, 
now, the transition has different characteristics.

\subsection{Why ferromagnetism occurs}

\subsubsection{The reason for ferromagnetism}

From the results presented above, clearly can be seen that
the observed ordered phase emerges because of the 
determinative decrease of the interaction energy. 
Furthermore, we know that the ordered phase is 
ferromagnetic. Consequently, 
the effort (or ''aim'') of the system to produce as much as
possible interaction energy decrease leads to 
ferromagnetism. Indeed, the decrease of the interaction
energy, in {\bf r}-space means also the decrease of
the number of double occupied sites, which, (see 
Sect. VIII.B, VIII.C.2), taking into account that the ferromagnetic state has the lowest double occupancy, 
leads to ferromagnetism.

\subsubsection{The aspect of the kinetic energy}

It is interesting to analyze why in the present case, in
order to find the emerging ordered phase, it is
enough to concentrate exclusively on the interaction energy
decrease without to have a look olso on possible 
modifications in the kinetic energy as well. The reason is 
related to the
effective flat band emergence. Namely, the system, 
starting from
a completely dispersive bare band structure, by interaction,
quenches the kinetic energy, exactly in order to be able to
concentrate only on the interaction part and fully take 
advantage
of its decreasing possibilities offered by the 
non-homogeneous $U_n$ values. The kinetic energy quench 
emerges practically at 
the kinetic energy value present when the interactions are 
turned on (see for example $\delta E_{kin}=2.6 \%$ in 
Sect. VIII.C.2).
This strategy is advantageous because by introducing
an effective flat band -- and using its huge degeneracy --,
the system easily can produce the 
ordering dictated by the interaction terms which are 
present in the system. Because of this reason, we expect
this mechanism to occur also in the case of other orderings
as well.

\subsection{Summary of the observed behavior and deduced
results}
 
a) In order to better understand the observed ferromagnetic
phase in pentagon chains at and above half filled 
upper band
and different $U_n$ values, first we determined the smallest
unit which provide a such type of behavior. This turned out
to be the $N_c=2$ cells system (see Sect. VIII.A). After 
this step we concentrated on the physical study of the 
$N_c=2$ case described with periodic boundary conditions.
 
b) From the results it turned out that 
the driving force of the transition is the determinative 
decrease in the interaction energy (in percentage
much higher in absolute
value than the increase in the kinetic energy). This is so
in pentagon chains with or without external links, 
antennas, or $\epsilon_n$ on-site one-particle potentials
(however these last three factors are able to increase the
$|\delta E_{int}|/|\delta E_{kin}|$ ratio), see 
Sect. VIII.C.1-2,5.
 
c) This effect, a) in order to occur, requires the presence
of different $U_n$ in the ground state wave function (see
Sect. VIII.C.3,5), b) emerges only in chains with a 
sufficient
degree of complexity (see Sect. VIII.D.4), c) is guaranted 
by a non-linear renormalization factor $p$ of the on-site
one-particle potentials $\epsilon_{n,R}$ 
(see Sect. VIII.C.6),
d) disappears when all $U_n$ become uniform (see
Sect. VIII.C.5), e) has been observed on the 
half (and above half) filled upper band (see Sect. IV.A-B), 
and f) differs semnificatively
from the Mielke-Tasaki type of behavior at the lowest band
\cite{Intr11}, or particle-hole 
transformed to the upper band (see Sect. VIII.D).
 
d) The strong decrease of the interaction energy is 
obtained by a sharp redistribution of the double occupancy
$d_{\bf j}=\langle \hat d_{\bf j} \rangle$
(comparative to the disordered phase, and constrained by the
existing sum rules, as N=constant, or physical conditions
as $\langle \hat n_{{\bf j},\sigma}\rangle \in [0,1]$), 
which produces high (small) $d_{\bf j}$ where $U_{\bf j}$ 
is small (high), see Figs.5,8. By its nature, such 
effect is non-existent at uniform $U$ since in this case a
$d_{\bf j}$ redistribution not affects the interaction 
energy (see for example Sect. VIII.C.5). One notes that 
this redistribution leads to a
minimum average number of double occupied
sites in ${\bf r}$-space, (see Sect. VIII.B.4).
 
e) The ``as much as possible interaction energy decrease''
process leads to ferromagnetism (see Sect. VIII.B.4 and E). 
The ferromagnetic phase not emerges at arbitrary 
interaction values and possesses lower bounds for 
$U_n$ (see Sect. VIII.C.4 and Fig.6).

f) The emerging ground state can be obtained solely by 
following exclusively only the interaction energy decrease 
possibilities, without to pay attention to the kinetic 
energy. This is because 
the system, by introducing an effective flat band, 
quenches the kinetic energy practically to the 
$E_{0,kin}$ value present before the interactions were 
turned on (see Sect. VIII.E.2). 
This is advantageous since by
taking advantage of the huge flat band
degeneracy, the system easily can introduce the 
ordering dictated by the interaction terms.

g) When the ferromagnetism occurs, automatically one has 1 
electron with fixed spin projection on all sites 
(see Sect. VIII.B.2). Furthermore,
once one has 1 electron with fixed spin projection 
on all sites,
the one particle interpretation of the results 
automatically leads to effective upper flat band (see 
Sect. VIII.B.3).

\section{Summary and conclusions}

We analyzed by rigorous techniques exact ground states of 
the pentagon chain in the high density limit, and showed 
that the interaction is able to generate an effective flat 
band from a dispersive bare band structure. By this effect 
transitions to ferromagnetic states, or correlated 
half-metallic states at high electron densities can be 
induced. These results prove by exact means the conjecture 
made previously at the two particle level \cite{brocks} 
that the Coulomb interaction is in principle able to 
stabilize 
magnetic order in acenes and thiophenes in the high density 
limit. The model considers site dependent on-site Coulomb 
interactions inside the unit cell, which account for
the particular environment and type of atom on a particular 
site inside the unit cell.  The method is based on a 
positive semidefinite operator technique, and the 
experimental 
realization, in principle can be achieved \cite{m2010}.
We also mention that this property is not restricted 
exclusively to pentagon chains, but can occur in other 
chain structures as well.

In the case of the Mielke-Tasaki type of flat band 
ferromagnetism emerging in a half filled lowest flat band 
\cite{Intr11}, both the flat band and the connectivity 
conditions (overlap of neighboring local Wannier functions) 
are necessary and result from the bare 
band structure. By contrast, in the here described process,
the interaction tunes a fully dispersive bare band 
structure to become partially flat and thereby produces 
ferromagnetism.

The deduced ground states for the pentagon chains were
proved to be unique and represent i) a nonsaturated
ferromagnetic state for a half filled upper band,
localized in the thermodynamic limit,
and ii) a correlated half metallic conducting state if the 
upper band is above half filling.
The correlation functions, relative to the 
${\bf r} \to \infty$
components characterizing the long-range ferromagnetic 
order, 
are diminished  at short distances, signaling the presence 
of short-range fluctuations. 

Physically, the appearance of 
ferromagnetism shows a new mechanism for the emergence of 
an ordered phase, described here in our knowledge for the 
first time. The characteristics of this process are as 
follows: Starting from a 
completely dispersive bare band structure, the interactions 
quench the kinetic energy, hence the ordered phase is 
obtained exclusively by a drastic decrease of the 
interaction energy. 
The Hubbard interaction being site dependent, this strong
decrease is obtained by a redistribution of the double 
occupancy such to attain small (high) double occupancy 
where the on-site Coulomb repulsion is high (small). The 
kinetic energy quench 
(i.e. the created upper effective flat band) is 
advantageous for
the system since enhance -- by its huge degeneracy --
the smooth 
transition possibility to an arbitrary ordered phase 
dictated and stabilized by the interactions present in the 
material. Because of this reason, we expect this mechanism 
to occur also in the case of other orderings as well.

We further note that a
given degree of complexity is needed
for the chain when the described ferromagnetism occurs, 
and the mechanism is
missing when the Hubbard interactions are homogeneous,
or the interactions are not present in the ground state wave
function.

\section*{Acknowledgements}

I mention that 
the presented results have been worked out
in a strong collaboration with Dieter Vollhardt and Arno 
Kampf partly in the frame of the Transregio TTR 80 
supported by the Deutsche Forschungsgemeinschaft. I would 
like to thank to these colleagues for many stimulating 
questions, observations, remarcs, suggestions, and 
discussions, which finally lead to the presented form of 
this material. In the same frame, I would like to thank to
Dieter Vollhardt for the continuous support that I have
received from his side during my work on this subject. 
Furthermore,
I acknowledge valuable discussions with Wolfgang Br\"utting
on organic materials and with Michael Sekania on exact 
diagonalization results. I also
gratefully acknowledge financial support by
the Alexander von Humboldt foundation and contracts
OTKA-K-100288 (Hungarian Research Fund), and
TAMOP 4.2.2/A-11/1/KONV-2012-0036 (cofinanced by EU and
European Social Fund). 

\appendix

\section{The structure of ${\bf F_n}({\bf k})$ for ${\bf n=2}$}

As described in connection with the derivation of 
Eqs. (\ref{EQU46}-\ref{EQU48}), the quantity $F_n({\bf k})$
can be obtained from $F_n({\bf k})=Det(F)$, where $F$ is
a $6\times 6$ matrix of the form
\begin{eqnarray}
F =
\left( \begin{array}{cccccc}
f_{1,1} & f_{1,2} & f_{1,3} & f_{1,4} & f_{1,5} & f_{1,6} \\
f_{2,1} & d_{1,1} & d_{1,2} & d_{1,3} & d_{1,4} & d_{1,5} \\
f_{3,1} & d_{2,1} & d_{2,2} & d_{2,3} & d_{2,4} & d_{2,5} 
\\ 
f_{4,1} & d_{3,1} & d_{3,2} & d_{3,3} & d_{3,4} & d_{3,5} \\
f_{5,1} & d_{4,1} & d_{4,2} & d_{4,3} & d_{4,4} & d_{4,5} \\
f_{6,1} & d_{5,1} & d_{5,2} & d_{5,3} & d_{5,4} & d_{5,5} 
\end{array} \right) .
\label{A6}
\end{eqnarray}
The matrix elements $f_{n,1}=f_{1,n}^*$ are defined 
above (\ref{EQU48}), and the matrix elements $d_{n,n'}$ 
are obtained from (\ref{EQU46}).
Here we present the coefficients $Y_{1,n}, Y_{2,n}$ which
appear in (\ref{EQU77}) in the expression for 
$F_{n}({\bf k})$ with $n=2$:
\begin{eqnarray}
Y_{1,2} &=& 2t^2 t_c \Big[(\frac{Q_1}{\sqrt{|t_h|}}-
\sqrt{|t_h|})^2(Q_3^2+\frac{t_f^2}{Q_3^2})-
Q_3^2(\frac{Q_1^2}{|t_h|}+2|t_h|) - \frac{2t_f^2Q_1}{
Q_3^2\sqrt{|t_h|}}(\frac{Q_1}{
\sqrt{|t_h|}}-\sqrt{|t_h|}) + \frac{t_f^2Q_1^2}{
Q_3^2|t_h|}\Big] 
\nonumber\\
&+& 2t^2t_c |t_h|Q_3^2 +
2t^2t_c \sqrt{|t_h|}\Big[(Q_3^2+\frac{t_f^2}{Q_3^2})(
\frac{Q_1}{\sqrt{|t_h|}}
-\sqrt{|t_h|}) -\frac{t_f^2Q_1}{Q_3^2\sqrt{|t_h|}}\Big],
\nonumber\\
Y_{2,2} &=& Q_1^2|t_c| ( Y_{2,2}^{(1)} +Y_{2,2}^{(2)} +
Y_{2,2}^{(3)}).
\label{A1}
\end{eqnarray}
Using the notation
\begin{eqnarray}
&&q_1=(\frac{Q_1^2}{|t_h|}+2|t_h|), \quad q_2=(2+
\frac{t^2}{Q_1^2}),
\nonumber\\
&&q_3=(Q_3^2+\frac{t_f^2}{Q_3^2}), \quad q_4=(\frac{Q_1}{
\sqrt{|t_h|}}-\sqrt{|t_h|}),
\label{A2}
\end{eqnarray}
the $Y_{2,2}$ terms in (\ref{A1}) are given by
\begin{eqnarray}
Y_{2,2}^{(1)} &=&-\frac{2t^2}{Q_1^2} \Big(q_1q_2q_3 + \frac{
2t_f^2Q_1}{Q_3^2\sqrt{|t_h|}}q_4
-\frac{t_f^2Q_1^2}{Q_3^2|t_h|}q_2 - \frac{t_f^2}{Q_3^2}q_1
- q_4^2 q_3\Big) 
\nonumber\\
&+& \Big[q_2\Big(q_1q_2q_3 + \frac{2t_f^2Q_1}{Q_3^2
\sqrt{|t_h|}}q_4 -\frac{t_f^2Q_1^2}{Q_3^2|t_h|}q_2
-\frac{t_f^2}{Q_3^2}q_1-q_3 q_4^2\Big) -q_4\Big(q_3q_2q_4
\nonumber\\
&+&
\frac{t_f^2}{Q_3^2}q_4 +
\frac{t_f^2Q_1}{Q_3^2\sqrt{|t_h|}}(1- q_2)- 
\frac{t_f^2}{Q_3^2}q_4-q_3q_4\Big)
+\Big(q_4^2q_3+\frac{t_f^2}{Q_3^2}q_1 +\frac{t_f^2Q_1^2}{
Q_3^2|t_n|}(1-2q_4)
\nonumber\\
&-&q_1q_3\Big)
+ \frac{t_f}{Q_3} (\frac{t_fQ_1}{Q_3\sqrt{|t_h|}}q_2q_4 +
\frac{t_f}{Q_3}q_1
- \frac{t_fQ_1}{Q_3\sqrt{|t_h|}}q_4 -\frac{t_f}{Q_3}
q_2q_1)\Big],
\label{A3}
\end{eqnarray}
\begin{eqnarray}
Y_{2,2}^{(2)} &=& \frac{t^2}{Q^2_1}\Big[(-\frac{t_f^2Q_1^2}{
Q_3^2|t_h|} + q_3q_1) +\sqrt{|t_h|}(-\frac{t_f^2Q_1}{
Q_3^2\sqrt{|t_h|}} +q_3q_4)\Big] - 2\Big[(q_1q_2q_3 +
\frac{2t_f^2Q_1}{Q_3^2\sqrt{|t_h|}}q_4
\nonumber\\
&-&\frac{t_f^2Q_1^2}{Q_3^2|t_h|}q_2
-\frac{t_f^2}{Q_3^2}q_1
-q_3q_4^2)+\sqrt{|t_h|}(q_2q_3q_4 + \frac{t_f^2Q_1}{Q_3^2
\sqrt{|t_h|}}(1-q_2)-q_3q_4)\Big],
\label{A4}
\end{eqnarray}
\begin{eqnarray}
Y_{2,2}^{(3)} &=& 2|t_h|\frac{t^2}{Q_1^2}(q_2q_3- 
\frac{t_f^2}{Q_3^2}) +\sqrt{|t_h|}
\frac{t^2}{Q_1^2}\Big(q_3q_4-\frac{t_f^2Q_1}{Q_3^2\sqrt{|t_h|}}
\Big)-2\sqrt{|t_h|}\Big(q_2q_3q_4
\nonumber\\
&+&\frac{t_f^2Q_1}{Q_3^2\sqrt{|t_h|}}(1-q_2)
- q_3q_4\Big)-2|t_h|\Big(q_2^2q_3+2\frac{t_f^2}{
Q_3^2}(1-q_2)-q_3\Big).
\label{A5}
\end{eqnarray}

\section{The uniqueness at ${\bf N}={\bf N}^*$}

\subsubsection{The kernel of $\hat H_G(\alpha,{\bf i},\sigma)$}

{\bf Lemma 1.}: For N particles, all states in
$ker(\hat H_G(\alpha,{\bf i},\sigma))$ can be written
in the form $|V_{G,\alpha,{\bf i},\sigma}\rangle = 
\hat G^{\dagger}_{\alpha,{\bf i},\sigma} \hat W^{\dagger}_{
G,\alpha}|0\rangle$, where $\hat W^{\dagger}_{G,\alpha}$ is
an arbitrary operator which introduces $(N-1)$ electrons 
into the system and preserves a nonzero norm for
$|V_{G,\alpha,{\bf i},\sigma}\rangle$. 

Proof: Consider an arbitrary state 
$|V'_G\rangle = \hat Y^{\dagger}_{G,\alpha} |0\rangle \in 
ker\Big(\hat H_G(\alpha,{\bf i},\sigma)\Big)$, where 
$\hat Y^{\dagger}_{G,\alpha}$ introduces $N$ particles into
the system. Therefore
\begin{eqnarray}
\hat H_G(\alpha,{\bf i},\sigma) |V'_G\rangle = 
\hat G_{\alpha,{\bf i},\sigma} \hat G^{\dagger}_{\alpha,
{\bf i},\sigma} \hat Y^{\dagger}_{G,\alpha} |0\rangle =0 
\label{EQU84}
\end{eqnarray}
necessarily holds by definition. Recalling $z_{\alpha} =
\hat G_{\alpha,{\bf i},\sigma} \hat G^{\dagger}_{\alpha,
{\bf i},\sigma} + \hat G^{\dagger}_{\alpha,{\bf i},\sigma} 
\hat G_{\alpha,{\bf i},\sigma} \ne 0$ defined above 
(\ref{EQU15}), one has
\begin{eqnarray}
|V'_G\rangle &=& \frac{z_{\alpha}}{z_{\alpha}} 
\hat Y_{G,\alpha}^{\dagger} |0\rangle = 
\frac{1}{z_{\alpha}} (\hat G^{\dagger}_{\alpha,{\bf i},
\sigma} \hat G_{\alpha,{\bf i},\sigma} +  \hat G_{\alpha,
{\bf i},\sigma} \hat G^{\dagger}_{\alpha,{\bf i},\sigma} )
\hat Y^{\dagger}_{G,\alpha} |0\rangle 
\nonumber\\
&=& \hat G^{\dagger}_{\alpha,{\bf i},\sigma} 
[\frac{1}{z_{\alpha}} \hat G_{\alpha,{\bf i},\sigma}
\hat Y_{G,\alpha}^{\dagger} ] |0\rangle =
\hat G^{\dagger}_{\alpha,{\bf i},\sigma} 
\hat W^{\dagger}_{G,\alpha} |0\rangle = |V_{G,\alpha,
{\bf i},\sigma}\rangle,
\label{EQU85}
\end{eqnarray}
where (\ref{EQU84}) has been used, and 
the notation $\hat W^{\dagger}_{G,\alpha}=
[(1/z_{\alpha}) \hat G_{\alpha,{\bf i},\sigma}
\hat Y_{G,\alpha}^{\dagger} ]$ has been introduced.
Since $\hat Y_{G,\alpha}^{\dagger}$ introduces $N$ particles
in the system, and $\hat G_{\alpha,{\bf i},\sigma}$ annihilates
one particle,
it results that $\hat W^{\dagger}_{G,\alpha}$
creates $(N-1)$ particles. In conclusion, Lemma 1. is proved. 

\subsubsection{The kernel of $\hat H_G$}

Based on the second line of (\ref{EQU83}) now one concentrates
on the kernel of $\hat H_G$, for which the following Lemma can be
formulated:

{\bf Lemma 2.}: All vectors from $ker(\hat H_G)$ 
[see (\ref{EQU83})]
in the 
presence of $N \geq 10N_c$ particles, can be written in 
the form 
\begin{eqnarray}
|V_G\rangle = [ \prod_{\sigma} \prod_{\alpha=1}^5 
\prod_{{\bf i}} \hat G^{\dagger}_{\alpha,{\bf i},
\sigma} ] \hat W^{\dagger}_G |0\rangle ,
\label{EQU86}
\end{eqnarray}
where $\hat W^{\dagger}_G$ is an arbitrary operator which 
introduces $(N-10N_c)$ particles in the system, and 
preserves a nonzero norm for $|V_G\rangle$.

In demonstrating Lemma 2, first one notes that 
$[ \prod_{\sigma} \prod_{\alpha=1}^5 \prod_{{\bf i}}
\hat G^{\dagger}_{\alpha,{\bf i},\sigma} ]$ introduces 
already $10N_c$ electrons into the system, that is why the 
specification $N \geq 10N_c$ is needed.

Based on expression of $\hat H_G$ from (\ref{EQU82}),
the $ker(\hat H_G)$ expression from (\ref{EQU83}), 
$ker(\hat H_G(\alpha,{\bf i},\sigma))$ from Lemma 1,
and the linear independence of the operators
$\hat G^{\dagger}_{\alpha,{\bf i},\sigma}$, the relation 
(\ref{EQU86}) is automatically obtained, hence Lemma 2
is demonstrated. We note that the linear independence of 
the operators $\hat G^{\dagger}_{\alpha,{\bf i},\sigma}$ 
is seen from the nonzero norm value of the vector
\begin{eqnarray}
|{\bar \Psi}_{G}\rangle = \prod_{\sigma} 
\prod_{{\bf i}}
\prod_{\alpha=1}^5 [\hat G^{\dagger}_{\alpha,{\bf i},\sigma}
]|0\rangle. 
\label{EQU87}
\end{eqnarray}
This is because
$\langle {\bar \Psi}_{G}|{\bar \Psi}_{G}\rangle 
= (\langle \Psi_G |\Psi_G \rangle )^2$ where $|\Psi_G\rangle$ is 
presented in (\ref{EQU42}), and as seen from 
(\ref{EQU48},\ref{EQU62}), the expression is nonzero. 
Since $|V_G\rangle$ in (\ref{EQU86}) 
contains $N$ particles, it results that $\hat W^{\dagger}_G$
introduces $N-10N_c$ particles into the system.

Consequently, Lemma 2. has been demonstrated.

\subsubsection{The connection between 
$|{\bar \Psi}_{G}\rangle$ and $ker(\hat H_P)$ }

Effectuating the product of operators in
$|{\bar \Psi}_{G}\rangle$ [see (\ref{EQU87})],
one obtains a huge number of 
additive $\hat T_{\nu}$ terms, 
\begin{eqnarray}
[ \prod_{\sigma} \prod_{m=1}^5 \prod_{{\bf i}}
\hat G^{\dagger}_{m,{\bf i},\sigma} ]|0\rangle = 
\sum_{\nu} \hat T_{\nu}|0\rangle,
\label{EQU88}
\end{eqnarray}
each term containing on different sites 
$5N_c$ electrons with $\sigma=\uparrow$ and $5N_c$ 
electrons with $\sigma=\downarrow$. Since totally $6N_c$
sites are present in the system, many of these 
$\hat T_{\nu}|0\rangle$ contributions contain
several empty sites. Consequently, the vector 
$|{\bar \Psi}_{G}\rangle$ it is not 
contained in $ker(\hat H_P)$, hence nor in $ker(\hat H')$. 
Based on this observation, in order to
introduce $|V_G\rangle$ from (\ref{EQU86}) also in 
$ker(\hat H_P)$, hence, according to the first line of
(\ref{EQU83}), also $ker(H')$, one must chose
a specific $\hat W^{\dagger}_G$  operator denoted by 
$\hat W^{\dagger}_G=\hat F^{\dagger}$,
which, at the fixed number of $N=11N_c=N^*$ particles 
creates the vector
\begin{eqnarray}
|V_{G,P}\rangle = [ \prod_{\sigma} \prod_{\alpha=1}^5 
\prod_{{\bf i}}
\hat G^{\dagger}_{\alpha,{\bf i},\sigma} ] 
\hat F^{\dagger} |0\rangle.
\label{EQU89}
\end{eqnarray}
This vector must has
in comparison to (\ref{EQU86}), the supplementary property 
that on each site of the lattice has at least one electron,
so $|V_{G,P}\rangle \in ker(\hat H_P)$, and consequently
$|V_{G,P}\rangle \in ker(\hat H')$. 
One analyzes below the expression of the $\hat F^{\dagger}$
operator at $N=N^{*}$ through five consecutive Lemmas. The 
study
starts by presenting a good choise for $\hat F^{\dagger}$ in
Lemma 3, the choise providing a state in $ker(\hat H')$ 
characterized by system total
spin $S=S_z=S_z^{Max}=N_c/2$. After this step Lemma 4 
demonstrates
that for fixed $S=S_z=S_z^{Max}=N_c/2$, this choise is 
unique.
Then, Lemma 5 demonstrates that $ker(\hat H')$ not 
contains vectors
which are belong to system total spin $S < N_c/2$. 
After this step, Lemma 6 shows that all spin sectors 
$S_z < S_z^{Max}$ at fixed $S=S_z^{Max}=N_c/2$ are 
contained in $ker(\hat H')$. Finally,
Lemma 7 concludes, that apart of the trivial $2S+1$ fold 
degeneracy related to the orientation of the total spin 
$S=N_c/2$, $|\Psi_g(N^*)\rangle$ is unique.

\subsubsection{The  $\hat F^{\dagger}$ operator}

{\bf Lemma 3.}:
Taking for $\hat F^{\dagger}$ in (\ref{EQU89}) at $N=N^*$
the expression
\begin{eqnarray}
\hat F^{\dagger}=\prod_{{\bf i}}
\hat c^{\dagger}_{{\bf i}+{\bf r}_{n_{\bf i}},\sigma}
\label{EQU90}
\end{eqnarray}
i.e. a product over $N_c$ creation operators with the same 
fixed spin index, one creation operator being taken from 
each cell, one 
obtains the vector (\ref{EQU31},\ref{EQU32}) which is
placed inside $ker(\hat H')$.

The proof follows the following steps. 

i) Given by Lemma 2., (\ref{EQU89}) with $\hat F^{\dagger}$
in (\ref{EQU90}), at $N=N^*$ is inside $ker(H_G)$. 

ii) The $\hat F^{\dagger}$ operator from (\ref{EQU90})
introduces $N_c$ electrons with fixed spin $\sigma$ in the 
system, one in each cell. After the action of the operator 
$\hat F^{\dagger}$ on the vacuum state,
5 empty sites remain in each cell. Each of
these 5 empty sites from the cell at ${\bf i}$ can be 
reached by the operators $\hat G^{\dagger}_{\alpha,{\bf i},
\sigma}$, $\alpha=1,2,3,4,5$, i.e.
$\prod_{\alpha=1}^5 \hat G^{\dagger}_{\alpha,{\bf i},
\sigma}$ can fill these empty sites up with 
electrons holding spin $\sigma$. Hence the remaining 
$5N_c$ empty sites of the system (after the action of 
$\hat F^{\dagger}$), will be filled up each with 
$\sigma$ spin electrons by the product 
$\prod_{{\bf i}}\prod_{\alpha=1}^5 
\hat G^{\dagger}_{\alpha,{\bf i},\sigma}$. Consequently, 
each site of the system will have one spin $\sigma$ 
electron, hence (\ref{EQU89}) transforms into (\ref{EQU32})
at the level of an unnormalized Hilbert space vector.

In other terms, (\ref{EQU89}) becomes of the form
$|V_{G,P}\rangle=\sum_{\nu} \hat T_{\nu}\hat F^{\dagger}
|0\rangle$. Because of $\hat c^{\dagger}_{{\bf i}+
{\bf r}_n,\sigma} \hat c^{\dagger}_{{\bf i}+{\bf r}_n,
\sigma}=0$, after the action of $\hat F^{\dagger}$, 
only those $\hat T_{\nu}$ terms survive, wich provide in
$|V_{G,P}\rangle$ one $\sigma$ spin electron on each site.  

iii) Because in (\ref{EQU32}), on each site of the system
one has at least one electron, namely that with the fixed 
spin $\sigma=\uparrow$, it results that the wave vector from
(\ref{EQU89}) with $\hat F^{\dagger}$ in
(\ref{EQU90}), is placed also inside $ker(\hat H_P)$. 

iv) Given now by the first line of (\ref{EQU83}), namely
$ker(\hat H') = ker(\hat H_G) \bigcap ker(\hat H_P)$, it 
results that (\ref{EQU89}) with $\hat F^{\dagger}$ taken
from (\ref{EQU90}) is placed in $ker(\hat H')$.

One notes, that in (\ref{EQU90}), the chosen position 
${\bf r}_{n_{\bf i}}$ in the cell at ${\bf i}$, has no 
importance. For each choise, (\ref{EQU89}) transforms in 
the same unnormalized form (\ref{EQU32}).

Consequently, Lemma 3. has been demonstrated. 

\subsubsection{Properties of $|V_{G,P}\rangle$}

The wave vector (\ref{EQU89}) containing $\hat F^{\dagger}$ from
(\ref{EQU90}) is a common eigenstate of ${\hat S}^2$ and $\hat S_z$
corresponding to eigenvalues $(N_c/2)[(N_c/2)+1]$ and $N_c/2$, hence
describes a state with total spin $S=N_c/2$, and projection 
$S_z=S=N_c/2$.

In order to prove, one uses ${\hat S}^2={\hat S}_z^2+(1/2)(
{\hat S}_{+} {\hat S}_{-}+{\hat S}_{-} {\hat S}_{+})$, where
$\hat S_z=\sum_{\bf j} {\hat S}_{{\bf j},z},
\hat S_{+}=\sum_{\bf j} {\hat S}_{{\bf j},+},
\hat S_{-}=\sum_{\bf j} {\hat S}_{{\bf j},-}$, the site
spin operators ${\hat S}_{{\bf j},z}, {\hat S}_{{\bf j},+},
{\hat S}_{{\bf j},-}$ being given in (\ref{EQU60}), and
$\sum_{\bf j}$ is made over all $6N_c$ sites.
From the first row of (\ref{EQU61}) one has 
${\hat S}_{-} {\hat S}_{+}|\Psi_g(N^*)\rangle=
{\hat S}_{-} {\hat S}_{+}|V_{G,P}\rangle=0$, and
\begin{eqnarray}
{\hat S}_z|V_{G,P}\rangle=(1/2)(6N_c-
\hat N_{\downarrow})|V_{G,P}\rangle = (N_c/2)
|V_{G,P}\rangle, 
\label{EQU98}
\end{eqnarray}
hence $S_z=N_c/2$ indeed holds. Furthermore one finds
\begin{eqnarray}
{\hat S}^2|V_{G,P}\rangle =[(\frac{N_c}{2})^2+\frac{1}{2}
{\hat S}_{+} {\hat S}_{-}]|V_{G,P}\rangle.
\label{EQU99}
\end{eqnarray}
But from (\ref{EQU99}) one deduces that
\begin{eqnarray}
{\hat S}_{+} {\hat S}_{-}|V_{G,P}\rangle &=&
\sum_{{\bf j}_1,{\bf j}_2}
\hat c^{\dagger}_{{\bf j}_1,\uparrow} \hat c_{{\bf j}_1,
\downarrow} \hat c^{\dagger}_{{\bf j}_2,\downarrow} 
\hat c_{{\bf j}_2,\uparrow}|V_{G,P}\rangle= 
\sum_{{\bf j}}\hat c^{\dagger}_{{\bf j},\uparrow} 
\hat c_{{\bf j},\downarrow} \hat c^{\dagger}_{{\bf j},
\downarrow} \hat c_{{\bf j},\uparrow}]|V_{G,P}\rangle
=\sum_{\bf j}\hat n_{{\bf j},\uparrow}(1-\hat n_{{\bf j},
\downarrow})|V_{G,P}\rangle
\nonumber\\
&=& \sum_{\bf j}(1-\hat n_{{\bf j},
\downarrow})|V_{G,P}\rangle = (6N_c- \hat N_{\downarrow})
|V_{G,P}\rangle= N_c |V_{G,P}\rangle.
\label{EQU100}
\end{eqnarray}
Consequently
\begin{eqnarray}
&&{\hat S}^2|V_{G,P}\rangle =\frac{N_c}{2}(\frac{N_c}{2}+1)
|V_{G,P}\rangle, \quad \Longrightarrow \quad
S=\frac{N_c}{2}.
\nonumber\\
&&\hat S_z |V_{G,P}\rangle = \frac{N_c}{2}|V_{G,P}\rangle, 
\quad \Longrightarrow \quad
S_z=S_z^{Max}=S=\frac{N_c}{2}.
\label{EQU101}
\end{eqnarray}
Hence, the statement of this subsection has been proved. Based on 
it, one shows below that other states corresponding to $S=S_z^{Max}=
N_c/2$ are not present in $ker(\hat H')$.

\subsubsection{The kernel $ker(\hat H')$ at $S=S_z^{Max}=N_c/2$}

{\bf Lemma 4}:
The kernel $ker(\hat H')$ at fixed $S=S_z^{Max}=N_c/2$ contains
only the unique vector $|V_{G,P}\rangle$.

In proving Lemma 4 one mentions that in the $|V_{G,P}\rangle$
vector from (\ref{EQU89}), the product
$[ \prod_{\sigma} \prod_{\alpha=1}^5 \prod_{{\bf i}}
\hat G^{\dagger}_{\alpha,{\bf i},\sigma} ]$ is needed to place
$|V_{G,P} \rangle$ in $ker(\hat H_G)$. This product alone provides
zero total spin. Hence, in order to obtain a state with total spin
$S=S_z=N_c/2$ in $ker(\hat H')$, all $N_c$ multiplicative 
components of 
$\hat F^{\dagger}$ must have the same spin index. Hence, in 
obtaining
a vector in $ker(\hat H')$ holding $S=S_z^{Max}=N_c/2$, only
the possibility from (\ref{EQU89}) with $\hat F^{\dagger}$ from
(\ref{EQU90}) remains. Here only the different placement 
possibilities of ${\bf r}_{n_{\bf i}}$ in the cell defined at 
${\bf i}$ [see (\ref{EQU90})] remains the unique
modification possibility if $S=S_z=N_c/2$ is fixed. But as 
mentioned at the end of the Section VII.A.5, in (\ref{EQU90}) 
all possible choises of ${\bf r}_{n_{\bf i}}$ in the cell 
defined at 
${\bf i}$ transform the wave vector (\ref{EQU89}) in the unique
(unnormalized) (\ref{EQU32}). Consequently $|V_{G,P}\rangle$ 
is the
unique vector which spans the $S=S_z=S_z^{Max}=N_c/2$ sector of
$ker(\hat H')$. Consequently, Lemma 4 has been demonstrated.

\subsubsection{The kernel $ker(\hat H')$ for $S < N_c/2$}  

One shows below that the kernel $ker(\hat H')$ not contains wave
vectors coresponding to total spin $S < N_c/2$. One starts by 
noting
that in order to remain in $ker(\hat H_G)$, the studied vector
needs the $[ \prod_{\sigma} \prod_{\alpha=1}^5 
\prod_{{\bf i}}\hat G^{\dagger}_{\alpha,{\bf i},\sigma} ]$ 
component. Consequently, the
unique modifications which lower the system total spin $S$ 
below
$N_c/2$ in (\ref{EQU89}) and leave the vector in 
$ker(\hat H_G)$,
are the spin-flips in $\hat F^{\dagger}$
from (\ref{EQU90}). Indeed, based on similar calculations as 
presented in (\ref{EQU99},\ref{EQU100}), one simply finds for 
$N_{fl}=n'$ spin flips in $\hat F^{\dagger}$ from 
(\ref{EQU90}) a 
resulting total system spin $S=(N_c-n')/2$ in the vector 
from (\ref{EQU89}).

{\bf Lemma 5.}:
By flipping at least one spin in (\ref{EQU90}), the Hilbert
space vector
\begin{eqnarray}
|V'_{G,P}\rangle = [ \prod_{\sigma} \prod_{\alpha=1}^5 
\prod_{{\bf i}} \hat G^{\dagger}_{\alpha,{\bf i},
\sigma} ] \hat F_1^{\dagger} |0\rangle ,
\label{EQU91}
\end{eqnarray}
where $\hat F_1^{\dagger}$ contains at least one flipped 
spin relative to (\ref{EQU90}), is not contained in 
$ker(H_P)$, hence, given by the first line of
(\ref{EQU83}) it is not contained in $ker(\hat H')$.

In proving Lemma 5. one uses the observation presented in
the last paragraph of Sect. IV. A.2 which shows that at 
least one reversed spin in $\hat F^{\dagger}$ from 
(\ref{EQU90}) eliminates $|V'_{G,P}\rangle$ from 
$ker(\hat H_P)$ hence $ker(\hat H')$. Consequently, Lemma 5
has been demonstrated. In order to enhance the 
understanding, we present in Appendix C a detailed study of
$1 \leq n' < N_c$ spin-flip cases in $\hat F^{\dagger}$ from
(\ref{EQU90}).

A direct consequence of this result is that $ker(\hat H')$ 
for $S < N_c/2$ total spin is an empty manifold.

\subsubsection{The kernel $ker(\hat H')$ for $S=N_c/2$ and 
$S_z < N_c/2$}  

According to Lemma 4 and Lemma 5, $ker(\hat H')$ not contains 
vectors with total system spin $S < N_c/2$, and for $S=S_z^{Max}=
N_c/2$ contains the unique vector $|V_{G,P}\rangle$ from 
(\ref{EQU89}) with $\hat F^{\dagger}$ from (\ref{EQU90}). 
Consequently, for $S=S_z^{Max}=N_c/2$, in $ker(\hat H')$ the 
unique vector is $|\Psi_g(N^{*}\rangle=|V_{G,P}\rangle$. It 
results that the ground state deduced in (\ref{EQU31},\ref{EQU32})
is unique in the $S=S_z^{Max}$ spin sector, where the unique
allowed $S$ value is $S=N_c/2$. One shows below that for 
$S=N_c/2$, $ker(\hat H')$ contains also the $S_z < S_z^{Max}=N_c/2$
components.  

{\bf Lemma 6}:
All $S_z\in [-S,+S]$, components of $|V_{G,P}\rangle =
|\Psi_g(N^*)\rangle$ describing
the $S=S_z^{Max}=N_c/2$ total system spin state are contained in 
$ker(\hat H')$.

For demonstration,
let us denote the $S_z < S_z^{Max}=N_c/2$ spin projection values
by $S_z=N_c/2-\bar n$, where $\bar n =1,2,...,N_c$ holds. In this 
case $|\Psi_g(N^*,\bar n)\rangle$ (the ground state in the
$S=N_c/2, \: S_z=N_c/2-\bar n$ spin sector) can be obtained 
from $|\Psi_g(N^*)\rangle$ by the successive
application of the total lowering operator ${\hat S}_{-}$ as 
follows
\begin{eqnarray}
|\Psi_g(N^*,\bar n)\rangle= ({\hat S}_{-})^{\bar n}
|\Psi_g(N^*)\rangle, \quad {\hat S}_{-}=\sum_{
{\bf i}} \sum_{n=1}^6 {\hat S}_{
{\bf i}+{\bf r}_n,-},
\label{B1}
\end{eqnarray}
where $\hat S_{{\bf j},-}$ for an arbitrary site ${\bf j}$ is given
in the last line of (\ref{EQU60}).

The fact that $|\Psi_g(N^*,\bar n)\rangle$ is indeed 
contained in the ground state manyfold $ker(\hat H')$ results
from the fact that $|\Psi_g(N^*)\rangle =
|\Psi_g(N^*,\bar n=0)\rangle \in ker(\hat H')$, and the following
two properties:

i) One has
\begin{eqnarray}
\hat P_{\bf j} \hat S_{{\bf j}',-}=  
\hat S_{{\bf j}',-}\hat P_{\bf j}(1-\delta_{
{\bf j},{\bf j}'}), 
\label{B2}
\end{eqnarray}
hence $\hat H_P|\Psi_g(N^*,\bar n)
\rangle=0$ holds. 

ii) One obtains
\begin{eqnarray}
\hat G^{\dagger}_{\alpha,{\bf i},\sigma}
{\hat S}_{-}={\hat S}_{-}\hat G^{\dagger}_{\alpha,{\bf i},
\sigma} -\delta_{\sigma,\uparrow} \hat G^{\dagger}_{\alpha,
{\bf i},-\sigma},
\label{B3}
\end{eqnarray}
consequently, $\hat H_G |\Psi_g(N^*,\bar n)\rangle=0$ also
holds. As a consequence, for all $\bar n \in [1,N_c]$,
$|\Psi_g(N^*,\bar n)\rangle$ is contained in $ker(\hat H')$.
In conclusion, Lemma 6. has been demonstrated.

As an observation one mentions that
taking into account (\ref{B3}) from where
${\hat S}_{-}\hat G^{\dagger}_{\alpha,{\bf i},\sigma}= 
\hat G^{\dagger}_{\alpha,{\bf i},\sigma}{\hat S}_{-} +
\delta_{\sigma,\uparrow} \hat G^{\dagger}_{\alpha,
{\bf i},-\sigma}$, and the equality 
$\hat G^{\dagger}_{\alpha,{\bf i},\sigma}
\hat G^{\dagger}_{\alpha,{\bf i},\sigma}=0$,
it results that [see (\ref{EQU86})]
\begin{eqnarray}
{\hat S}_{-} [(\prod_{\sigma} \prod_{{\bf i}} 
\prod_{\alpha=1}^5 \hat G^{\dagger}_{\alpha,{\bf i},\sigma})
\hat W_G^{\dagger}] |0\rangle &=&
[(\prod_{\sigma} \prod_{{\bf i}}
\prod_{\alpha=1}^5 \hat G^{\dagger}_{\alpha,{\bf i},\sigma})
({\hat S}_{-} \hat W_G^{\dagger})] |0\rangle 
\nonumber\\
&=&
[(\prod_{\sigma} \prod_{{\bf i}} 
\prod_{\alpha=1}^5 \hat G^{\dagger}_{\alpha,{\bf i},\sigma})
{\hat W}_{G,1}^{\dagger}] |0\rangle,
\label{B4}
\end{eqnarray}
where 
\begin{eqnarray}
{\hat W}_{G,1}^{\dagger}= {\hat S}_{-}\hat W_G^{\dagger}.
\label{B5}
\end{eqnarray}
As can be seen, $ker(H_G)$ remains the same for all $S_z \leq 
S_z^{Max}$, and the genuine cause for the emergence of the
$S=N_c/2$ ferromagnetic state is the requirement to have the 
ground state placed also in $ker(\hat H_P)$.

\subsubsection{The kernel $ker(\hat H')$ in the light of the obtained
results}

The summary of all obtained results relating $ker(\hat H')$ is
contained in Lemma 7:

{\bf Lemma 7}: The kernel $ker(\hat H')$ at $N=N^{*}$ is 
$2S+1=2(N_c/2)+1=N_c+1$ dimensional and contains the linear 
independent wave
vectors $|\Psi_g(N^*,\bar n)\rangle$, where 
$\bar n=0,1,2,...N_c$.
Consequently, at $N=N^*$, apart from the trivial $2S+1$ fold
degeneracy related to the orientation of the total spin,
the ground state $|\Psi_g(N^*)\rangle$  is unique.

Indeed, acording to Lemma 5, states with $S < N_c/2$ are not 
contained in $ker(\hat H')$. Furthermore, based on Lemma 4,
at $S=S_z^{Max}=N_c/2$, 
the kernel $ker(\hat H')$ contains the
unique $|V_{G,P}\rangle=|\Psi_g(N^*)\rangle$ vector. Finally,
Lemma 6 shows that all $S=N_c/2$, $S_z < S_z^{Max}$ spin sectors
of $|\Psi_g(N^*)\rangle$ are also contained in $ker(\hat H')$.
Consequently, the $|\Psi_g(N^*)\rangle$ state holding the total
system spin $S=N_c/2$, apart from the trivial $2S+1$ fold
degeneracy related to the orientation of the total spin, is 
unique. Hence Lemma 7 has been demonstrated.


\section{Study of $ker(\hat H')$  for  $S<N_c/2$}

\subsection{Introduction and motivations}

Here we provide details of a supplementary proof of 
Lemma 5. Namely, we
show with the use of three Corollaries, that for arbitrary
number of spin flips in $\hat F^{\dagger}$, (\ref{EQU90}),
the resulting vector of the form (\ref{EQU89}) is never
contained in $ker(\hat H')$. We start with one spin flip,
$n'=1$ (Corollary I). Then the number of spin flips is
increased to $1 < n' \leq N_c/2$ (Corollary II), finally 
we analyze the case $N_c/2 < n' < N_c$ (Corollary III).

\subsection{One spin flip in $\hat F^{\dagger}$}
 
{\it Corollary I.}:
Starting from $\hat F^{\dagger}$, (\ref{EQU90}), and
adding one flipped spin in an arbitrary cell ${\bf i}_1$
we obtain the quantity
\begin{eqnarray}
\hat F^{\dagger}_1= \hat c^{\dagger}_{{\bf i}_1 + 
{\bf r}_{n_{{\bf i}_1}},-\sigma} \prod_{{\bf i}=1, 
{\bf i} \ne {\bf i}_1}^{N_c} 
\hat c^{\dagger}_{{\bf i}+{\bf r}_{n_{\bf i}},\sigma}.
\label{EQU92}
\end{eqnarray}
The vector $|V'_{G,P}\rangle$, (\ref{EQU91}), constructed 
with (\ref{EQU92}) is then not contained in 
$ker(\hat H_P)$ and hence also not in $ker(\hat H')$.

{\it Proof of Corollary I}:
When $\hat F^{\dagger}_1$ acts on the vacuum state in
(\ref{EQU91}), one obtains 5 empty sites in the cell
$I_{{\bf i}_1}$ defined at the site ${\bf i}_1$,
and a site, say ${\bf i}_1+{\bf r}_{n_1}$ with one $-\sigma$
electron. Hence by acting with 
$\prod_{\alpha=1}^5 \hat G^{\dagger}_{\alpha,
{\bf i}_1,\sigma}$ on the state thus obtained, the product
$\prod_{\alpha=1}^5 \hat G^{\dagger}_{\alpha,
{\bf i}_1,\sigma}$ can put a spin $\sigma$ electron on 
${\bf i}_1+{\bf r}_{n_1}$, creating here a double 
occupacion. In this case, the above mentioned product 
further introduces four spin $\sigma$ electrons on other 
four sites in $I_{{\bf i}_1}$. Since there are six sites
per cell, this leaves one empty site in this cell, say at
${\bf i}_1+{\bf r}_{n_{1,e}}$.

Altogether, the vector 
\begin{eqnarray}
|\chi_{\sigma}\rangle = [ \prod_{{\bf i}} 
\prod_{\alpha=1}^5 \hat G^{\dagger}_{\alpha,{\bf i},
\sigma} ] \hat F^{\dagger}_1 |0\rangle ,
\label{EQU93}
\end{eqnarray}
has $\sigma$ electrons on $6N_c -2$ sites, one double 
occupied site at ${\bf i}_1+{\bf r}_{n_1}$, and one 
empty site at ${\bf i}_1+{\bf r}_{n_{1,e}}$. The latter two
sites are located inside the cell $I_{{\bf i}_1}$.

Since  $|V'_{G,P}\rangle = [ \prod_{{\bf i}} 
\prod_{\alpha=1}^5 \hat G^{\dagger}_{\alpha,{\bf i},
-\sigma} ] |\chi_{\sigma}\rangle$,
the product $[ \prod_{{\bf i}} \prod_{\alpha=1}^5
\hat G^{\dagger}_{\alpha,{\bf i},-\sigma} ]$ introduces
into $|\chi_{\sigma}\rangle$ additional $5N_c$ many
$-\sigma$ electrons on $6N_c-1$ possible sites (the 
position of the double occupation is excluded). Since 
the number of possible sites is much higher than the 
number of additional $-\sigma$ electrons, 
$|V'_{G,P}\rangle$ will include at least one term 
containing an empty site at ${\bf i}_1+{\bf r}_{n_{1,e}}$.

As a consequence $|V'_{G,P}\rangle$ from (\ref{EQU91}) 
constructed with (\ref{EQU92}) will not be contained in 
$ker(\hat H_P)$, and therefore also not in $ker(\hat H')$.

Thus Corollary I. has been proved.

\subsection{$1 < n' \leq N_c/2$ spin flips in 
$\hat F^{\dagger}$}

{\it Corollary II.}: 
Starting from $\hat F^{\dagger}$, (\ref{EQU90}), and
adding $1 < n' \leq N_c/2$ many flipped spins in arbitrary 
cells located at ${\bf i}_1, {\bf i}_2, ..., {\bf i}_{n'}$,
one obtains
\begin{eqnarray}
\hat F^{\dagger}_1= \: [ \: \prod_{\beta=1}^{n'} 
\hat c^{\dagger}_{{\bf i}_{
\beta} + {\bf r}_{n_{{\bf i}_{\beta}}},-\sigma} \: ] \: 
[ \: 
\prod_{{\bf i}=1, {\bf i} \ne {\bf i}_1,...,{\bf i}_{n'}}^{
N_c} \hat c^{\dagger}_{{\bf i}+{\bf r}_{n_{\bf i}},\sigma} 
\: ].
\label{EQU94}
\end{eqnarray}
The vector $|V'_{G,P}\rangle$, (\ref{EQU91}), constructed 
with (\ref{EQU94}) contains terms with empty sites. 
Consequently is not contained in $ker(\hat H_P)$, and 
hence also not in $ker(\hat H')$.

{\it Proof of Corollary II.}:
To prove Corollary II, we observe that now the vector
\begin{eqnarray}
|\chi_{\sigma}^{(n')}\rangle = [ \prod_{{\bf i}} 
\prod_{\alpha=1}^5 \hat G^{\dagger}_{\alpha,{\bf i},\sigma}
] \hat F^{\dagger}_1 |0\rangle ,
\label{EQU95}
\end{eqnarray}
constructed with $\hat F^{\dagger}_1$ from (\ref{EQU94}), 
will -- in contrast to (\ref{EQU93}) used for Corollary I
-- contain terms containing $n'$ double occupations in $n'$ 
many cells, $I_{{\bf i}_1},I_{{\bf i}_2},...,
I_{{\bf i}_{n'}}$, i.e., on the sites ${\bf i}_1+
{\bf r}_{n_{1}}, {\bf i}_2+{\bf r}_{n_{2}}, ...,
{\bf i}_{n'}+{\bf r}_{n_{n'}}$. Furthermore, these terms
will also contain $n'$ many empty sites in the same cells,
i.e. at positions ${\bf i}_1+{\bf r}_{n_{1,e}}, {\bf i}_2
+{\bf r}_{n_{2,e}}, ..., {\bf i}_{n'}+{\bf r}_{n_{n',e}}$.

In this situation the vector from (\ref{EQU91}) takes
the form
\begin{eqnarray}
|V'_{G,P}\rangle = [ \prod_{{\bf i}} 
\prod_{\alpha=1}^5 \hat G^{\dagger}_{\alpha,{\bf i},
-\sigma} ] |\chi^{(n')}_{\sigma}\rangle ,
\label{EQU96}
\end{eqnarray}
where  again $[ \prod_{{\bf i}} \prod_{\alpha=1}^5
\hat G^{\dagger}_{\alpha,{\bf i},-\sigma} ]$ introduces 
into $|\chi^{(n')}_{\sigma}\rangle$ $5N_c$ many $-\sigma$
electrons, but now on $6N_c-n' \geq 5N_c + N_c/2 > 5N_c$ 
sites. As can be seen, at least $N_c/2$ many sites are not
filled, so that at least one of the $n'$ empty sites,
$n' \leq N_c/2$, will remain in (\ref{EQU96}), and hence 
also in (\ref{EQU91}). Since (\ref{EQU91}) contains terms 
with at least one empty site, it follows that 
$|V'_{G,P}\rangle$ is not contained in $ker(\hat H_P)$, 
and hence also not in $ker(\hat H')$.

Thus Corollary II. has been proved.

\subsection{$N_c/2 < n' < N_c$ spin flips in 
$\hat F^{\dagger}$}

{\it Corollary III.}:
When $\hat F^{\dagger}_1$ in (\ref{EQU91}) contains 
$N_c/2 < n' < N_c$ many flipped spins the resulting
vector $|V'_{G,P}\rangle$, (\ref{EQU91}), contains terms 
with empty sites. Consequently it is not contained in 
$ker(\hat H_P)$ and hence also not in $ker(\hat H')$.

{\it Proof of Corollary III.}:
The proof automatically results from Corollary I. and 
Corollary II. by interchanging the spins, i.e. 
$\sigma \to -\sigma$, $-\sigma \to \sigma$.

Thus, Corollary III. has been proved.


\section{Details of the calculation of different energy
expressions at $N_c$=2}
\subsection{Calculations in terms of the non-interacting wave 
function}

At $U_n=0$, after deducing the bare band structure by
diagonalizing the matrix $\tilde M$ 
[see (\ref{EQU7}),(E9),(F6)], we must express besides
its eigenvalues (i.e. the bare band structure), also
the corresponding eigenvectors.
This means that for all bands $b=1,2,..., N_b$, for the
eigenvalues $\epsilon_{b,{\bf k},\sigma}$, we express also
the corresponding eigenfunctions $|\Psi_{b,{\bf k},\sigma}
\rangle$. Note that if ${\bf a}$ denotes the Bravais 
vector of the chain, at $N_c=2$ only two $k={\bf k}{\bf a}$
values exist, namely $k=0,\pi$. 

In the non-interacting case $\langle \hat n_{{\bf i}+
{\bf r}_n,\sigma} \rangle=\langle \hat n_{{\bf i}+
{\bf r}_n,-\sigma} \rangle$ is deduced based on the
relation
\begin{eqnarray}
\langle \hat n_{{\bf i}+{\bf r}_n,\sigma} \rangle =
\sum_{k=0,\pi} \sum_{b=1}^{N_b-1} \langle \Psi_{b,k,\sigma}|
\hat n_{{\bf i}+{\bf r}_n,\sigma} |\Psi_{b,k,\sigma}
\rangle + \langle \Psi_{b=N_b,k_m,\sigma}|
\hat n_{{\bf i}+{\bf r}_n,\sigma} |\Psi_{b=N_b,k_m,\sigma}
\rangle,
\label{eqNA1}
\end{eqnarray}
where $k_m$ represents the momentum value $0$ or $\pi$
at which $\epsilon_{b=N_b,k_m,\sigma}$ is minimum at fixed
$b=N_b$. We note that $N_b=6$ for the pentagon case
with external 
links (see Fig.5), $N_b=4$ for the pentagon case without 
external links (see Fig.8), while in the studied triangular case
$N_b=2$ (see Fig.9). 

Effectively the calculation is performed in 
${\bf k}$ space using the Fourier transformed Fermi 
operators $\hat c_{n,{\bf k},\sigma}$ obtained via
$\hat c_{{\bf i}+{\bf r}_n,\sigma}=(1/\sqrt{N_c})
\sum_{\bf k} e^{-i{\bf k}({\bf i}+{\bf r}_n)} \hat c_{n,
{\bf k},\sigma}$. As a result, in the averages from 
the right side of (\ref{eqNA1}), 
instead of $\hat n_{{\bf i}+{\bf r}_n,\sigma}$ only the
expression
$(1/2)(\hat c^{\dagger}_{n,0,\sigma}\hat c_{n,0,\sigma} +
\hat c^{\dagger}_{n,\pi,\sigma} \hat c_{n,\pi,\sigma})$
appears, and one obtains
\begin{eqnarray}
\langle \hat n_{{\bf i}+{\bf r}_n,\sigma} \rangle &=& 
\frac{1}{2} \Big[
\sum_{k=0,\pi} \sum_{b=1}^{N_b-1} \langle \Psi_{b,k,\sigma}|
(\hat c^{\dagger}_{n,0,\sigma}\hat c_{n,0,\sigma} +
\hat c^{\dagger}_{n,\pi,\sigma} \hat c_{n,\pi,\sigma})
|\Psi_{b,k,\sigma}\rangle 
\nonumber\\
&+& \langle \Psi_{b=N_b,k_m,\sigma}|
(\hat c^{\dagger}_{n,0,\sigma}\hat c_{n,0,\sigma} +
\hat c^{\dagger}_{n,\pi,\sigma} \hat c_{n,\pi,\sigma})
|\Psi_{b=N_b,k_m,\sigma}\rangle \Big],
\label{eqNA2}
\end{eqnarray}

Once $\langle \hat n_{{\bf i}+{\bf r}_n,\sigma} \rangle$
has been deduced, using (\ref{eqNA2}),
the interaction energy $E_{0,int}$ becomes
\begin{eqnarray}
E_{0,int}=\sum_{all \: {\bf j}}U_{\bf j} \langle \hat n_{
{\bf j},\sigma} \hat n_{{\bf j},-\sigma}\rangle =
\sum_{all \: {\bf j}}U_{\bf j} \langle \hat n_{
{\bf j},\sigma} \rangle \langle  \hat n_{{\bf j},-\sigma}
\rangle =\sum_{all \: {\bf j}}U_{\bf j} \langle \hat n_{
{\bf j},\sigma} \rangle^2.
\label{eqNA3}
\end{eqnarray}

The kinetic energy $E_{0,kin}$ is deduced based on the
obtained $\epsilon_{b,k,\sigma}$ eigenvalues 
providing the bare band structure 
\begin{eqnarray}
E_{0,kin}=\sum_{\sigma} \Big[
\sum_{k=0,\pi} \sum_{b=1}^{N_b-1}\epsilon_{b,k,\sigma} +
\epsilon_{b=N_b,k_m,\sigma}\Big].
\label{eqNA4}
\end{eqnarray}
Finally, the total energy becomes
$E_{0,g}=E_{0,kin}+E_{0,int}$.

\subsection{Calculations in terms of the exact interacting 
wave function}

In the case of the half filled upper band exact ground state
one has for the ground state wave function
\begin{eqnarray}
|\Psi_g\rangle =\prod_{\bf k} \prod_{n=1}^{N_b}
\hat c^{\dagger}_{n,{\bf k},\sigma} \prod_{\alpha=1}^{
N_{\alpha}} \hat G^{\dagger}_{\alpha,{\bf k},-\sigma} 
|0\rangle,
\label{eqNA5}
\end{eqnarray}
where $\sigma$ is fixed, and $N_{\alpha}=N_b-1$ represents 
the number of different $\hat G_{\alpha,{\bf k},\sigma}$ 
operators. Besides $|\Psi_g\rangle$, the corresponding
ground state energy $E_g$ is known, being provided by the
method used [see below (D9)]. 

In order to calculate $E_{int}$ one needs the value of
$\langle \hat n_{{\bf i}+{\bf r}_n,\downarrow} \rangle$.
The calculation is done in ${\bf k}$ space, and as in the 
case of (\ref{eqNA2}), one findes
\begin{eqnarray}
\langle \hat n_{{\bf i}+{\bf r}_n,\downarrow} \rangle =
\langle \frac{1}{2}(\hat c^{\dagger}_{n,0,\downarrow}
\hat c_{n,0,\downarrow} + \hat c^{\dagger}_{n,\pi,
\downarrow} \hat c_{n,\pi,\downarrow})\rangle,
\label{eqNA6}
\end{eqnarray}
where now, the expectation value must be deduced in terms 
of (\ref{eqNA5}). One finds as a result
\begin{eqnarray}
\langle \hat n_{{\bf i}+{\bf r}_n,\downarrow} \rangle = 1 -
\frac{1}{2} \sum_{k=0,\pi} \frac{F_n(k)}{B(k)},
\label{eqNA7}
\end{eqnarray}
where $F_n(k)$ and $B(k)$, as a generalization of the
results from Appendix A, are given by
\begin{eqnarray}
&&B(k)=\langle 0| \prod_{\alpha=1}^{N_{\alpha}} 
\hat G_{\alpha,k,\downarrow}|\prod_{\alpha=1}^{N_{\alpha}}
\hat G^{\dagger}_{\alpha,k,\downarrow}|0\rangle,
\nonumber\\
&&F_n(k)=\langle 0|\hat G_{N_{\alpha},k,\downarrow}...
\hat G_{2,k,\downarrow}\hat G_{1,k,\downarrow}|
\hat c_{n,k,\downarrow} \hat c^{\dagger}_{n,k,\downarrow}|
\hat G_{1,k,\downarrow}\hat G_{2,k,\downarrow}...\hat G_{
N_{\alpha},k,\downarrow}|0\rangle.
\label{eqNA8}
\end{eqnarray}
The effective calculation is done via 
$B(k)=Det(\tilde M_B)$ and $F_n(k)=Det(\tilde F)$. 
Here the $N_{\alpha}\times N_{\alpha}$
matrix $\tilde M_B=\{d_{\alpha,\alpha'}\}$ has the 
matrix elements
$d_{\alpha,\alpha'}=\{\hat G_{\alpha,{\bf k},-\sigma},
\hat G^{\dagger}_{\alpha',{\bf k},-\sigma}\}$. In the same 
time, the $N_b\times N_b$ matrix $\tilde F$, is constructed
from the matrix $\tilde M_B$ by extending it with a new 
first row $(F_{1,1},F_{1,2},...,F_{1,N_b})$ and new column
with elements $F_{\nu,1}, \nu=1,...,N_b$. Here one has
$F_{\nu,1}=F_{1,\nu}^*$, $F_{1,1}=1$, and $F_{1,1+\nu}=
\{\hat c_{n,k,-\sigma}, \hat G^{\dagger}_{\nu,k,-\sigma} 
\}$. Furthermore, for $m,m' \geq 2$ one has 
$F_{m,m'}=d_{m-1,m'-1}$. Exemplification for $N_b=6$ is present in Appendix D [see also (A1)].

Once $\langle \hat n_{{\bf i}+{\bf r}_n,\downarrow}\rangle$
has been deduced, taking into account that for an arbitrary
${\bf j}$ one has $\hat n_{{\bf j},\uparrow}|\Psi_g\rangle=
|\Psi_g\rangle$, one finds
\begin{eqnarray}
E_{int}=\sum_{all \: {\bf j}}U_{\bf j} \langle \hat n_{
{\bf j},\sigma} \hat n_{{\bf j},-\sigma}\rangle =
\sum_{all \: {\bf j}}U_{\bf j} \langle \hat n_{{\bf j},
\downarrow}\rangle,
\label{eqNA9}
\end{eqnarray}
where for average, (\ref{eqNA7}) has to be used.

When $E_{int}$ has been deduced and $E_g$ is known, one has
for the kinetic energy $E_{kin}=E_g-E_{int}$. We note that
the deduced $E_g$ values are present below (\ref{EQU13}) for the
pentagon case with external links ($E_g=C_g$), in (E12) for the
pentagon case without external links, and in (F11) for the 
studied triangle case.


\section{The pentagon chain without external links, 
antennas and $\epsilon_n$ values}

\subsection{The chain}

We analyze now two pentagon cells without external links, 
antennas, and on-site one-particle potentials $\epsilon_n$,
with periodic boundary conditions as shown in Fig.7.

The number of cells
$N_c=2$, the number of sites per cell (see Fig.7) is 4,
consequently the number of bands $N_b=4$, the
total number of sites $N_{\Lambda}=2*4=8$, and the maximum 
possible number of electrons in the system is $N_{Max}=16$
(in a given band, with both spin indices, the total 
possible number of electrons is 4).
One has in the chain $N=7*Nc=14$ electrons which 
corresponds to half filled upper band. Consequently in the
noninteracting case one has 
$N^{(0)}_{\uparrow}=N^{(0)}_{\downarrow}=7$. In the
interacting case at half filling, in the interacting ground
state one obtains
$N_{\uparrow}=8, N_{\downarrow}=6$ in the whole system.

\subsection{The Hamiltonian in ${\bf r}$ space}

The Hamiltonian becomes (notations from Fig.7):
\begin{eqnarray}
\hat H_0 &=& \sum_{\sigma} \sum_{{\bf i}=1}^{N_c} \: \big\{
\: \big[ \: t ( \hat c^{\dagger}_{{\bf i}+{\bf r}_4,\sigma}
\hat c_{{\bf i}+{\bf r}_1,\sigma} +
\hat c^{\dagger}_{{\bf i}+{\bf a},\sigma} \hat c_{{\bf i}+
{\bf r}_4,\sigma}) +
t_h \hat c^{\dagger}_{{\bf i}+{\bf r}_2,\sigma} 
\hat c_{{\bf i}+{\bf r}_3,\sigma} 
\nonumber\\
&+& t' ( \hat c^{\dagger}_{{\bf i}+{\bf r}_1,\sigma} 
\hat c_{{\bf i} +{\bf r}_2,\sigma} +
\hat c^{\dagger}_{{\bf i}+{\bf r}_3,\sigma} 
\hat c_{{\bf i}+{\bf a},\sigma}) + H.c. \big]
\nonumber\\
&+& \epsilon_4 \hat n_{{\bf i}+{\bf r}_4,\sigma} +
+ \epsilon_2 (\hat n_{{\bf i}+{\bf r}_2,\sigma} + 
\hat n_{{\bf i}+{\bf r}_3,\sigma}) + \epsilon_1 
\hat n_{{\bf i}+{\bf r}_1,\sigma}) 
\big\} ,
\label{E9}
\end{eqnarray}
where $N_c$ represents the number of celles. There are 
totally $N_{\Lambda}=4 N_c$ sites into the system, and the 
total filling corresponds to $N_{tot}=8 N_c$. 
One further has $\epsilon_2=\epsilon_3$, and for convenience
one takes ${\bf r}_1=0$.
 
The interacting part of the Hamiltonian is taken as 
\begin{eqnarray}
\hat H_U = \sum_{{\bf i}=1}^{N_c} [
U_4 \hat n_{{\bf i}+{\bf r}_4,\uparrow} \hat n_{{\bf i}
+{\bf r}_4,\downarrow}+ U_2 (\hat n_{{\bf i}+{\bf r}_2,
\uparrow} \hat n_{{\bf i}+{\bf r}_2,\downarrow}+
\hat n_{{\bf i}+{\bf r}_3,\uparrow} \hat n_{{\bf i}
+{\bf r}_3,\downarrow}) +
U_1 \hat n_{{\bf i},\uparrow} \hat n_{{\bf i},\downarrow}],
\label{E10}
\end{eqnarray}
where one has $U_2=U_3$. The total Hamiltonian becomes
\begin{eqnarray}
\hat H= \hat H_0 + \hat H_U.
\label{E11}
\end{eqnarray} 

\subsection{$\hat H_0$ in momentum space}

For the Fourier sum of the introduced Fermi operators one 
uses as before
$\hat c_{{\bf i}+{\bf r}_n,\sigma}= (1/\sqrt{N_c}) 
\sum_{{\bf k}=1}^{N_c}
e^{-i{\bf k}{\bf i}} e^{-i {\bf k}{\bf r}_n} 
\hat c_{n,{\bf k},\sigma}$,
where ${\bf k}$ is directed along the line of the chain 
($x$-axis), and one has
$|{\bf k}|= k = 2 m \pi/(a N_c)$, $m=0,1,2,...,N_c-1$, 
$|{\bf a}|=a$ being the
lattice constant. The noninteracting part of the 
Hamiltonian becomes
\begin{eqnarray}
\hat H_0 &=& \sum_{\sigma} \sum_{{\bf k}=1}^{N_c} \: 
\big\{ \: \big[ \: t \big (
\hat c^{\dagger}_{4,{\bf k},\sigma} \hat c_{1,{\bf k},
\sigma} e^{i {\bf k}{\bf r}_4} +
\hat c^{\dagger}_{1,{\bf k},\sigma} \hat c_{4,{\bf k},
\sigma} e^{i {\bf k}({\bf a}-{\bf r}_4)} \big) +
t_h \hat c^{\dagger}_{2,{\bf k},\sigma} \hat c_{3,{\bf k},
\sigma} e^{i {\bf k}({\bf r}_2-{\bf r}_3)}  
\nonumber\\
&+& t' \big( 
\hat c^{\dagger}_{1, {\bf k},\sigma} \hat c_{2, {\bf k},
\sigma} e^{-i {\bf k}{\bf r}_2} +
\hat c^{\dagger}_{3, {\bf k},\sigma} \hat c_{1, {\bf k},
\sigma} e^{i {\bf k}({\bf r}_3-{\bf a})} \big)  + H.c. \big]
\nonumber\\ 
&+& \epsilon_4 \hat n_{4,{\bf k},\sigma} + 
\epsilon_2 (\hat n_{2, {\bf k},\sigma} + 
\hat n_{3, {\bf k},\sigma}) +
\epsilon_1 \hat n_{1, {\bf k},\sigma} \big\} .
\label{E12}
\end{eqnarray}
From (\ref{E12}), taking real hopping matrix elements, 
one finds
\begin{eqnarray}
\hat H_0 &=& \sum_{\sigma} \sum_{{\bf k}=1}^{N_c} \: 
\big\{ \: \big[ \:
2t \cos \frac{{\bf k} {\bf a}}{2} e^{i {\bf k}({\bf r}_4
- {\bf a}/2)}
\hat c^{\dagger}_{4,{\bf k},\sigma} \hat c_{1,{\bf k},
\sigma} +
2t \cos \frac{{\bf k} {\bf a}}{2} e^{-i {\bf k}({\bf r}_4
- {\bf a}/2)}
\hat c^{\dagger}_{1,{\bf k},\sigma} \hat c_{4,{\bf k},
\sigma} 
\nonumber\\
&+& t_h e^{i {\bf k}({\bf r}_2- {\bf r}_3)}
\hat c^{\dagger}_{2,{\bf k},\sigma} \hat c_{3,{\bf k},
\sigma} +
t_h e^{-i {\bf k}({\bf r}_2- {\bf r}_3)}
\hat c^{\dagger}_{3,{\bf k},\sigma} \hat c_{2,{\bf k},
\sigma} 
\nonumber\\
&+& t' e^{-i {\bf k}{\bf r}_2}
\hat c^{\dagger}_{1,{\bf k},\sigma} \hat c_{2,{\bf k},
\sigma} + t' e^{i {\bf k}{\bf r}_2}
\hat c^{\dagger}_{2,{\bf k},\sigma} \hat c_{1,{\bf k},
\sigma} 
\nonumber\\
&+& t' e^{i {\bf k}({\bf r}_3- {\bf a})}
\hat c^{\dagger}_{3,{\bf k},\sigma} \hat c_{1,{\bf k},
\sigma} +
t' e^{-i {\bf k}({\bf r}_3- {\bf a})}
\hat c^{\dagger}_{1,{\bf k},\sigma} \hat c_{3,{\bf k},
\sigma} \big] 
\nonumber\\
&+& \epsilon_4 \hat n_{4,{\bf k},\sigma} + 
\epsilon_2 (\hat n_{2, {\bf k},\sigma} + \hat n_{3, 
{\bf k},\sigma}) +
\epsilon_1 \hat n_{1, {\bf k},\sigma} \big\} .
\label{E13}
\end{eqnarray}
Using the length notations $b=|{\bf r}_3-{\bf r}_2|, 
a=|{\bf a}|$, where ${\bf a}$ is the Bravais vector, one 
has in the exponents present in 
(\ref{E13}) the expressions
\begin{eqnarray}
&&{\bf k}({\bf r}_4-\frac{{\bf a}}{2}) =0, \quad
{\bf k}({\bf r}_2-{\bf r}_4)= - k b, \quad
{\bf k}{\bf r}_2= \frac{k(a-b)}{2}, 
\nonumber\\
&&{\bf k}({\bf r}_3-{\bf a})= -\frac{k(a-b)}{2}, \quad
{\bf k}{\bf a}= ka .
\label{E14}
\end{eqnarray}
From Eqs.(\ref{E13}-\ref{E14}) one obtains for the 
noninteracting part
\begin{eqnarray}
\hat H_0 &=& \sum_{\sigma} \sum_{k = 1}^{N_c} \: \big\{ \: 
\big[ \: 
2t \cos \frac{ak}{2} ( \hat c^{\dagger}_{4,{\bf k},\sigma} 
\hat c_{1,{\bf k},\sigma} + \hat c^{\dagger}_{1,{\bf k},
\sigma} \hat c_{4,{\bf k},\sigma} ) + 
t_h e^{- i k b} \hat c^{\dagger}_{2,{\bf k},\sigma} 
\hat c_{3,{\bf k},\sigma} + 
\nonumber\\
&+& t_h e^{i k b} \hat c^{\dagger}_{3, {\bf k},\sigma} 
\hat c_{2, {\bf k},\sigma} + t'e^{- i \frac{k(a-b)}{2}} 
\hat c^{\dagger}_{1, {\bf k},\sigma} 
\hat c_{2, {\bf k},\sigma} + t'e^{i \frac{k(a-b)}{2}} 
\hat c^{\dagger}_{2, {\bf k},\sigma} \hat c_{1, {\bf k},
\sigma} + t'e^{- i \frac{k(a-b)}{2}}
\hat c^{\dagger}_{3, {\bf k},\sigma}\hat c_{1, {\bf k},
\sigma} +
\nonumber\\ 
&+& t'e^{ i \frac{k(a-b)}{2}} \hat c^{\dagger}_{1, {\bf k},
\sigma} \hat c_{3, {\bf k},\sigma} \big]  +
\epsilon_4 \hat n_{4,{\bf k},\sigma} + \epsilon_1 
\hat n_{1, {\bf k},\sigma} + \epsilon_2 (\hat n_{2, 
{\bf k},\sigma} + \hat n_{3, {\bf k},\sigma}) \big\} . 
\label{E15}
\end{eqnarray}
The band structure is obtained by diagonalizing $\hat H_0$.
One has
\begin{eqnarray}
\hat H_0= \sum_{\sigma} \sum_{k=1}^{N_c} (\hat c^{
\dagger}_{4,k,\sigma},\hat c^{\dagger}_{1,k,\sigma}, 
\hat c^{\dagger}_{2,k,\sigma}, \hat c^{\dagger}_{3,k,
\sigma} ) \tilde M
\left( \begin{array}{c}
\hat c_{4,k,\sigma} \\
\hat c_{1,k,\sigma} \\
\hat c_{2,k,\sigma} \\
\hat c_{3,k,\sigma} \\
\end{array} \right) ,
\label{E16}
\end{eqnarray}
where $\tilde M$, based on (\ref{E15}) has the form
\begin{eqnarray}
\tilde M =
\left( \begin{array}{cccc}
\epsilon_4 & 2t \cos \frac{ka}{2} & 0 & 0 \\
2t \cos \frac{ka}{2} & \epsilon_1 & t' e^{- i \frac{k(a-b)}{2}} & 
t' e^{i \frac{k (a-b)}{2}} \\
0 & t' e^{i \frac{k(a-b)}{2} } & \epsilon_2 & t_h e^{-i k b} \\
0 & t' e^{- i \frac{k(a-b)}{2}} & t_h e^{i k b} & 
\epsilon_2 
\end{array} \right) .
\label{E17}
\end{eqnarray}
The energies are deduced from (\ref{E17}) by 
diagonalizing 
$\tilde M$. The orthonormalized eigenvectors of  
$\tilde M$ provide the 
diagonalized canonical Fermi operators. For the 
diagonalized energies 
$\epsilon$ one finds
\begin{eqnarray}
0 &=& 2 \{ (\epsilon_4-\epsilon) t_h t'^2 - t^2
[(\epsilon_2-\epsilon)^2 - t_h^2] \} \cos a k 
\nonumber\\
&+& \{ [(\epsilon_2 - \epsilon)^2 - t_h^2][(\epsilon_4
-\epsilon)(\epsilon_1-\epsilon) - 2 t^2] - 2 t'^2 
(\epsilon_4-\epsilon)(\epsilon_2-\epsilon) \} .
\label{E18}
\end{eqnarray}

\subsection{$\hat H$ transformed in positive semidefinite
form}

One introduces three block operators
\begin{eqnarray}
&&\hat G_{1,{\bf i},\sigma}= a_{1,4} \hat c_{{\bf i}+
{\bf r}_4,\sigma} + a_{1,1} \hat c_{{\bf i},\sigma} + 
a_{1,2} \hat c_{{\bf i}+{\bf r}_2,\sigma} ,
\nonumber\\
&&\hat G_{2,{\bf i},\sigma}= a_{2,4} \hat c_{{\bf i}+
{\bf r}_4,\sigma} + a_{2,2} \hat c_{{\bf i}+{\bf r}_2,
\sigma} + a_{2,3} \hat c_{{\bf i}+{\bf r}_3,\sigma} ,
\nonumber\\  
&&\hat G_{3,{\bf i},\sigma}= a_{3,4} \hat c_{{\bf i}+
{\bf r}_4,\sigma} + a_{3,3} \hat c_{{\bf i}+{\bf r}_3,
\sigma} + a_{3,5} \hat c_{{\bf i}+{\bf a},
\sigma} .
\label{E19}
\end{eqnarray}
Using now $\hat P_{\bf i}=\hat n_{{\bf i},\uparrow} 
\hat n_{{\bf i},\downarrow} -(\hat n_{{\bf i},\uparrow} 
+ \hat n_{{\bf i},\downarrow}) +1$, and
$\hat P^{(m)}= \sum_{{\bf i}=1}^{N_c} \hat P_{{\bf i}+
{\bf r}_m}$, where $m=1,2,3,4$ is the sublattice index,
the transformed Hamiltonian becomes
\begin{eqnarray}
&&\hat H= \sum_{\sigma}\sum_{{\bf i}=1}^{N_c} \sum_{m=1}^4
\hat G_{m,{\bf i},\sigma} \hat G^{\dagger}_{m,{\bf i},
\sigma} + \sum_{m=1}^4 U_m \hat P^{(m)} + E_g,
\nonumber\\
&&E_g = - N_c (U_4+U_1+2U_2 + 2K )+ p \hat N,
\label{E20}
\end{eqnarray}
where the matching conditions are
\begin{eqnarray}
&&-t=a^*_{1,4} a_{1,1}=a^*_{3,5} a_{3,4}, \quad
-t'=a^*_{1,1} a_{1,2}=a^*_{3,3} a_{3,5}, \quad
-t_h = a^*_{2,2} a_{2,3},
\nonumber\\
&&0 = a^*_{1,4} a_{1,2} + a^*_{2,4} a_{2,2}, \quad
0 = a^*_{2,4} a_{2,3} + a^*_{3,4} a_{3,3},
\nonumber\\
&&p-(U_4+\epsilon_4) = |a_{1,4}|^2 + |a_{2,4}|^2 + 
|a_{3,4}|^2,
\nonumber\\
&&p-(U_2+\epsilon_2)= |a_{1,2}|^2 + |a_{2,2}|^2 =  
|a_{2,3}|^2 + |a_{3,3}|^2,
\nonumber\\
&&p-(U_1+\epsilon_1)= |a_{1,1}|^2 + |a_{3,5}|^2 ,
\label{E21}
\end{eqnarray}
and $K$ becomes of the form
\begin{eqnarray}
K = (|a_{1,4}|^2+|a_{1,1}|^2 +|a_{1,2}|^2)+(|a_{2,4}|^2+
|a_{2,2}|^2 + |a_{2,3}|^2)+(|a_{3,4}|^2+|a_{3,3}|^2 +
|a_{3,5}|^2) .
\label{E22}
\end{eqnarray}
Introducing the notations
\begin{eqnarray}
\epsilon_{4,X}=p-U_4-\epsilon_4,\quad 
\epsilon_{2,X}=p-U_2-\epsilon_2, \quad
\epsilon_{1,X}=p-U_1-\epsilon_1,
\label{E23}
\end{eqnarray}
one finds
\begin{eqnarray}
K=\epsilon_{4,X}+\epsilon_{1,X}+2\epsilon_{2,X}
\label{E24}
\end{eqnarray}
and one obtains
\begin{eqnarray}
&&p=(U_2+\epsilon_2+|t_h|)+\frac{1}{2}[\sqrt{A^2+8{t'}^2}
+A],
\nonumber\\
&&A=(U_1-U_2)+(\epsilon_1-\epsilon_2)-|t_h|,
\label{E25}
\end{eqnarray}
while $U_4$ given by the matching conditions becomes
\begin{eqnarray}
U_4 = p -\epsilon_4 - \frac{t^2}{t'^2 |t_h|}[(p-U_2-
\epsilon_2)^2 - t^2_h].
\label{E26}
\end{eqnarray}
Besides the condition (\ref{E26}) the solutions of the
matching equations require
\begin{eqnarray}
t_h=-|t_h|, \: \epsilon_{1,X},\epsilon_{4,X} > 0,
\: \epsilon_{2,X} > |t_h|, \: U_1,U_2,U_4 > 0,
\label{E27}
\end{eqnarray}
and the block operators from (\ref{E19}) become
\begin{eqnarray}
&&\hat G_{1,{\bf i},\sigma} = e^{i\phi_1} 
\sqrt{\epsilon_{2,X}-|t_h|}\frac{|t|}{
|t'|} [ \: \:\hat c_{{\bf i}+{\bf r}_4,\sigma} - 
\frac{t'^2}{t(\epsilon_{2,X}-|t_h|)}
\hat c_{{\bf i},\sigma} + \frac{t'}{t} \hat c_{{\bf i}+
{\bf r}_2,\sigma} ] ,
\nonumber\\
&&\hat G_{2,{\bf i},\sigma} = e^{i\phi_2} \frac{|t|}{|t'|} 
\frac{\epsilon_{2,X}
-|t_h|}{\sqrt{|t_h|}} [ \: \: \hat c_{{\bf i}+{\bf r}_4,
\sigma} +\frac{t' |t_h|}{t(\epsilon_{2,X}-|t_h|)} 
( - \hat c_{{\bf i}+{\bf r}_2,\sigma} +
\hat c_{{\bf i}+{\bf r}_3,\sigma} ) ] ,
\nonumber\\  
&&\hat G_{3,{\bf i},\sigma} = e^{i\phi_3} \sqrt{
\epsilon_{2,X}-|t_h|}\frac{|t|}{
|t'|} [ \: \:\hat c_{{\bf i}+{\bf r}_4,\sigma} - 
\frac{t'^2}{t(\epsilon_{2,X}-|t_h|)} \hat c_{{\bf i}+
{\bf a},\sigma} + \frac{t'}{t} 
\hat c_{{\bf i}+{\bf r}_3,\sigma} ] , 
\label{E28}
\end{eqnarray}
where $\phi_m$, $m=1,2,3$ are arbitrary phases.

\subsection{The ground state wave function}

The ground state wave function at $N=7N_c$ (half filled 
upper band) becomes
\begin{eqnarray}
|\Psi_g\rangle = \prod_{\sigma} \prod_{m=1}^3 \prod_{
{\bf i}=1}^{N_c} \hat G^{\dagger}_{m,{\bf i},\sigma} 
\hat F^{\dagger} |0\rangle , \quad \hat F^{\dagger} =
\prod_{{\bf i}=1}^{N_c} \hat c^{\dagger}_{{\bf i}+{\bf r}_{
m_{\bf i}},\sigma},
\label{E29}
\end{eqnarray}
where $\sigma$ is fixed. In ${\bf k}$-space, in 
non-normalized form it becomes
\begin{eqnarray}
|\Psi_g\rangle = \prod_{{\bf k}=1}^{N_c} \Big[ \Big( 
\prod_{n=1}^4 \hat c^{\dagger}_{n,{\bf k},\uparrow} \Big)
\Big(\prod_{m=1}^3\hat G^{\dagger}_{m,{\bf k},\downarrow}
\Big)|0\rangle.
\label{E30}
\end{eqnarray}
Based on (\ref{E28}), taking the unimportant phases
$\phi_m=0$ for all $m$, one has
\begin{eqnarray}
&&\hat G_{1,{\bf k},\sigma} = 
\sqrt{\epsilon_{2,X}-|t_h|}\frac{|t|}{
|t'|} [ \: \:\hat c_{4,{\bf k},\sigma}e^{-i{\bf k}
{\bf r}_1} - \frac{t'^2}{t(\epsilon_{2,X}-|t_h|)}
\hat c_{1,{\bf k},\sigma} + \frac{t'}{t} \hat c_{2,{\bf k},
\sigma}e^{-i{\bf k}{\bf r}_2} ] ,
\nonumber\\
&&\hat G_{2,{\bf k},\sigma} = \frac{|t|}{|t'|} 
\frac{\epsilon_{2,X}
-|t_h|}{\sqrt{|t_h|}} [ \: \: \hat c_{4,{\bf k},
\sigma} +\frac{t' |t_h|}{t(\epsilon_{2,X}-|t_h|)} 
( - \hat c_{2,{\bf k},\sigma}e^{-i{\bf k}{\bf r}_2} +
\hat c_{3,{\bf k},\sigma}e^{-i{\bf k}{\bf r}_3} ) ] ,
\nonumber\\  
&&\hat G_{3,{\bf k},\sigma} = \sqrt{
\epsilon_{2,X}-|t_h|}\frac{|t|}{
|t'|} [ \: \:\hat c_{4,{\bf k},\sigma}e^{-i{\bf k}{
\bf r}_4} - 
\frac{t'^2}{t(\epsilon_{2,X}-|t_h|)} \hat c_{1,{\bf k},
\sigma}e^{-i{\bf k}{\bf a}} + \frac{t'}{t} 
\hat c_{3,{\bf k},\sigma} ] . 
\label{E31}
\end{eqnarray}
The ground state wave function is ferromagnetic as in the 
case of the pentagon chain with external links.

The renormalized on-site one-particle potentials which
give the effective upper flat band become
\begin{eqnarray}
\epsilon_{4,R}=\epsilon_4+U_4-p,\quad 
\epsilon_{2,R}=\epsilon_2+U_2-p, \quad
\epsilon_{1,R}=\epsilon_1+U_1-p.
\label{E32}
\end{eqnarray}


\section{The simple triangular chain}

\subsection{The chain}

We analyze now two triangle cells without external links, 
antennas, and on-site one-particle potentials $\epsilon_n$,
with periodic boundary conditions as shown in Fig.9.

The number of cells
$N_c=2$, the number of sites per cell is 2,
consequently the number of bands $N_b=2$, the
total number of sites $N_{\Lambda}=2*2=4$, and the maximum 
possible number of electrons in the system is $N_{Max}=4N_c
=8$
(in a given band, with both spin indices, the total 
possible number of electrons is 4).
One has in the chain $N=3*Nc=6$ electrons which 
corresponds to half filled upper band. Consequently in the
noninteracting case one has 
$N^{(0)}_{\uparrow}=N^{(0)}_{\downarrow}=3$. In the
interacting case which is ferromagnetic one obtains
$N_{\uparrow}=4, N_{\downarrow}=2$ in the whole system.

\subsection{The Hamiltonian in ${\bf r}$ space}

Even if the calculations are done with $\epsilon_n=0$, one
presents below the full Hamiltonian which becomes 
(notations from Fig.9):
\begin{eqnarray}
\hat H_0 &=& \sum_{\sigma} \sum_{{\bf i}=1}^{N_c} \: \big\{
\: \big[ \: t ( \hat c^{\dagger}_{{\bf i}+{\bf r}_1,\sigma}
\hat c_{{\bf i},\sigma} +
\hat c^{\dagger}_{{\bf i}+{\bf a},\sigma} \hat c_{{\bf i}+
{\bf r}_1,\sigma}) +
t' \hat c^{\dagger}_{{\bf i}+{\bf a},\sigma} 
\hat c_{{\bf i},\sigma} + H.c. \big] 
\nonumber\\
&+& \epsilon_1 \hat n_{{\bf i}+{\bf r}_1,\sigma} +
+ \epsilon_2 \hat n_{{\bf i},\sigma} \big\} ,
\label{EE9}
\end{eqnarray}
where $N_c$ represents the number of celles and ${\bf r}_2
=0$. The interacting part of the Hamiltonian is taken as 
\begin{eqnarray}
\hat H_U = \sum_{{\bf i}=1}^{N_c} [
U_1 \hat n_{{\bf i}+{\bf r}_1,\uparrow} \hat n_{{\bf i}
+{\bf r}_1,\downarrow}+ U_2 \hat n_{{\bf i},\uparrow} 
\hat n_{{\bf i},\downarrow} ],
\label{EE10}
\end{eqnarray}
the total Hamiltonian being
\begin{eqnarray}
\hat H= \hat H_0 + \hat H_U.
\label{EE11}
\end{eqnarray} 

\subsection{$\hat H_0$ in momentum space}

For the Fourier sum one uses
$\hat c_{{\bf i}+{\bf r}_n,\sigma}= (1/\sqrt{N_c}) 
\sum_{{\bf k}=1}^{N_c}
e^{-i{\bf k}{\bf i}} e^{-i {\bf k}{\bf r}_n} 
\hat c_{n,{\bf k},\sigma}$,
where ${\bf k}$ is directed along the line of the chain 
($x$-axis), and one has
$|{\bf k}|= k = 2 m \pi/(a N_c)$, $m=0,1,2,...,N_c-1$, 
$|{\bf a}|=a$ being the
lattice constant. The noninteracting part of the 
Hamiltonian becomes
\begin{eqnarray}
\hat H_0 &=& \sum_{\sigma} \sum_{{\bf k}=1}^{N_c} \: 
\big\{ \: \big[ \: t \big (
\hat c^{\dagger}_{1,{\bf k},\sigma} \hat c_{2,{\bf k},
\sigma} e^{i {\bf k}{\bf r}_1} +
\hat c^{\dagger}_{2,{\bf k},\sigma} \hat c_{1,{\bf k},
\sigma} e^{-i{\bf k}{\bf r}_1}+ \hat c^{\dagger}_{2,
{\bf k},\sigma} \hat c_{1,{\bf k},\sigma}
e^{i {\bf k}({\bf a}-{\bf r}_1)} + \hat c^{\dagger}_{1,
{\bf k},\sigma} \hat c_{2,{\bf k},\sigma}
e^{-i {\bf k}({\bf a}-{\bf r}_1)} \big) 
\nonumber\\
&+& t' \big(\hat c^{
\dagger}_{2,{\bf k},\sigma} \hat c_{2,{\bf k},\sigma}
e^{+i{\bf k}{\bf a}}+\hat c^{
\dagger}_{2,{\bf k},\sigma} \hat c_{2,{\bf k},\sigma}
e^{-i{\bf k}{\bf a}} \big) + 
\epsilon_1 \hat n_{1,{\bf k},\sigma} + 
\epsilon_2 \hat n_{2,{\bf k},\sigma} \big]. 
\label{EE12}
\end{eqnarray}
This can be written as
\begin{eqnarray}
\hat H_0= \sum_{\sigma} \sum_{k=1}^{N_c} (\hat c^{
\dagger}_{4,k,\sigma},\hat c^{\dagger}_{1,k,\sigma}, 
\hat c^{\dagger}_{2,k,\sigma}, \hat c^{\dagger}_{3,k,
\sigma} ) \tilde M
\left( \begin{array}{c}
\hat c_{4,k,\sigma} \\
\hat c_{1,k,\sigma} \\
\hat c_{2,k,\sigma} \\
\hat c_{3,k,\sigma} \\
\end{array} \right) ,
\label{EE16}
\end{eqnarray}
where $\tilde M$ is a $2\times 2$ matrix, which, based on 
(\ref{EE12}) has the form
\begin{eqnarray}
\tilde M =
\left( \begin{array}{cc}
\epsilon_1 & t e^{i{\bf k}{\bf r}_1}(1+e^{-i{\bf k}{\bf a}})
 \\
 t e^{-i{\bf k}{\bf r}_1}(1+e^{+i{\bf k}{\bf a}}) &
 \epsilon_2 +2t' \cos {\bf k}{\bf a}  
\end{array} \right) .
\label{EE17}
\end{eqnarray}
One obtains the diagonalized energies from (\ref{EE17}) by 
diagonalizing 
$\tilde M$. The orthonormalized eigenvectors of  
$\tilde M$ provide the 
diagonalized canonical Fermi operators. For the 
diagonalized energies 
$\lambda$ one finds the equation
\begin{eqnarray}
[\lambda^2+\epsilon_1 \epsilon_2 -\lambda (\epsilon_1+
\epsilon_2)-2t^2] +2 [t' (\epsilon_1-\lambda)-t^2]\cos k 
=0,
\label{EE18}
\end{eqnarray}
where $k={\bf k}{\bf a}$ holds.

From (\ref{EE18}) one finds flat bands where simultaneously
\begin{eqnarray}
&&t' (\epsilon_1-\lambda)-t^2=0,
\nonumber\\
&&\lambda^2+\epsilon_1 \epsilon_2 -\lambda (\epsilon_1+
\epsilon_2)-2t^2=0
\label{E18a}
\end{eqnarray}
holds. These equalities provide the flat band conditions as
\begin{eqnarray}
&&\lambda=\epsilon_1-\frac{t^2}{t'},
\nonumber\\
&&\frac{t^2}{t'}+(\epsilon_2-\epsilon_1)-2t'=0.
\label{E18b}
\end{eqnarray}

\subsection{$\hat H$ transformed in positive semidefinite
form}

One introduces one block operator (for prefactors see Fig.9)
\begin{eqnarray}
\hat G_{{\bf i},\sigma}= a_{1} \hat c_{{\bf i}+
{\bf r}_1,\sigma} + a_{2} \hat c_{{\bf i},\sigma} + 
a_{3} \hat c_{{\bf i}+{\bf a},\sigma}.
\label{EE19}
\end{eqnarray}
Using now $\hat P_{\bf i}=\hat n_{{\bf i},\uparrow} 
\hat n_{{\bf i},\downarrow} -(\hat n_{{\bf i},\uparrow} 
+ \hat n_{{\bf i},\downarrow}) +1$, and
$\hat P^{(m)}= \sum_{{\bf i}=1}^{N_c} \hat P_{{\bf i}+
{\bf r}_m}$, where $m=1,2$ is the sublattice index,
the transformed Hamiltonian becomes
\begin{eqnarray}
&&\hat H= \sum_{\sigma}\sum_{{\bf i}=1}^{N_c}
\hat G_{{\bf i},\sigma} \hat G^{\dagger}_{{\bf i},
\sigma} + \sum_{m=1}^2 U_m \hat P^{(m)} + E_g,
\nonumber\\
&&E_g = - N_c (U_1+U_2 + 2K )+ p \hat N,
\label{EE20}
\end{eqnarray}
where the matching conditions read
\begin{eqnarray}
&&-t=a^*_{1} a_{2}=a^*_{3} a_{1}, \quad
-t'=a^*_{3} a_{2}, 
\nonumber\\
&&p-(U_1+\epsilon_1) = |a_{1}|^2,
\nonumber\\
&&p-(U_2+\epsilon_2)= |a_{2}|^2 + |a_{3}|^2,
\label{EE21}
\end{eqnarray}
and $K$ becomes of the form
\begin{eqnarray}
K = |a_{1}|^2+|a_{2}|^2 +|a_{3}|^2).
\label{EE22}
\end{eqnarray}
The solutions of the matching equations become ($t'<0$ is 
required)
\begin{eqnarray}
&&a_1=\frac{|t|}{\sqrt{|t'|}},\quad 
a_2=a_3=- sign(t) \sqrt{|t'|}, \quad
K=\frac{t^2}{|t'|}+2|t'|,
\nonumber\\
&&p=U_1+\epsilon_1+\frac{t^2}{|t'|}=U_2+\epsilon_2+2|t'|.
\label{EE23}
\end{eqnarray}
The block operator is given by
\begin{eqnarray}
\hat G_{{\bf i},\sigma}=- sign(t) \sqrt{|t'|} (\hat c_{{\bf i},\sigma}+\hat c_{{\bf i}+{\bf a},\sigma})+ \frac{|t|}{
\sqrt{|t'|}} \hat c_{{\bf i}+{\bf r}_1,\sigma},
\label{EE24}
\end{eqnarray}
which in ${\bf k}$ space reads 
\begin{eqnarray}
\hat G_{{\bf k},\sigma}=- sign(t) \sqrt{|t'|} (1+e^{-i
{\bf k}{\bf a}})\hat c_{2,{\bf k},\sigma} +
\frac{|t|}{\sqrt{|t'|}} e^{-i{\bf k}{\bf r}_1}
\hat c_{1,{\bf k},\sigma} .
\label{EE25}
\end{eqnarray}

\subsection{The ground state wave function}

The ground state wave function at $N=3N_c$ (half filled 
upper band) becomes
\begin{eqnarray}
|\Psi_g\rangle = \prod_{\sigma} \prod_{{\bf i}=1}^{N_c} 
\hat G^{\dagger}_{{\bf i},\sigma} 
\hat F^{\dagger} |0\rangle , \quad \hat F^{\dagger} =
\prod_{{\bf i}=1}^{N_c} \hat c^{\dagger}_{{\bf i}+{\bf r}_{
m_{\bf i}},\sigma},
\label{EE29}
\end{eqnarray}
where $\sigma$ is fixed. In ${\bf k}$-space, in 
non-normalized form it becomes
\begin{eqnarray}
|\Psi_g\rangle = \prod_{{\bf k}=1}^{N_c} \Big[ \Big( 
\prod_{n=1}^2 \hat c^{\dagger}_{n,{\bf k},\uparrow} \Big)
\hat G^{\dagger}_{{\bf k},\downarrow} \Big]|0\rangle.
\label{EE30}
\end{eqnarray}
The ground state wave function is ferromagnetic.
The conditions for the emergence of the solution are
[see the last line of (F14)]
\begin{eqnarray}
U_2=U_1+(\epsilon_1-\epsilon_2)+\frac{t^2}{|t'|}-2|t'|, 
\quad t' < 0 .
\label{E30a}
\end{eqnarray}
The renormalized on-site one-particle potentials which
give the effective upper flat band become
\begin{eqnarray}
\epsilon_{1,R}=\epsilon_1+U_1-p=-\frac{t^2}{|t'|},\quad 
\epsilon_{2,R}=\epsilon_2+U_2-p=-2|t'|.
\label{EE32}
\end{eqnarray}
The second equalities above have been obtained from the
second line of (F14).


\end{document}